\documentclass[a4paper, 11pt]{article}
\pdfoutput=1 % if your are submitting a pdflatex (i.e. if you have
\usepackage{jcappub} % for details on the use of the package, please
\usepackage[T1]{fontenc} % if needed

\usepackage{array}
\usepackage{enumitem}
\usepackage{graphicx}
\usepackage{epstopdf, epsfig}
\usepackage{subcaption}
\usepackage{tikz}
\usetikzlibrary{intersections}
\usepackage{pgfplots}
\usepackage{ environ}
\usepackage{framed}
\usepackage{bm}
\usepackage[us,12hr]{datetime}
\usepackage{algorithmic}
\usepackage[ruled]{algorithm2e}
\usepackage[british]{babel}

\title{Cosmic Web \& Caustic Skeleton: non-linear Constrained Realizations -- 2D case studies}

\author[a]{Job Feldbrugge}
\author[b]{Rien van de Weygaert}

\affiliation[a]{Higgs Centre for Theoretical Physics, University of Edinburgh, Edinburgh, Scotland, EH8 9YL}
\affiliation[b]{Kapteyn Astronomical Institute, University of Groningen, Groningen, The Netherlands}

\emailAdd{Job.Feldbrugge@ed.ac.uk}

\abstract{
\noindent 
The cosmic web consists of a complex configuration of voids, walls, filaments, and clusters, which formed under the gravitational collapse of Gaussian fluctuations. Understanding under what conditions these different structures emerge from simple initial conditions, and how different cosmological models influence their evolution, is central to the study of the large-scale structure. Here, we present a general formalism for setting up initial random density and velocity fields satisfying non-linear constraints for specialized $N$-body simulations. These allow us to link the non-linear conditions on the eigenvalue and eigenvector fields of the deformation tensor, as specified by caustic skeleton theory, to the current-day cosmic web. By extending constrained Gaussian random field theory, and the corresponding Hoffman-Ribak algorithm, to non-linear constraints, we probe the statistical properties of the progenitors of the walls, filaments, and clusters of the cosmic web. Applied to cosmological $N$-body simulations, the proposed techniques pave the way towards a systematic investigation of the evolution of the progenitors of the present-day walls, filaments, and clusters, and the embedded galaxies, putting flesh on the bones of the caustic skeleton. The developed non-linear constrained random field theory is valid for generic cosmological conditions. For ease of visualization, the case study presented here probes the two-dimensional caustic skeleton.

}

\graphicspath{
{./figures/}}

\begin{document}
%\pagecolor{yellow!30!orange}
\maketitle

%%%%%%%%%%%%%%%%%%%%%%%%%%%%%%%%%%%%%%%%%%%%%%%%%%%%%%%%%%%%%%%%

\section{Introduction}
In this study, we extend and elaborate our analytical model for the formation of the cosmic web \cite{Zeldovich:1970,Bond:1996,Weygaert:2008}, and its complex multiscale network of filaments and sheets, by the Caustic Skeleton model \cite{Feldbrugge:2018}. The formalism, expressions, and details were introduced and extensively described in \cite{Feldbrugge:2018}. It entails an exact and computationally practical set of expressions for inferring the full two- or three-dimensional skeleton of the cosmic web\footnote{The caustic conditions underlying the model that are derived in \cite{Feldbrugge:2018} are applicable for any $N$-dimensional space.} emerging from any cosmological primordial field of linear density and velocity perturbations. It extends the early studies by the Soviet mathematician Arnold and collaborators \cite{Arnold:1982a,Arnold:1982b}, that studied the caustic structure predicted by the Zel'dovich formalism \cite{Zeldovich:1970}.

While the Caustic Skeleton details the structural outline of the nodes, filamentary branches, and sheetlike membranes that constitute the skeleton of the cosmic web, in the present study we develop the formalism to detail the mass distribution in and around the caustic skeleton. To be able to predict the mass distribution in and around filaments and walls, including the density profiles along as well as perpendicular to the filamentary ridges and sheetlike membranes, we develop a fully non-linear constrained field formalism. It is an extension and elaboration of the linear constrained field formalism introduced by \cite{Bertschinger:1987}, and further developed in \cite{Hoffman:1991} and \cite{Weygaert:1996}.

We subsequently implement the (non-linear) caustic conditions for the various structural features of the cosmic web. It allows us to explore in detail, and to great depth, the processes in and around the evolving features of the cosmic web and the hierarchical buildup of the connections between the various components of the cosmic web. The present study is the first in a series detailing the evolution of the mass distribution in and around the cosmic web. For reasons of presentation and clarity, the present study is limited to the two-dimensional universe. 

%%%%%%%%%%%%%%%%%%%%%%%%%%%%%%%%%%%%%%%%%%%%%%%%%%%%%%%%%%%%%%%%
\subsection{The cosmic web} 
The cosmic web is the largest complex structural pattern known in nature and is the fundamental spatial organization of matter, gas, and galaxies on scales of a few up to a hundred Megaparsec \cite{Shandarin:1989,Bond:1996,Weygaert:2008}. Galaxies, intergalactic gas, and dark matter exist in a wispy weblike arrangement of dense compact clusters, elongated filaments, and sheetlike membranes surrounding
near-empty void regions. The prominent elongated filamentary features form the dense boundaries around the large tenuous walls and define a network pervading the entire universe. The filaments connect up at massive, compact clusters located at the nodes of the network, and together with the walls, they surround the vast, underdense, and near-empty voids. Around 50\% of dark matter and 50\% of galaxies, is residing in filaments \cite[see e.g.][]{Cautun:2014,Ganeshaiah:2019}. The complex connectivity and intrinsic multiscale character reflect the primordial conditions out of which the wealthy structure and variety of objects in the Universe have emerged through gravitational evolution.

Cosmological observations are producing ever more detailed maps of the spatial galaxy distribution, while a range of observational data is accessing the neutral and ionized gas distribution in the \textit{cosmic web}, and even that of the dark matter distribution \cite{Hossen:2022, Kovacs:2022}. In the coming years a large array of major observational redshift surveys, in particular, those of Euclid, DESI, the Vera Rubin observatory, and SKA -- will map the weblike organization of galaxies over unprecedented large cosmic volumes. Given this wealth of data, it is of crucial importance to prepare for the systematic scientific exploration of the cosmic web and study its dependence on the cosmological setting in which it forms.  As the structure, dynamics, and connectivity of the cosmic web depend sensitively on the underlying cosmology, these new methods will exploit the geometry and topology of the cosmic web to infer constraints on the underlying physics and cosmological parameters. 

The evolution of the complex multiscale structure of the cosmic web is the product of the interplay of a range of non-linear processes. $N$-body simulations, numerically simulating cosmological structure formation, are very useful to obtain a good impression of the resulting structure. However, even though several up-to-date projects provide a large number of simulations that offer a good representation of many aspects of the structure formation process \cite{ABACUSSUMMIT:2021, Quijote:2020}, they are usually demanding, expensive, and do not have the flexibility to infer the required information on all relevant cosmological structure formation aspects. A full understanding of the outcome makes it necessary to invoke complementary analytical models and descriptions for the interpretation of the results. This will provide substantially more profound and versatile insight into the physical processes driving the formation and evolution of the cosmic web.

Given the importance of basing the analysis of cosmic web observations on a good analytical model for its non-linear evolution, the present study involves the further development of the Caustic Skeleton model of the cosmic web that we introduced in \cite{Feldbrugge:2018}. 

%%%%%%%%%%%%%%%%%%%%%%%%%%%%%%%%%%%%%%%%%%%%%%%%%%%%%%%%%%%%%%%%
\subsection{Phase-space evolution}
The process of the formation and evolution of structure in the Universe is driven by the gravitational growth of tiny primordial density and velocity perturbations. Inhomogeneities in the gravitational force field lead to the displacement of mass out of the lower-density areas towards higher-density regions. Complex structures arise at the locations where different mass streams meet up, along which gravitational collapse sets in. While reaching this stage the matter distribution starts to develop non-linearities, and we see the emergence of complex structural patterns, the Cosmic Web.  In the current universe, we see this happening at Megaparsec scales. 

Of key importance for understanding the emergence of non-linear structure in the Universe is an insight into the structure of the mass flows accompanying the structure formation process. The first recognizable features to emerge in the cosmic matter distribution are the flattened wall-like and elongated filamentary features, along with the large underdense void regions that assume most of the cosmic volume between these features. Ultimately these merge into a pervasive weblike network. Their formation is the result of the accumulation of different mass streams at specific locations, with walls and filaments forming at locations where various mass streams meet up.  Insight into the process and the formation of the cosmic web is therefore attained by assessing the multi-stream nature of the mass flows. To analyze the multi-stream structure of the mass distribution, we turn to the Lagrangian description of structure growth. It takes into account its full six-dimensional phase-space structure \cite[see][for key contributions on this.]{Shandarin:2010,Shandarin:2011,Shandarin:2012,Abel:2012,Falck:2012}. 

The first stages of cosmic structure formation are remarkably accurately described by the analytical framework of the Zel'dovich approximation \cite{Zeldovich:1970}. Its first-order Lagrangian description has proven to work extremely well until non-linear gravitational evolution induces contraction and collapse, and the appearance of multi-stream regions. This goes along with the emergence of the first nontrivial wall-like and elongated features in the cosmic matter distribution. It not only represents a remarkably accurate description of the evolving mass distribution up to the appearance of multi-stream regions, but it also suggests the morphology of cellular or weblike arrangement of matter in the universe and hence predicted the later detection of the weblike organization of galaxies and gas in the Universe \cite{ Einasto:1978, Lapparent:1986, Colless:2003, Huchra:2012, Granett:2012}. 

%%%%%%%%%%%%%%%%%%%%%%%%%%%%%%%%%%%%%%%%%%%%%%%%%%%%%%%%%%%%%%%%
\subsection{Multistreams \& caustics}
Further insight into the non-linear evolution of the mass distribution after the emergence of multi-stream regions is obtained by concentrating on the features that form at the locations where the different gravitationally directed mass streams meet up. These form the structural spine around which matter subsequently assembles and virializes into recognizable physical objects. In terms of six-dimensional phase-space, it corresponds to the local folding of the phase-space sheet along which matter -- in particular the gravitationally dominant dark matter component -- has distributed itself. As seen from the six-dimensional phase-space perspective, the features in the cosmic web are the projections of a three-dimensional mass sheet in a six-dimensional phase-space onto the spatial (Eulerian) volume. There where different mass streams cross at a particular location, the sheet is folded and the projection of these folds is seen as a structural feature in the cosmic web.

The Caustic Skeleton model tracks the evolution of the dark matter sheet in six-dimen\-sional \textit{phase}-\textit{space} and marks the locations where the fluid forms caustics, and structural singularities in the evolving mass distribution. We derived the complete set of analytical expressions for where and which class of singularities would appear \cite{Feldbrugge:2018}. These non-linear {\it caustic conditions} are fully characterized by the eigenvalues and the eigenvector fields of the deformation tensor. Dynamically, this is related to the instrumental role of the tidal forces shaping the anisotropic features in the cosmic web \cite[see e.g.][]{Weygaert:2008}. Indeed, one may trace the major aspects of the spatial outline of the cosmic web in the primordial tidal field structure, specifically that of its eigenvalues (see figure~\ref{fig:Initial_Conditions}).

Crucial for the spatial pattern outlined by the caustic features, and their subsequent hierarchical evolution is that not only the spatial structure of the {\it eigenvalue} fields but also that of the {\it eigenvector} fields are of decisive importance \cite{Feldbrugge:2018}. Conventionally, the key role of the latter has almost been entirely ignored. It is the incorporation of the eigenvector fields that has enabled us to develop a fully analytical theory of the Cosmic Web. 

The Caustic Skeleton model highlights the place and time where gravitational collapse turns non-linear, and where and when walls, filaments, and clusters emerge. Moreover, these caustic mark the locations where these structures merge to form the interconnected structure which we observe today. For the application of the \textit{Caustic Skeleton}, we need to augment it with a model for large-scale structure formation, specifically in the form of a (primordial) deformation field. The prime example of this is the Zel'dovich approximation. The inferred caustic features define the \textit{caustic skeleton}. It marks the location and spatial outline of the caustic features in the initial (Lagrangian) field. The skeleton of the cosmic web in Eulerian space using the Lagrangian map $\bm{x}_t$ to Eulerian space. Following the identification of the various caustic varieties and caustic points in Lagrangian space, the application of the map $\bm{x}_t$ will produce the corresponding weblike structure in Eulerian space. 

\bigskip
Surprisingly, these caustics are intimately related to the occurrence of degenerate critical points, which were in the last century classified by catastrophe theory. Thom famously classified the stable critical points of families of functions into a finite set \cite{Thom:1975}, popularized by Zeeman \cite{Zeeman:1972,Zeeman:1976}. Arnold extended catastrophe theory to the classification of caustics and connected them to the ADE Coxeter groups classification (also famously occurring in the classification of semi-simple Lie algebras) \cite{Arnold:1972,Arnold:1976,Arnold:1984}. In the three-dimensional setting, these are the $A$ varieties $A_2,A_3,A_4,$ and $A_5$, the $D$ umbilic varieties $D_4$ and $D_5$. The $E$ variety only occurs stably in higher-dimensional settings.

By considering the morphological and dynamical nature of the various ADE singularities, one may directly see that each of the caustic classes can be identified with morphological/structural features in the cosmic web \cite{Feldbrugge:2018}. That is, in three dimensions, the $A_3$ caustics can be identified with the walls, the $A_4$ ones with the filaments, and the $A_5$ with the nodes. A highly interesting finding is that also $D_4$ umbilic singularities are to be identified with filamentary structures, implying there to be two classes of filaments \cite{Feldbrugge:2018}. The locations of the caustic singularities trace out a Lagrangian skeleton of the emerging cosmic web and establish their connectivity (see the discussion in \cite{Hidding:2014}). Important is also to realize that this analytical framework allows us to establish a complete analytical framework for studying their connectivity, defining the weaving of the cosmic web \cite{Zeldovich:1970,Bond:1996,Weygaert:1996,Aragon:2010a,Aragon:2010b,Cautun:2014}.

%%%%%%%%%%%%%%%%%%%%%%%%%%%%%%%%%%%%%%%%%%%%%%%%%%%%%%%%%%%%%%%%
\subsection{Caustic skeleton \& non-linear constraints}
Caustic skeleton theory models the cosmic web in terms of a skeleton consisting of several singular features (sheets, curves, and points). However, redshift surveys and lensing surveys show the large-scale structure in terms of the spatial distribution of galaxies and the underlying dark matter. For the purpose of relating the observed or observable matter and galaxy distribution to that of the underlying complex phase-space structure, it is of key importance to expand the caustic formalism so that it will be able to generate realizations resembling observed structures. For such {\it reconstruction} purposes, a considerable range of advanced formalisms and techniques have been developed that seek to infer the primordial conditions out of which these evolved. For the caustic formalism this 

In addition to this {\it cosmography} instrument, a second incentive for such an extension of the caustic formalism is that of {\it laboratory tool}. The latter allows us a systematic exploration of the impact of the presence of caustic features on the surrounding matter distribution, the relationship between different cosmic structures, and a wide range of astrophysical questions on the nature and state of galaxies and gas in and around various morphological and structural structures of the cosmic web. Examples are that of the early constrained simulations of voids and filaments \cite{Weygaert:1993,Haarlem:1993}. It calls for the development of a {\it constrained field formalism} for the non-linear and non-Gaussian conditions that exist in the cosmic web. 

The central issue and task of the present study are therefore to bridge this gap by extending the Caustic Skeleton formalism with the possibility to model the mass distribution around the various caustic features of the cosmic web. It entails the development of a formalism that generates a realization of the implied mass distribution given the constraint of the presence and location of one (or even more) of the caustic singularities. The complication is that it involves constraints on the deformation tensor, in terms of the corresponding eigenvalues and eigenvectors. Even for the primordial perturbation field, these are non-Gaussian, let alone for the more generic non-linear stages that our Caustic Skeleton model seeks to model.

\bigskip
In line with the above, the present study seeks to develop and present a non-linear constrained random field formalism that allows us to generate realizations of the initial density, velocity, and potential field around various specified caustic singularities. This will allow us to infer the analytic properties of the initial density perturbation and gravitational potential in the vicinity of the progenitors of the walls, filaments, and clusters of the cosmic web. In a sense, it allows us to put flesh on the bones of the caustic skeleton by running a set of $N$-body simulations on these initial conditions. In turn, this will allow us to infer the formation histories and present-day properties of the various elements of the caustic skeleton. 

We concentrate on the development and exposition of the formal -- mathematical -- method to generate non-linear constrained realizations. The formalism is an extension of the one defined for linear constraints on Gaussian initial conditions, as defined in the Gaussian random field theory of \cite{Bertschinger:1987}, \cite{Rybicki:1992} and \cite{Hoffman:1991}. In particular, it follows the application of this formalism to generate constrained realizations implied by conditions imposed on the initial density, velocity, gravity, and/or tidal fields, as outlined by \cite{Weygaert:1996}.  

\bigskip
While to good approximation linearity may still be presumed to hold on large scales, allowing the inference of the mass distribution from the mapping of the velocity flows in our Local Universe \cite{Bertschinger:1989,Dekel:1999,Courtois:2012,Hoffman:2015}, and to investigate the formation and evolution of the cosmic web in our Local Universe. Well-known in this context are the CLUES simulations \cite{Sorce:2016}. An important aspect of these procedures is to incorporate the strength/reliability of the observational data in the significance of the obtained reconstruction, for which Wiener filtering has become a widely used instrument \cite{Zaroubi:1995,Erdogdu:2004}. Linear constrained Gaussian random field theory technique can also be rephrased as an optimization problem. With this perspective, linear constraints can be implemented for a large number of constraints resembling a local patch of Lagrangian space \cite[see for example the splice method][]{Cadiou:2021}. 

A range of avenues has been taken to extend the ability to impose linear constraints to that of the far more ubiquitous situation of constraints that pertain to the non-linear cosmic matter distribution, or that are a non-linear function of observable quantities. One option that has many applications in e.g. geophysics is that of (lognormal) Kriging, in which one exploits the information available at irregularly placed locations by taking into account the spatial correlations \cite{Platen:2011}. There have also been attempts to augment the linear Gaussian field formalism by invoking a (first order) model for the displacement of mass elements in order to follow these along their path backward in time \citep{Doumler:2013}.

However, because of the restricted validity to regions that are not (yet) multistream, it cannot be applied for our purpose of modeling the mass distribution around the intrinsically multistream filamentary features. Given the fact that the majority of matter, halos and galaxies reside in filaments \citep{Cautun:2014,Ganeshaiah:2019}, more sophisticated higher-order schemes will be necessary. One option is to incorporate higher-order Lagrangian schemes for modeling the orbits of galaxies and/or paths of mass elements. The Least Action Principle algorithm introduced by \cite{Peebles:1989} is one such approach. Sophisticated higher-order Lagrangian formalisms have been proposed by \cite{Mohayaee:2006, Lavaux:2008, Hada:2018, Shi:2018, Zhu:2018}. For example, \cite{Mohayaee:2006} and \cite{Lavaux:2008} used the Monge-Amp\`ere-Kantorovitch algorithm to follow the orbits of objects back in time, culminating in a compelling reconstruction of the local velocity field corresponding to the 2MRS survey. 

Another approach has become an important cosmological tool over the past 15 years. This concerns the ability to translate the observed spatial distribution of galaxies within a given volume of the (local) Universe towards its implied primordial Gaussian density and velocity perturbations out of which these galaxies and structures emerged through gravitational growth. To this end, advanced and sophisticated Bayesian reconstruction schemes have been developed, which involve elaborate and computationally demanding  MCMC sampling procedures \cite{Kitaura:2008,Kitaura:2009, Jasche:2010,Leclercq:2015,Hess:2016, Bos:2016, Leclercq:2017, McAlpine:2022}. For large galaxy surveys in the local Universe, such as the 2MRS and SDSS surveys, impressive results have been obtained that allow us to study the properties and evolution of its detailed mass distribution. Telling examples are the 2MRS constrained simulations of the Local Universe by Kitaura and collaborators \cite{Hess:2013} (also see \cite{Hidding:2016}) and the recent detailed Sibelius simulation \cite{McAlpine:2022}. These simulations can be used to address a variety of problems, including BAO measurements and the removal of redshift space distortions. 

\bigskip
Notwithstanding the success of the developed techniques, there remains a major challenge in regimes where the imposed constraints concern mass elements that are not located in multistream regions. Most methods are still restricted to a regime in which mass elements have not yet passed through multistream regions. For our purpose, it is of key importance to be able to model the mass distribution around the intrinsically multistream caustic structures. Moreover, we also need a computationally less demanding non-linear constraint procedure for the purpose of exploring the structure and mass distribution in and around these caustic features in the cosmic web. In other words, a laboratory tool that allows us to zoom in on arbitrary aspects of the cosmic web. Hence, the introduction and presentation of a fully non-linear constrained random field procedure. 

Within the context of developing this non-linear constrained field theory and its application to the caustic skeleton model, and its extension with non-linear constrained field theory, we observed that the study of the cosmic web has a remarkable correspondence to statistical field theory for the study of phase transition in thermodynamics. Both theories describe the emergent behavior of a large number of particles described by a Euclidean path integral \cite{Feynman:1965}. We, therefore, expect that the well-established path integral techniques will be beneficial to the further systematic and analytic exploration of the intricate cosmic web.

\begin{figure}
  \centering
  \begin{subfigure}[b]{\textwidth}
  \includegraphics[width=\textwidth, trim={4cm 0 4cm 0},clip]{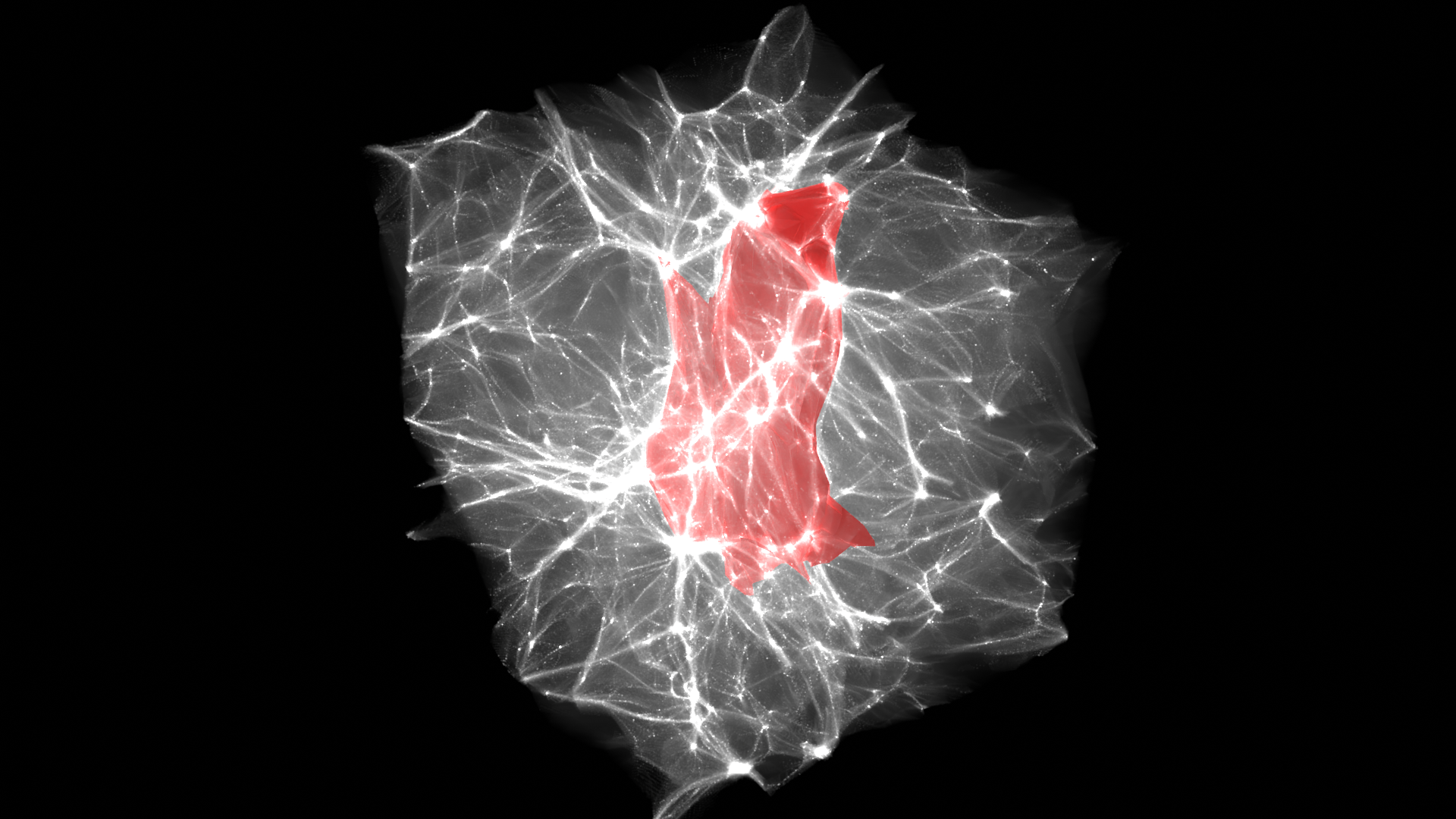}
  \caption{Cusp constraint}
  \end{subfigure}
  \begin{subfigure}[b]{\textwidth}
  \includegraphics[width=\textwidth, trim={4cm 0 4cm 0},clip]{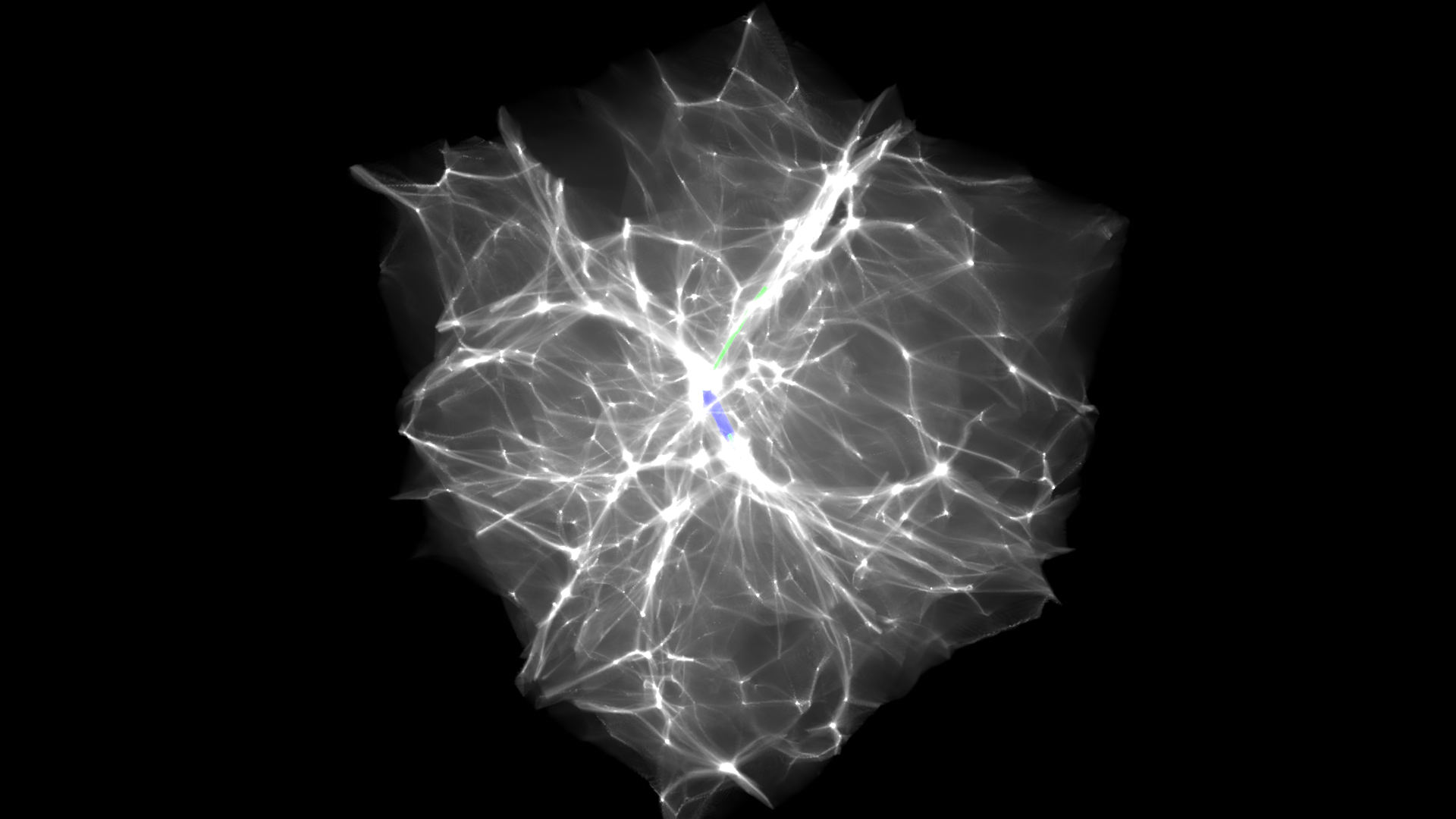}
  \caption{Umbilic constraint}
  \end{subfigure}
  \caption{Two examples of a three-dimensional caustic constrained dark matter $N$-body simulation at growing mode $b_+=0.8$. \textit{Top:} a cusp constraint at the center of the simulation (the red sheet), and \textit{Bottom:} an umbilic constraint at the center of the simulation (the blue line). The constrained simulation in the right panel also includes a swallowtail filament (the green line). Note the existence of two different classes of filaments.}
  \label{fig:3DExample}
  \end{figure}

\bigskip
For the illustration of the potential and performance of our non-linear constrained random field formalism, here we limit ourselves to case studies of the two-dimensional caustic skeleton. The principal purpose of the case study illustrations is to elucidate the workings of formalism and its possible and potential applications. For this, $2$D case studies are usually more transparent. It paves the path toward a systematic study of the different features of the three-dimensional cosmic web, the subject of a range of upcoming contributions \cite{Feldbrugge:2022}. See figure \ref{fig:3DExample} for an example of a three-dimensional dark matter $N$-body simulation on constrained initial conditions associated with a cusp wall and an umbilic filament at the center of the simulation. We observe that the cusp caustic is located in the region which we would visually identify as a wall bounding a void. The umbilic filament is located at a bridge between two dense clusters.

%%%%%%%%%%%%%%%%%%%%%%%%%%%%%%%%%%%%%%%%%%%%%%%%%%%%%%%%%%%%%%%%
\subsection{Outline} 
First, in section \ref{sec:Eulerian_space} we provide a visual justification of the caustic skeleton by illustrating the evolution of the skeleton using a dark matter-only two-dimensional $N$-body simulation. In particular, we show that the skeleton based on the Zel'dovich approximation applies to the current cosmic web, where the Zel'dovich approximation itself breaks down. The Caustic Skeleton formalism is rooted in Lagrangian fluid dynamics, for which section \ref{sec:Caustic_Skeleton_Theory} provides a concise summary. We subsequently extend constrained random field theory to include non-linear constraints in section \ref{sec:GRF}. Using this formalism, we implement the caustic conditions and evaluate the mean field of the different elements of the two-dimensional caustic skeleton in section \ref{sec:caustic_skeleton_constraints}. In section \ref{sec:composite_constraints}, we extend non-linear constrained random field theory to multiple caustic conditions at different locations. This paves the way for a systematic study of the interplay between the walls, filaments, and clusters of the cosmic web. Finally, in section \ref{sec:conclusion} we summarize the results and discuss possible applications.

\begin{figure}
\centering
\begin{subfigure}[b]{0.49\textwidth}
\includegraphics[width=\textwidth]{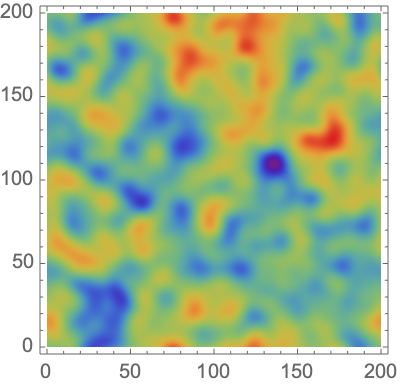}
\end{subfigure}~
\begin{subfigure}[b]{0.49\textwidth}
\includegraphics[width=\textwidth]{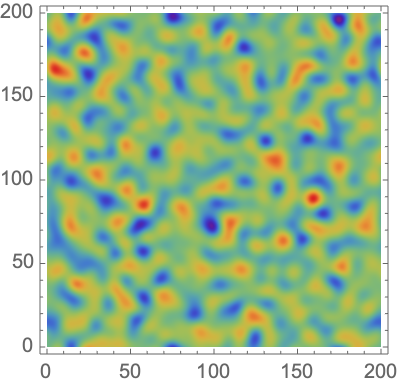}
\end{subfigure}\\
\begin{subfigure}[b]{0.49\textwidth}
\includegraphics[width=\textwidth]{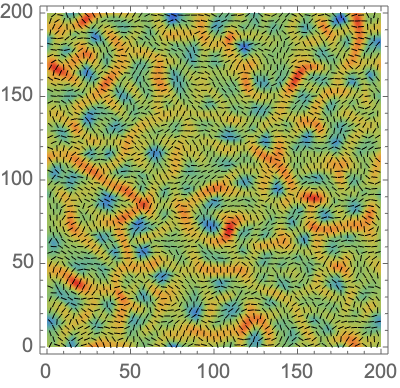}
\end{subfigure}~
\begin{subfigure}[b]{0.49\textwidth}
\includegraphics[width=\textwidth]{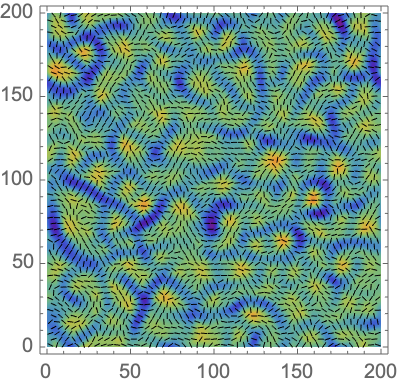}
\end{subfigure}
\caption{A Gaussian random field and the corresponding eigenvalue and eigenvector fields. \textit{Upper left:} the initial gravitational potential $\phi_0$. \textit{Upper right:} the corresponding density perturbation field $\delta$. \textit{Lower left:} the first eigenvalue and eigenvector fields $\mu_1,\bm{v}_1$. \textit{Lower right:} the second eigenvalue and eigenvector fields $\mu_2,\bm{v}_2$.}
\label{fig:Initial_Conditions}
\end{figure}

%%%%%%%%%%%%%%%%%%%%%%%%%%%%%%%%%%%%%%%%%%%%%%%%%%%%%%%%%%%%%%%%
\section{The caustic skeleton \& cosmic web illustrated}
\label{sec:Eulerian_space}
In this section, we describe the gravitational collapse of Gaussian fluctuations, the resulting emergence of an intricate cosmic network, known as the cosmic web, and the accurate and profound connection with the formation of caustic singularities in the mass distribution. The cosmic web -- consisting of voids, walls, filaments, and clusters -- is intimately tied to the development of multi-stream regions, which in turn are most naturally described in terms of caustics. These caustics capture the formation histories of the multi-stream regions and trace the spine of the cosmic web. Moreover, we point out a strong correspondence between the geometry of the eigenvalue fields of the initial deformation tensor (governed by the gravitational potential) and the geometry of the web at late times. 

In a first step towards elucidating and appreciating the intimate connection between the cosmic mass distribution in the cosmic web and its analytically predicted spine by the Caustic Skeleton model, we provide a visual impression of the primordial inhomogeneous matter distribution and corresponding inhomogeneous and anisotropic gravitational force field.

The present study limits itself to the treatment of the gravitationally evolving dark matter distribution in a two-dimensional setting. The development of the dark mark distribution in the corresponding four-dimensional phase-space can be seen as that of a gradually deforming two-dimensional {\it dark matter sheet} embedded in 4-D phase-space. For a visually direct assessment of the relation between the structural components of the cosmic web and the caustic singularities arising in the matter distribution, we confine our attention to the patterns observed in Eulerian space. 

The starting point for the emergence of the cosmic web is the primordial field of random Gaussian density and velocity fluctuations. The cosmic web is the product of the subsequent gravitationally propelled evolution. At each cosmic epoch, the mass distribution attains a weblike morphology at the scale at which the matter distribution evolves away from its initial linear evolution and non-linear features start to appear as mass concentrations decouple from the Hubble expansion and start to contract gravitationally. At this quasi-linear phase, the migration streams involved in the buildup of the structure start to cross, and we see the emergence of multi-stream regions at the locations of the emerging non-linear features. 

%%%%%%%%%%%%%%%%%%%%%%%%%%%%%%%%%%%%%%%%%%%%%%%%%%%%%%%%%%%%%%%%
\subsection{Caustics in the cosmological matter distribution}
\label{sec:primordial}
The cosmic mass distribution that we discuss in this section concerns the dark matter distribution in a two-dimensional realization in a box of $200 \mbox{Mpc}$ length. The cosmological background is a flat Einstein-de Sitter universe with a dark matter density $\Omega_m=1$ and Hubble constant $H_0=71\text{ Mpc/s/km}$. The structure evolves gravitationally out of a primordial Gaussian random field with a power-law power spectrum. The corresponding gravitational potential $\phi_0$ has a power-law spectrum $P_{\phi}(k) \propto k^{n_s}$, with a high-frequency cutoff at a smoothing scale $R_s$. For the present reference model the power law index $n_s=-1$ and the cutoff scale
$R_s=2.5 \text{ Mpc}$. The primordial field is generated on a $512^2$ grid.

The map of the Gaussian random potential field is shown in the top lefthand frame in figure~\ref{fig:Initial_Conditions}, with the corresponding map of the density field $\delta$ in the top righthand frame of the same figure. We use the Zel'dovich approximation's analytical expression for the deformation tensor to infer on the basis of the caustic conditions the nature of the caustic singularities that will arise in the evolving mass distribution \cite{Feldbrugge:2018}. The caustic conditions refer to the mass distribution in Lagrangian space, and they yield the Lagrangian location and outline of the various caustic singularities present in the generated mass distribution.  Table~\ref{table:caustics-cosmciweb} guides the morphological identity of the various caustic classes, for the two-dimensional situation. The full identification for the two- and three-dimensional situations can be found in \cite{Feldbrugge:2018}.

\begin{figure}
\centering
\begin{subfigure}[b]{0.49\textwidth}
\includegraphics[width=\textwidth]{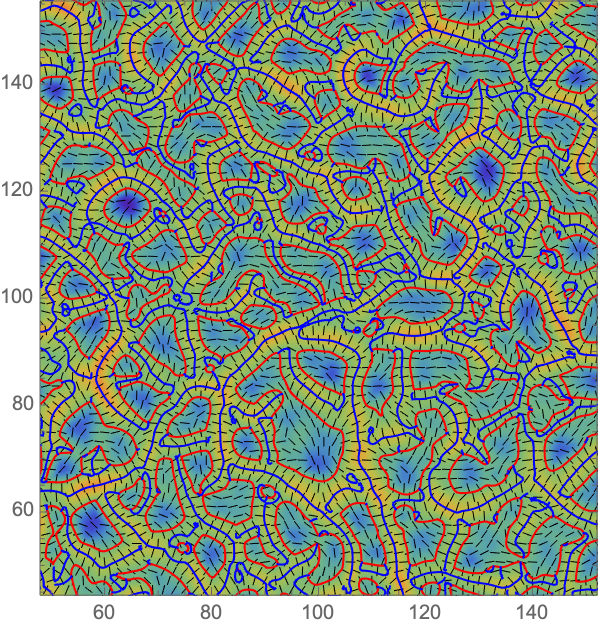}
\caption{$\lambda_1,\bm{v}_1$}
\end{subfigure}
\begin{subfigure}[b]{0.49\textwidth}
\includegraphics[width=\textwidth]{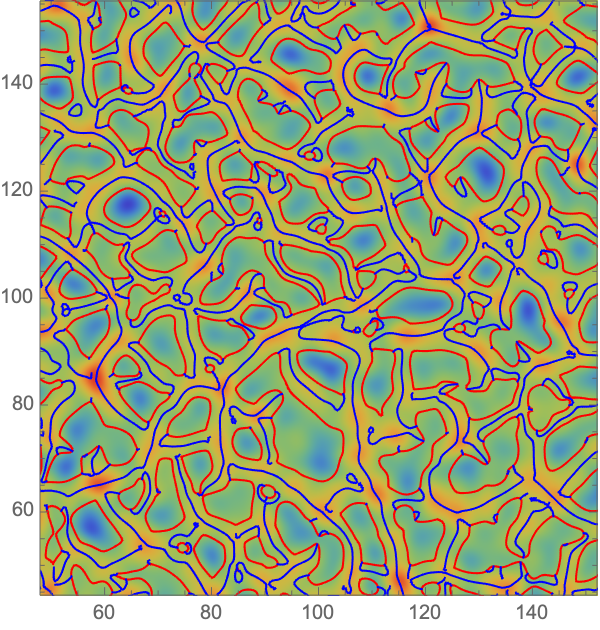}
\caption{$\delta$}
\end{subfigure}\\[0.5cm]
\begin{subfigure}[b]{0.49\textwidth}
\includegraphics[width=\textwidth]{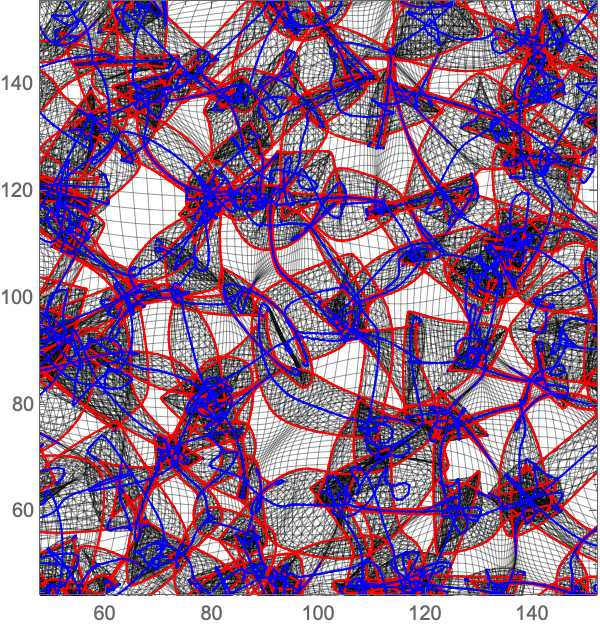}
\caption{Zel'dovich approximation}
\end{subfigure}~
\begin{subfigure}[b]{0.49\textwidth}
\includegraphics[width=\textwidth]{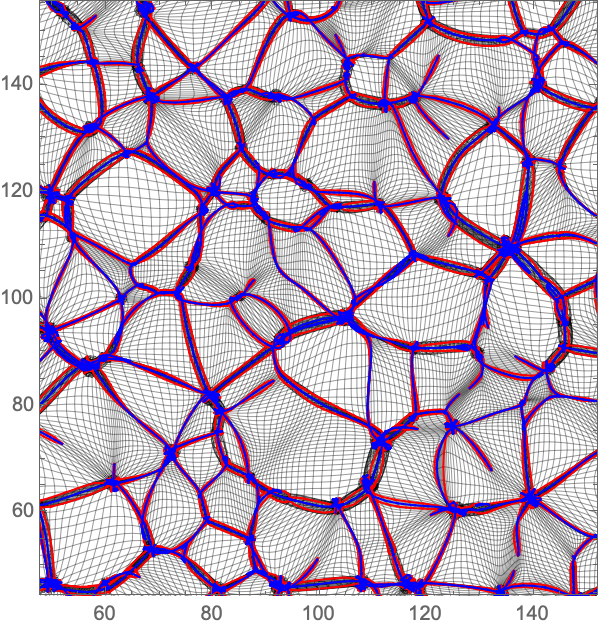}
\caption{$N$-body simulation}
\end{subfigure}
\caption{The caustic skeleton consisting of the $A_2$ fold curves (red) and $A_3$ cusp curve (blue) in Lagrangian and Eulerian space. \textit{Upper left (Lagrangian space):} the initial eigenvalue field $\lambda_1$ with the caustic skeleton. \textit{Upper right (Lagrangian space):} the initial density perturbation field, with the caustic skeleton superimposed. \textit{Lower left (Eulerian space):} the Zel'dovich approximation with the caustic skeleton at scale factor $a=1$ (Eulerian space). \textit{Lower right (Eulerian space):}  the $N$-body simulation with the caustic skeleton at scale factor $a=1$.}
\label{fig:Eulerian}
\end{figure}

The primordial mass distribution is evolved to the current epoch ($z=0$). For the purpose of the present study, the evolution is followed in two ways. A two-dimensional dark matter $N$-body simulation \cite{Hidding:2020} is used to follow the fully non-linear evolution of the mass distribution. This resulting non-linear structure is discussed in section~\ref{sec:nbody}. In the Zel'dovich approximation, the Eulerian location and environment of these caustic features are obtained by using the corresponding expression for the displacement of a mass element. In the $N$-body simulation, their location follows from the simulation displacement of the corresponding mass elements (see figure~\ref{fig:Eulerian} and figure~\ref{fig:Eulerian_Evolution}).

\begin{table}
\centering
{\scriptsize
\begin{tabular}{ |l | l | l | l |}
  \hline
  \ && \\
\textbf{Caustic Singularity} & \textbf{Symbol} & \textbf{2D cosmic web}\\
\ && \\
\hline
\ && \\
Fold & $A_2$ & shell-crossing \\
Cusp & $A_3$ & filament \\
Swallowtail &$A_4$ &  cluster\\
Elliptic/hyperbolic & $D_4^{\pm}$ & cluster\\
Morse point & $A_3^+$ & creation/annihilation point\\
Morse point & $A_3^-$ & merger point\\
\ && \\
\hline
\end{tabular}
}
\caption{Elements of the two-dimensional caustic skeleton and their caustic conditions.}
\label{table:caustics-cosmciweb}
\end{table}

The direct comparison between the evolving $N$-body particle distribution allows us to assess the position and role of the various caustics within the cosmic mass distribution, and in particular that in the weblike network in which it has organized itself.

%%%%%%%%%%%%%%%%%%%%%%%%%%%%%%%%%%%%%%%%%%%%%%%%%%%%%%%%%%%%%%%%
\subsection{Primordial deformation field and cosmic web}
At the heart of the caustic conditions in \cite{Feldbrugge:2018} is the observation that the caustic singularities that are emerging in the matter distribution, and the structural weblike components surrounding them, are determined by the deformation and tidal field induced by the inhomogeneous mass distribution \cite{Zeldovich:1970,Weygaert:1996,Weygaert:2008,Hidding:2014}.

The key role of the eigenvalues of the deformation tensor in establishing the morphological nature of cosmic structure has been acknowledged since the seminal studies by \cite{Zeldovich:1970} and \cite{Doroshkevich:1970}. It formed the basis for the expectation that the cosmic mass distribution would be organized in a cellular pattern \cite[also see][]{Shandarin:1989,Shandarin:2009}. The instrumental role of the corresponding eigenvectors in establishing the overall structural pattern of the cosmic web, and its various components, was hardly acknowledged. The explicit demonstration by \cite{Feldbrugge:2018} revealed their importance in outlining the weblike structure.

The intimate connection between the deformation's eigenvalues and eigenvectors and the subsequent formation of the cosmic web is clearly revealed in the bottom panels of figure~\ref{fig:Initial_Conditions}. The panels provide a telling illustration of the central role of the deformation/tidal field tensor in establishing the spatial structure of the cosmic web that will emerge from these initial conditions as a result of gravitationally driven evolution. The map in the bottom lefthand panel represents the first eigenvalue $\lambda_1$ field and the bottom lefthand panel the second eigenvalue $\lambda_2$ field. The corresponding eigenvectors $\bm{v}_1$ and $\bm{v}_2$ are superimposed by means of black solid bars oriented along the eigenvector direction, depicted at a discrete number of grid locations. With respect to the eigenvector fields, we note that at initial times (\textit{i.e.}, in Lagrangian space), the eigenvector fields are normal, $\bm{v}_{t,1}\cdot \bm{v}_{t,2}=0$, since the deformation tensor is symmetric. The maps reveal that to a considerable extent a pervasive weblike network can already be recognized in the primordial deformation field. The eigenvalue and eigenvector maps are distinctly non-Gaussian fields, marking a highly structured pattern, with a high level of spatial coherence.

%%%%%%%%%%%%%%%%%%%%%%%%%%%%%%%%%%%%%%%%%%%%%%%%%%%%%%%%%%%%%%%%
\subsection{Multistream regions and caustic skeleton}
Induced by the matter streams set into motion by the primordial inhomogeneous force field, spatially steered and directed by the corresponding deformation tensor orientation, matter starts to aggregate at locations where multiple streams meet up. The outline of these locations can already be recognized by the typical cellular patterns in the spatial primordial deformation eigenvalue distribution. Dependent on the precise nature and geometry of these multi-stream regions, \textit{i.e.}, the way in which the cosmic matter phase-space sheet is folded, we observe the formation and evolution of different characteristics components of the cosmic web. The nature of the complex spatial folding of the phase-space sheet in and around the flow field singularities determines the identity of the structural component of the cosmic web that arises in the spatial matter distribution.

The caustic conditions \cite{Feldbrugge:2018} allow us to infer exactly the location of the singularities in the matter distribution, the caustics. These conditions pertain to the nature and location of singularities in Lagrangian space. It means that at any cosmic epoch, the caustic conditions identify the mass elements that are included in a caustic singularity, for any of the possible caustic classes. Locating these mass elements at any subsequent cosmic epoch, by projecting them to their Eulerian position, yields the spatial outline of the spine of the cosmic web at that epoch. This spine is the assembly of the caustic membranes, ridges, and nodes around which matter organizes itself into the cosmic web. In other words, the set of analytical expressions for the caustic conditions represents a fully analytical description of the caustic skeleton of the cosmic web, and hence of its formation and evolution. Around this, a fully analytical theory of cosmic web formation can be formulated. The present study entails a major step in this program. 

\bigskip
The top panels of figure~\ref{fig:Eulerian} present a telling illustration of the intimate connection between the eigenvalue and eigenvector fields of the deformation tensor and the spatial outline of the various caustic singularities. The eigenvector field is represented by black directional bars along the direction of eigenvectors, depicted at a discrete number of grid locations. The top lefthand panel of figure~\ref{fig:Eulerian} captures the geometry of the first eigenvalue $\lambda_1$ field, along with the corresponding eigenvector $\bm{v}_1$ field.   Also interesting for our discussion is the observation that the eigenvector fields exhibit an interesting correlation with the eigenvalue fields. Roughly speaking, the eigenvector fields are normal to the ridges of the corresponding eigenvalue fields.

Both the $A_2$ fold (red) and $A_3$ cusp caustic curves (blue) are superimposed on the eigenvalue and eigenvector maps in figure~\ref{fig:Eulerian}. The red fold curves provide a reasonable approximation of the boundaries of multi-stream regions, which they enclose. Particularly interesting for understanding the connectivity of the cosmic web is the stringy network of (blue) cusp curves. At late times, they outline an interconnected network that bisects the multi-stream regions and forms knots at the clusters of the cosmic web. In other words, the ridges outlined by the cusp curves define the filamentary spine of the cosmic web. In addition, we see how blue (Lagrangian) regions bounded by the red fold curves are the progenitors of the voids: they delineate the mass elements that will remain in mono-stream regions. 

\begin{figure}
\centering
\begin{subfigure}[b]{0.49\textwidth}
\includegraphics[width=\textwidth]{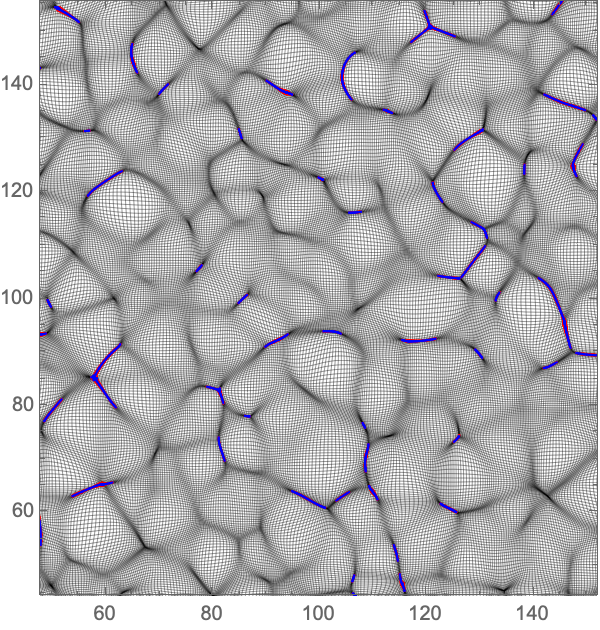}
\caption{$a=0.2$}
\end{subfigure}~
\begin{subfigure}[b]{0.49\textwidth}
\includegraphics[width=\textwidth]{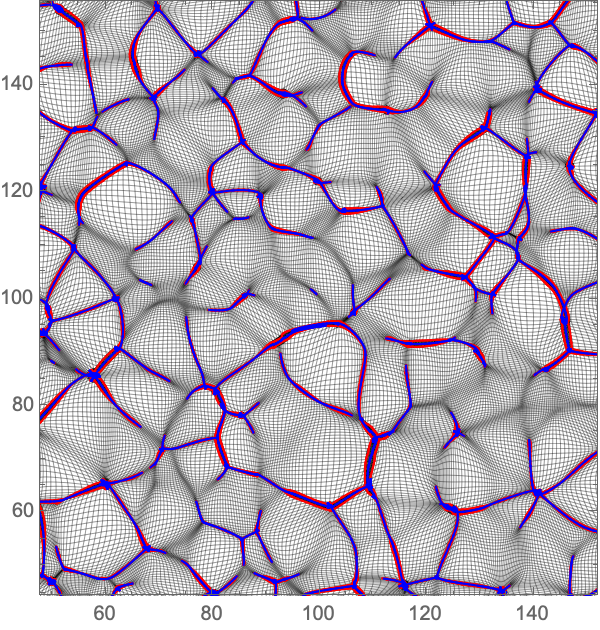}
\caption{$a=0.4$}
\end{subfigure}\\[0.5cm]
\begin{subfigure}[b]{0.49\textwidth}
\includegraphics[width=\textwidth]{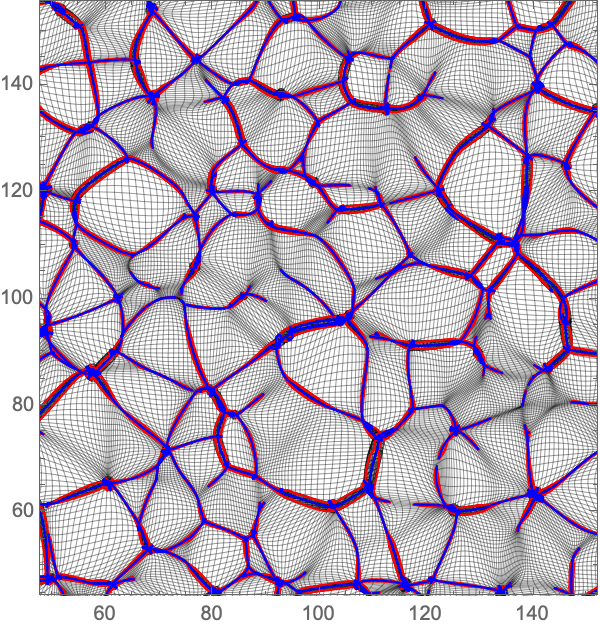}
\caption{$a=0.6$}
\end{subfigure}~
\begin{subfigure}[b]{0.49\textwidth}
\includegraphics[width=\textwidth]{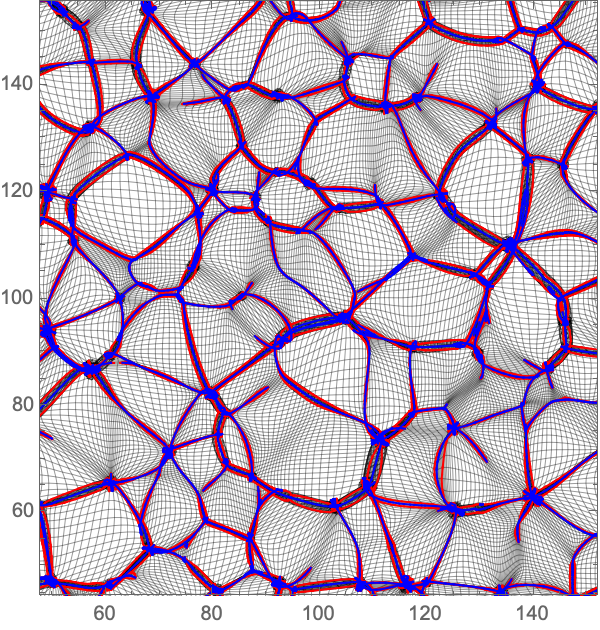}
\caption{$a=0.8$}
\end{subfigure}
%\begin{subfigure}[b]{0.48\textwidth}
%\includegraphics[width=\textwidth]{Evolution_100}
%\end{subfigure}
\caption{The evolution of the caustic skeleton consisting of the fold curves (red) and cusp curve (blue) in an $N$-body simulation in an Einstein-de Sitter universe. The skeleton for scale factor $a=1$ can be seen in the lower right panel of figure \ref{fig:Eulerian}.}
\label{fig:Eulerian_Evolution}
\end{figure}

\bigskip
The top righthand panel of figure~\ref{fig:Eulerian} compares the primordial density field with the caustic skeleton. At large scales, the eigenvalue field closely resembles the initial density perturbation $\delta_0 \propto \nabla^2\phi_0$. At these scales, the caustic skeleton follows the geometry of the super-level sets of the density perturbation. However, on intermediate to small scales, noticeable and significant qualitative differences appear. 

The red overdense regions in the primordial density field experience an inflow of matter. These lead to the emergence of the first caustics in and around these locations. Meanwhile, in and around the underdense blue regions we see the outflow of matter and the appearance of the first voids. When comparing the primordial density field with the primordial second eigenvalue field, we observe that the peaks and troughs in both fields largely coincide. However, significant differences and deviations between the density and eigenvalue field occur at the medium field levels. The eigenvalue fields emphasize line-like features that connect the high-density regions. These elongated features in the first eigenvalue field are the progenitors of the most prominent components of the cosmic web, the filaments. 

Within the context of Lagrangian fluid dynamics, we may appreciate the central role of the deformation field in establishing the weblike pattern of the cosmic web. Lagrangian fluid dynamics emphasizes the role of gradients $\nabla \bm{s}_t$ in the displacement field $\bm{s}_t$, whose eigenvalues describe the compression and dilation along the principal axes of mass elements. At early times, the deformation tensor $\nabla \bm{s}_t$ is proportional to the Hessian of the gravitational potential, $\nabla \bm{s}_t \propto \mathcal{H}\phi_0$. From the eigenequation, it follows that the eigenvalue fields are the quadratic (or cubic) roots of a polynomial whose coefficients depend on the first-order gradients of the displacement field. As a result, the eigenvalue fields are distinctly non-linear. In fact, this non-linearity captures part of the non-linear dynamics of the corresponding gravitational evolution and collapse. It also translates into the distinct non-Gaussian stochastic character of the eigenvalue fields (and corresponding eigenvector fields), in turn, a reflection of the distinctly non-Gaussian weblike pattern that we already find back in the primordial eigenvalue fields (see figure~\ref{fig:Initial_Conditions}, bottom panels).

%%%%%%%%%%%%%%%%%%%%%%%%%%%%%%%%%%%%%%%%%%%%%%%%%%%%%%%%%%%%%%%%
\subsection{The Eulerian caustic skeleton}
\label{sec:nbody}
To develop a visual appreciation of the roles of the different caustics in the observed large-scale structure of the Universe, the most direct assessment is that based on the implied structure in Eulerian space. In the current study, the evolution is followed in two ways. The Zel'dovich approximation offers a -- first-order -- analytical expression for the displacement of all mass elements. By displacing the mass elements accordingly, we obtain the Eulerian location and outline of the various caustic features present in the evolved matter distribution.  A two-dimensional dark matter $N$-body simulation \cite{Hidding:2020} is used to follow the fully non-linear evolution of the mass distribution. In the $N$-body simulation, their location follows from the simulation displacement of the corresponding mass elements. The bottom panels of figure~\ref{fig:Eulerian} show the implied Eulerian structure of two descriptions for the non-linear evolution of structure. The bottom lefthand panel shows the structure implied by the Zel'dovich approximation, at the current cosmic epoch $z=0$, while the bottom righthand panel shows the structure
obtained through an $N$-body simulation. 

Both the Zel'dovich modeling as well as the $N$-body simulation involve a $512^2$ particle simulation, in a box of size $200 \mbox{Mpc}$, and evolve the primordial mass distribution specified in section~\ref{sec:primordial}. 

The bottom lefthand panel of figure~\ref{fig:Eulerian} illustrates immediately the failure of the Zel'dovich approximation to represent the evolving cosmic mass distribution. The approximation linearly extrapolates the initial flow of the mass elements and ignores any further change in the gravitational interactions. As a result, it leads to overshooting after shell-crossing. Nonetheless, even though the Zel'dovich approximation does not accurately predict the location of the multi-stream regions and density fields at advanced evolutionary stages, it does manage to accurately predict the locations and times at which mass elements undergo shell-crossing and form caustics in Lagrangian space at these late times.

The comparison with the resulting weblike matter distribution of the $N$-body simulation, in the bottom righthand panel, however, reveals the high level of accuracy with which the predicted caustic skeleton delineates the mass distribution. The fold curves ($A_2$, red) provide a reasonable approximation of the boundaries of the multi-stream regions. Meanwhile, the cusp curves ($A_3$, blue) neatly bisect the multi-stream regions and form knots in the clusters of the cosmic web. With respect to the latter, we should appreciate that a cluster rarely consists of a single cluster caustic. Instead, the related caustics are correlated and combine to form at the knots an intertwined and intricate set of multi-stream curves. Filaments, on the other hand, experience a more organized development. As a result, they are usually associated with a single and unique cusp curve. 

%%%%%%%%%%%%%%%%%%%%%%%%%%%%%%%%%%%%%%%%%%%%%%%%%%%%%%%%%%%%%%%%
\subsection{The evolving caustic skeleton}
The evolving structure of the cosmic web, and its diverse structural components, can be followed systematically by following the corresponding development of the different caustic features through cosmic time. Figure \ref{fig:Eulerian_Evolution} provides a timeline for the resulting development of the $A_2$ fold curves and the $A_3$ filamentary cusp curves of the cosmic web at 4 cosmic epochs ($a=0.2$, $a=0.4$, $a=0.6$ and $a=0.8$). On the mass distribution followed by an $N$-body simulation (black), we find superimposed the red fold lines, indicating the location of the mass elements entering a multi-stream region, and the blue cusp lines, indicating the filamentary spine of the cosmic web. 

The multi-stream regions of the cosmic web, and the corresponding caustic features, form and evolve gradually. At early times, the {\it dark matter sheet} consists of a single single-stream region. As the mass elements undergo gravitational contraction and collapse, we observe the formation of several Zel'dovich pancakes corresponding to the maxima of the eigenvalue field $\lambda_1$ (see also figure~\ref{fig:Eulerian}, upper left panel). These multi-stream regions grow and connect at the saddle points of the eigenvalue fields to form the web-like structure we observe in redshift surveys (see the upper right and lower panels).

From the evolutionary sequence in figure~\ref{fig:Eulerian_Evolution} we observed that fully contracted filaments move coherently, while the large voids expand and the smaller voids contract. The latter is a key aspect of the hierarchical buildup of the void population \cite[see][]{Sheth:2004}. Also, the filamentary ridge of the cosmic web builds up hierarchically, with several filaments merging while forming more massive filaments. By concentrating on the evolution of the (blue) cusp curves, we can follow this process in detail. Filaments merge at $A_4$ swallowtail junctions, revealing the details of the process in which the connectivity of the various cosmic web elements is established. Because of the explicit analytical expressions for this process in terms of the corresponding caustic condition for $A_4$ caustics, we are handed a full and explicit analytical description of the hierarchical establishment of the complex connectivity of the cosmic web.

%%%%%%%%%%%%%%%%%%%%%%%%%%%%%%%%%%%%%%%%%%%%%%%%%%%%%%%%%%
\section{The caustic skeleton in Lagangian space}\label{sec:Caustic_Skeleton_Theory}
While a Lagrangian fluid evolves, it may develop singular features known as caustics. These caustic are fully classified by Lagrangian catastrophe theory and play an important role in the development of the cosmic web. In this section, we give an elementary description of Lagrangian fluid dynamics and summarize the highlights of caustic skeleton theory in two dimensions. In the present section, we develop the mathematical formalism underlying the Caustic Skeleton. It summarizes the extensive -- Lagrangian -- formalism outlined in \cite{Feldbrugge:2018}. Here we present the specific formulation of the caustic conditions in terms of the 2D eigenvalues and eigenvectors of the deformation tensor, providing the language for the development of the non-linear constrained random field formalism for the various caustic classes.

%%%%%%%%%%%%%%%%%%%%%%%%%%%%%%%%%%%%%%%%%%%%%%%%%%%%%%%%%%%%%%%%
\subsection{Eulerian vs. Lagrangian perspective}
The gravitational evolution, contraction, and collapse of density fluctuations in an expanding universe may be described from various perspectives. Most treatments follow the Eulerian perspective. The evolution of density, velocity and gravitational potential of the matter fluid are considered in terms of a localized description of these quantities. The -- local mean -- density, velocity, and potential, evolve as described by the equation for the conservation of mass, the equation of motion through the Euler equation, and the Poisson equation for the potential corresponding to the density distribution. The Eulerian formalism is concise and leads to a reasonably accurate description of the mean flow in a large fraction of space.

Because the Eulerian description concerns local mean quantities, it is not able to describe the multi-stream nature of cosmic flow fields. To be able to describe cosmic flow fields consisting of several mass streams, each with its distinct mass content and velocity, a Lagrangian formalism offers a powerful and natural alternative for the description of the dynamics of the evolving cosmic matter distribution. 

%%%%%%%%%%%%%%%%%%%%%%%%%%%%%%%%%%%%%%%%%%%%%%%%%%%%%%%%%%%%%%%%
\subsection{Lagrangian fluid dynamics}
The Lagrangian formalism models the matter fluid in terms of a collection of mass elements uniformly filling space at the initial time. Subsequently, it follows the path and evolving physical properties of the individual mass elements. Identifying each mass element by its initial -- Lagrangian -- position $\bm{q}$, its path is specified in terms of its displacement $\bm{s}_t(\bm{q})$. The evolving spatial distribution of mass elements is then captured by the Lagrangian map $\bm{x}_t:\mathcal{L}\to \mathcal{E}$,
\begin{align}
\bm{x}_t(\bm{q}) = \bm{q} + \bm{s}_t(\bm{q})\,.
\end{align}
The Lagrangian map describes the position of a mass element starting from the position $\bm{q}$ in Lagrangian space $\mathcal{L}$ to the final position $\bm{x}_t(\bm{q})$ at time $t$ in Eulerian space $\mathcal{E}$, with the displacement field $\bm{s}_t(\bm{q})= \bm{x}_t(\bm{q}) - \bm{q}$. Each mass element can flow, stretch, and rotate while conserving its mass. The density of a Lagrangian fluid is derived from the Lagrangian map through the conservation of mass, \textit{i.e.},
\begin{align}
\rho_t(\bm{x})
&= \sum_{\bm{q} \in A_t(\bm{x})} \frac{\bar{\rho}}{|\det \nabla \bm{x}_t(\bm{q})|}\nonumber\\
&= \sum_{\bm{q} \in A_t(\bm{x})} \frac{\bar{\rho}}{|1+\mu_{t,1}(\bm{q})||1+\mu_{t,2}(\bm{q})|}\,,
\label{eq:density}
\end{align}
with the mean initial density $\bar{\rho}$, and the eigenvalue fields $\mu_{t,1}$ and $\mu_{t,2}$ of the deformation tensor $\nabla \bm{s}_t$, defined by the eigenequation
\begin{align}
\nabla \bm{s}_t(\bm{q}) \bm{v}_{t,i}(\bm{q}) = \mu_{t,i}(\bm{q}) \bm{v}_{t,i}(\bm{q})\,,
\label{eq:EigenvalueAndEigenvector}
\end{align} 
with the corresponding eigenvector fields $\bm{v}_{t,1}$, and $\bm{v}_{t,2}$. Each stream contributes to the density, as the sum runs over the points that reach $\bm{x}$ in time $t$, \textit{i.e.},
\begin{align}
A_t(\bm{x}') = \{\bm{q}\,|\,\bm{x}_t(\bm{q})=\bm{x}'\}\,.
\end{align} 
Lagrangian fluid dynamics is well-known to be an effective tool to model the cosmic web, as demonstrated by the well-known Zel'dovich approximation \cite{Zeldovich:1970}, second-order Lagrangian perturbation theory ($2$LPT) \cite{Buchert:1992, Buchert:1993a, Buchert:1993b, Buchert:1994a, Buchert:1994b, Bouchet:1995}, and $N$-body simulations \cite{Springel:2005, illustris:2014, eagle:2015, ABACUSSUMMIT:2021, Quijote:2020}. 

%%%%%%%%%%%%%%%%%%%%%%%%%%%%%%%%%%%%%%%%%%%%%%%%%%%%%%%%%%%%%%%%
\subsection{The phase-space sheet}
A two-dimensional Lagrangian fluid forms a two-dimensional sheet $M_t=\{(\bm{q},\bm{x}_t(\bm{q})) \, |\, \bm{q}\in \mathcal{L}\}$ in four-dimensional \textit{phase-space} $\mathcal{L}\times \mathcal{E}$, consisting of the initial and the final positions of the mass elements. \footnote{Note that there exists a direct correspondence between this definition of phase-space and the more conventional phase-space consisting of the position and momentum of a particle in Hamiltonian mechanics.} 
\begin{figure}
\centering
\begin{subfigure}[b]{0.49\textwidth}
\begin{tikzpicture}
%\draw[help lines, color=gray!30, dashed] (-0.9,-0.9) grid (4.9,4.9);
\draw[->, thick] (-0.4,0)--(5,0) node[right]{$\mathcal{L}$};
\draw[->, thick] (0,-0.4)--(0,5) node[above]{$\mathcal{E}$};
\draw[-,ultra thick, red] (0,0)--(4.9,4.9);
\filldraw[black] (2,2) circle (2pt);
\filldraw[black] (0,2) circle (2pt);
\filldraw[black] (3,3) circle (2pt);
\filldraw[black] (0,3) circle (2pt);
\draw[->] (1.8,2) -- node[above] {$\varphi$} ++ (-1.6,0);
\draw[->] (2.8,3) -- node[above] {$\varphi$} ++ (-2.6,0);
\end{tikzpicture}
\end{subfigure}~
\begin{subfigure}[b]{0.49\textwidth}
\begin{tikzpicture}
%\draw[help lines, color=gray!30, dashed] (-0.9,-0.9) grid (4.9,4.9);
\draw[->, thick] (-0.4,0)--(5,0) node[right]{$\mathcal{L}$};
\draw[->, thick] (0,-0.4)--(0,5) node[above]{$\mathcal{E}$};
\draw[|-|,ultra thick, blue] (0,0.9)--(0,2.85);
\draw [-, ultra thick, red] plot [smooth, tension=1] coordinates { (0,0) (1.75,2.75)  (3.5,1) (4.9,4.9)};
\filldraw[black] (0,2) circle (2pt);
\filldraw[black] (0.925,2) circle (2pt);
\filldraw[black] (2.575,2) circle (2pt);
\filldraw[black] (4.15,2) circle (2pt);
\filldraw[black] (4.725,4.) circle (2pt);
\draw[->] (3.95,2) -- node[above] {$\varphi$} ++ (-3.75,0);
\draw[->] (4.525,4.) -- node[above] {$\varphi$} ++ (-4.325,0);
\end{tikzpicture}
\end{subfigure}
\caption{Projection of phase-space to Eulerian space. The left panel illustrates the initial dark matter sheet in phase-space. The arrow represents the projection to Eulerian space $\varphi$. \textit{Left:} the initial dark matter sheet for which the mapping $\varphi$ is one-to-one. \textit{Right:} an evolved dark matter sheet consisting of a single- and a triple-stream region (in blue). Each point in the triple-stream region corresponds to three points on the dark matter sheet. The triple-stream region is bounded by a fold caustic at which the mass elements shell-cross.}\label{fig:Phase-Space}
\end{figure}
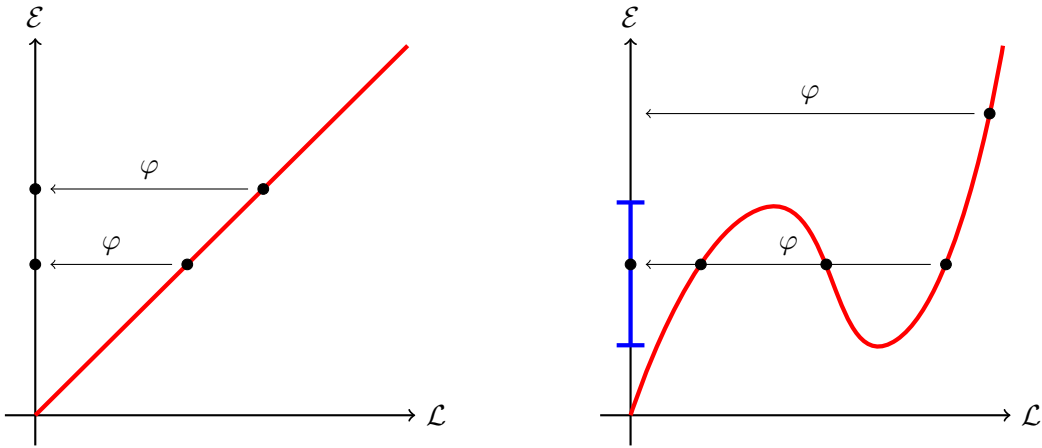

The cosmic web emerges in the projection of the phase-space sheet $M_t$ onto the final position $\varphi:\mathcal{L}\times \mathcal{E} \to \mathcal{E}$, defined by the map $(\bm{q},\bm{x}) \mapsto\bm{x}$ (see figure \ref{fig:Phase-Space} for an illustration). Initially, the displacement field vanishes, $\bm{s}_0(\bm{q})=\bm{0}$, and the phase-space manifold is diagonal $M_0=\{(\bm{q},\bm{q})\,|\,\bm{q} \in \mathcal{L}\}$. The mapping $\varphi$ is one-to-one, corresponding to a single-stream region (see the left panel of figure\ \ref{fig:Phase-Space}). When the fluid evolves, the phase-space manifold can develop complex configurations at which the mapping $\varphi$ becomes many-to-one, \textit{i.e.}, a final position can be reached from multiple initial positions.

A connected region, in Eulerian space, with $n$ possible initial positions is known as an $n$-stream region (see the triple-stream region in the right panel of figure \ref{fig:Phase-Space}). In these multi-stream regions, gravitational collapse becomes non-linear and virialized structures start to form. The multi-stream regions partition the universe into regions with distinct formation histories, based on the dynamics of gravitational collapse. The boundary of these multistream regions is delineated by the mass elements that are just undergoing
\textit{shell-crossing} as they collapse. These mass elements consitute the \textit{fold caustic}. 

At a fold, the orientation of the mass elements flips as they experience \textit{shell-crossing} and meet up with mass elements in
another stream. This happens as the deformation tensor becomes singular, \textit{i.e.}
\begin{equation}
  \det \nabla \bm{x}_t = 0 \,,
\end{equation}
\noindent while the density \eqref{eq:density} spikes to infinity. Within these multistream regions, we see the emergence of high-order caustics. These are associated with the membranes, filaments, and clusters of the cosmic web (see next section). The single-stream regions correspond colloquially to the voids, where the evolution of the cosmic web is well described by Eulerian fluid dynamics. A variety of recent studies have directed their attention to the phase-space structure of the cosmic matter distribution, and the corresponding identity of the various structural components of the cosmic web. This, in turn, provided a profound means of analyzing $N$-body simulations and identifying the multi-stream regions in these. \cite{Hahn:2007, Abel:2012, Shandarin:2012,  Feldbrugge:2014b, Ramachandra:2015, Ramachandra:2017, Shandarin:2019, Shandarin:2021}.

%%%%%%%%%%%%%%%%%%%%%%%%%%%%%%%%%%%%%%%%%%%%%%%%%%%%%%%%%%%%%%%%
\subsection{Caustics in Lagrangian fluids}
The study of the cosmic web in terms of multi-stream regions and Lagrangian catastrophe theory has a rich history, predating the developments of $N$-body simulations. Amazingly, by analyzing the geometry of the caustics emerging in Lagrangian fluid dynamics, Y.\ Zel'dovich, V.\ Arnol'd and collaborators successfully predicted the qualitative features of the cosmic web \cite{Arnold:1982a, Arnold:1982b, Shandarin:1983, Rozhanskii:1984, Shandarin:1989}. The large-scale structure is described in terms of a network of caustics, analogous to the light patterns emerging on the sea bed or the bottom of a swimming pool \cite{Berry:1977, Berry:1980, Feldbrugge:2019}. 

However, these studies were mainly restricted to two-dimensional models, as several technical problems inhibited the full treatment of the three-dimensional cosmic web. In a recent publication \cite{Feldbrugge:2018}, these problems have been solved by deriving a complete set of equations, the {\it caustic conditions}, specifying the occurrence of \textit{shell-crossing} in an evolving fluid in any $N$-dimensional space. These conditions follow from considering the \textit{shell-crossing condition} that describes how a submanifold in $\mathcal{L}$ can develop a non-differentiable feature under the mapping $\bm{x}_t$. Given a submanifold $L \subset \mathcal{L}$, the mapping $\bm{x}_t(L)$ develops a non-differentiable point at time $t$ in $\bm{q}_c$ when there exists a non-zero vector tangent vector $\bm{T}$ in the tangent space of $L$ at $\bm{q}_c$ for which 
\begin{align}
(1+\mu_{t,i}(\bm{q}_c))\bm{v}_{t,i}^*(\bm{q}_c) \cdot \bm{T}=0
\label{eq:shellCrossingCondition}
\end{align}
for all $i=1,2$, with $\bm{v}_{i,t}^*$ the dual eigenvector field defined by the relation $\bm{v}_{t,i}\cdot \bm{v}_{t,j}^* = \delta_{ij}$. The shell-crossing condition indicates the key role of the eigenvalue $\mu_{t,i}$ and eigenvector fields $\bm{v}_{t,i}$ of the deformation tensor $\nabla \bm{s}_t$ over the density field in the development of multi-stream regions.

An important observation is that, even at early times, the eigenvalue fields are distinctly non-Gaussian. As illustrated in figure~\ref{fig:Initial_Conditions}, already the initial eigenvalue fields show the web-like structure of the late-time multi-stream regions and mass distribution. In this context, the most outstanding aspect is that of the prominent role of eigenvector fields. These are often ignored, while they turn out to play a central role in the conditions for shell-crossing and the definition of higher-order caustics, and thus in the tracing of the filaments and clusters of the cosmic web. 

The condition for shell-crossing leads to a set of \textit{caustic conditions} describing the properties of the displacement field in the vicinity of the caustic that stably occur in nature. These caustic conditions ~\eqref{eq:shellCrossingCondition} have the advantage of enabling an efficient and fast numerical computation. Earlier versions of caustic conditions derived from Lagrangian catastrophe theory, \citep{Arnold:1972, Arnold:1976, Poston:1978, Gilmore:1981, Kravtsov:1983, Arnold:1984, Arnold:2012a, Arnold:2012b} were more difficult to implement, which mostly limited their application to two-dimensional space.

An inventory and summary of the different elements of the caustic skeleton are listed in table \ref{table:caustics}. For a complete contemporary description of caustic skeleton theory, we refer to \cite{Hidding:2014} and \cite{Feldbrugge:2018}. An illustration of the caustic skeleton, and its various caustic constituent, is presented in figure~\ref{fig:caustics_Examples_Big}. In four rows, the caustic skeleton in four different regions of size $[50\text{ Mpc}\times 50 \text{ Mpc}]$ is shown. The caustic skeleton in the figure is evaluated, in Lagrangian space, by means of the analytical expression for the deformation tensor in the Zeldovich approximation (see sect.~\ref{sec:Zeldovich}). The lefthand column shows the resulting caustic skeleton in Lagrangian space, superimposed on the (initial, \textit{i.e.}, Lagrangian) density field. The second and third column shows the corresponding caustic skeleton in Eulerian space. The second and third columns differ in the mapping of the mass elements from Lagrangian to Eulerian space. The panels in the second column involve the displacement by means of the analytical first-order expression of the Zeldovich approximation, and the panels in the third column involve the full displacement of the corresponding $N-$body simulation. 

The four regions in the figure center around a different caustic feature. The top row centers around a (filamentary) cusp caustic, the second row around a swallowtail (point) caustic, the third row around an elliptic umbilic (point) caustic, and the fourth row around a hyperbolic umbilic caustic. The $A_2$ fold lines (red) delineate the boundaries of the multi-stream regions. The $A_3$ cusp curves (blue) bisect the multi-stream regions and trace the filamentary ridges of the skeleton. A key observation is that the swallowtail and umbilic caustics (2nd, 3rd and 4th row) individually connect only to three filamentary features. In section~\ref{sec:caustic_skeleton_constraints} of the present study, we will demonstrate that this is a generic property of the caustic skeleton).

 The skeleton in Eulerian space shown in these zoom-in panels is the one resulting from the Lagrangian-Eulerian mapping through the displacement specified by Zeldovich approximation. The ballistic first-order expression for the displacement in the Zeldovich approximation means that it is not surprising that the fold curve does not exactly match the shell-crossing regions seen in the $N$-body simulation (\textit{cf.} the Zeldovich vs.\ $N$=body panels in figure~\ref{fig:caustics_Examples_Big}). Nonetheless, it is quite telling that the caustics of the Zeldovich approximation are very close to the caustics emerging in the fully non-linear model. An interesting detail is an observation that the cluster caustics, \textit{i.e.}, the swallowtail and umbilic caustics, are rarely isolated. They show strong clustering and together form a more intricate multi-stream structure. In this way, a cluster may connect to more than three filaments,  even though the individual swallowtail and umbilic caustics connect to only three filaments.

 %%%%%%%%%%%%%%%%%%%%%%%%%%%%%%%%%%%%%%%%%%%%%%%%%%%%%%%%%%%%%%%%
\subsection{Caustic classes \& caustic conditions} 
\begin{figure}
\centering
\begin{subfigure}[b]{0.28\textwidth}
\includegraphics[width=\textwidth]{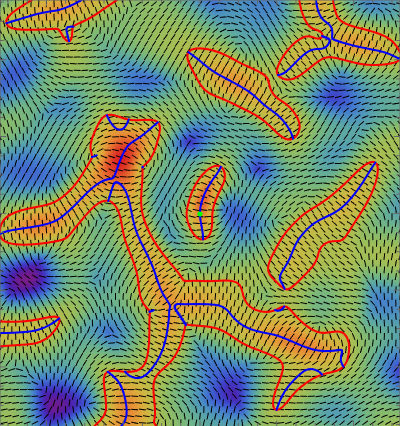}
\end{subfigure}~
\begin{subfigure}[b]{0.28\textwidth}
\includegraphics[width=\textwidth]{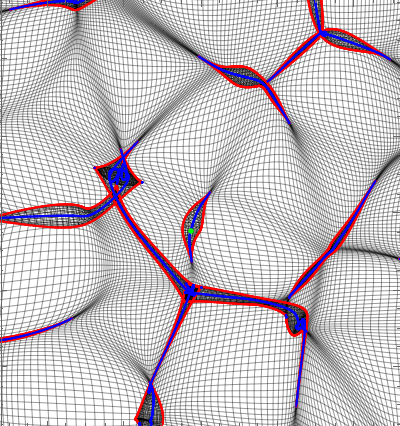}
\end{subfigure}~
\begin{subfigure}[b]{0.28\textwidth}
\includegraphics[width=\textwidth]{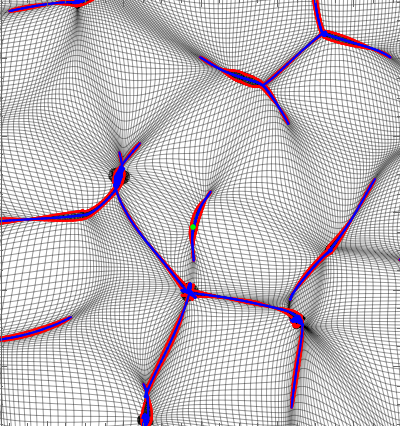}
\end{subfigure}\\ \vspace{0.2\baselineskip}
%%%%%%%%%%
%%%%%%%%%%
\begin{subfigure}[b]{0.28\textwidth}
\includegraphics[width=\textwidth]{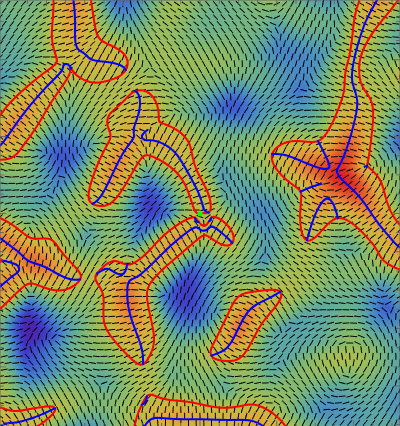}
\end{subfigure}~
\begin{subfigure}[b]{0.28\textwidth}
\includegraphics[width=\textwidth]{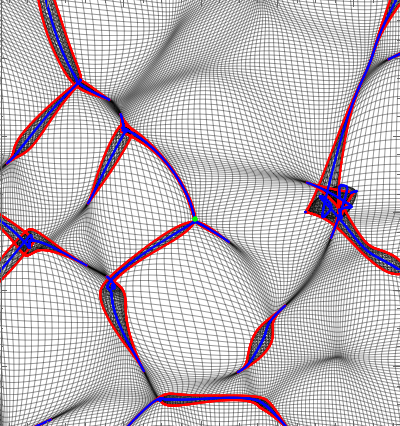}
\end{subfigure}~
\begin{subfigure}[b]{0.28\textwidth}
\includegraphics[width=\textwidth]{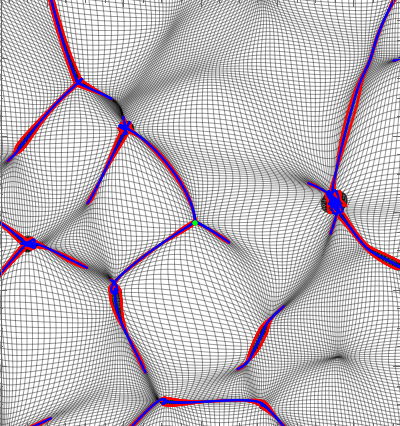}
\end{subfigure}\\ \vspace{0.2\baselineskip}
%%%%%%%%%%
%%%%%%%%%%
\begin{subfigure}[b]{0.28\textwidth}
\includegraphics[width=\textwidth]{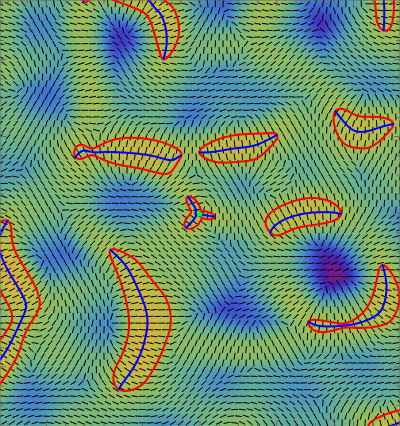}
\end{subfigure}~
\begin{subfigure}[b]{0.28\textwidth}
\includegraphics[width=\textwidth]{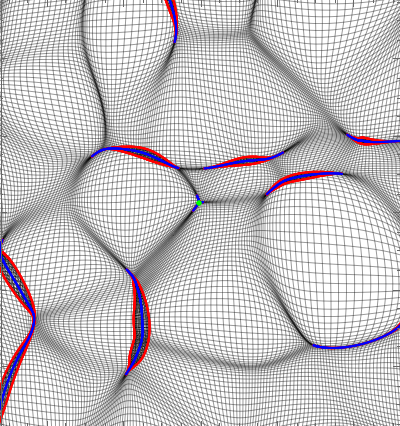}
\end{subfigure}~
\begin{subfigure}[b]{0.28\textwidth}
\includegraphics[width=\textwidth]{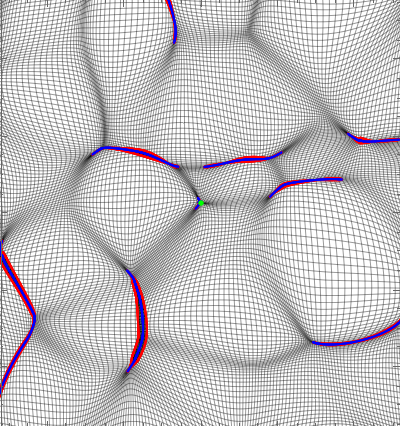}
\end{subfigure}\\ \vspace{0.2\baselineskip}
%%%%%%%%%%
%%%%%%%%%%
\begin{subfigure}[b]{0.28\textwidth}
\includegraphics[width=\textwidth]{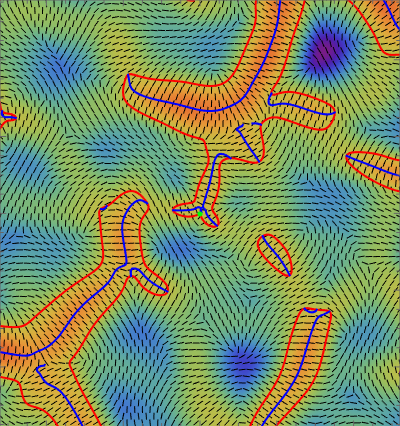}
\end{subfigure}~
\begin{subfigure}[b]{0.28\textwidth}
\includegraphics[width=\textwidth]{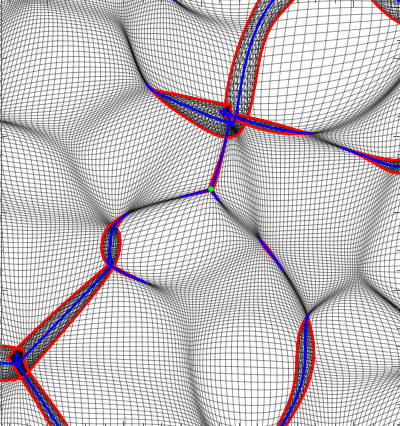}
\end{subfigure}~
\begin{subfigure}[b]{0.28\textwidth}
\includegraphics[width=\textwidth]{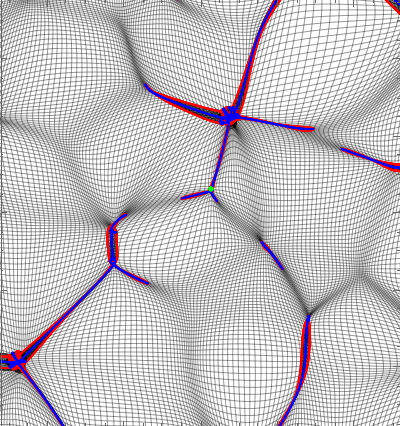}
\end{subfigure}
\caption{Elements of the caustic skeleton consisting of the fold curves (red), cusp curves (blue) and swallowtail and umbilic caustics (green dots)
  in Lagrangian and Eulerian space in a $[50\text{ Mpc}\times 50 \text{ Mpc}]$ box. \textit{From top to bottom:} the cusp, the swallowtail, the elliptic, and the hyperbolic caustic. \textit{Left:} the caustic skeleton in Lagrangian space. \textit{Centre:} the Zel'dovich approximation. \textit{Right:} the corresponding $N$-body simulation \cite{Hidding:2020}.}\label{fig:caustics_Examples_Big}
\end{figure}

When applying the shell-crossing condition to Lagrangian space, $L=\mathcal{L}$, we obtain the condition for the fold caustic $A_2$,
\begin{align}
1+\mu_{t,i}(\bm{q}_c)=0
\end{align}
for $i=1$ or $2$, marking the shell-crossing region at which the density spikes. The fold curve is independent of the eigenvector fields. The fold curve bounds the different multi-stream regions (see the red curves in figure~\ref{fig:caustics_Examples_Big}).

\bigskip
The fold curve develops non-differentiable points in the cusp caustic $A_3$, when the fold curve in Lagrangian space is parallel to the corresponding eigenvector field, \textit{i.e.}, 
\begin{align}
1+\mu_{t,i}(\bm{q}_c)=0\,,\\
\bm{v}_{t,i}(\bm{q}_c) \cdot \nabla \mu_{t,i}(\bm{q}_c)=0\,,\label{eq:cuspCondition}
\end{align}
The intersection of the red and blue curves in figure \ref{fig:caustics_Examples_Big} provides a visual illustration, having the tangent vector of the red line is parallel to the eigenvector field. The term $\bm{v}_{t,i} \cdot \nabla \mu_{t,i}$ arises in the shell-crossing condition \eqref{eq:shellCrossingCondition} with $L=A_2$, from the observation that the tangent vector of the fold curve $\bm{T}$ is normal to the gradient of the eigenvalue field $\nabla \mu_{t,i}$. Over time, the cusp point defines the cusp curve which is associated with the filaments of the cosmic web. The blue curve in the first row of figure~\ref{fig:caustics_Examples_Big} is an example of a fold curve and shows its relation to the filaments of the cosmic web. 

\bigskip
In turn, the cusp curve develops a non-differentiable point known as a swallowtail caustic $A_4$ when the cusp curve in Lagrangian space is parallel to the eigenvector field, \textit{i.e.},
\begin{align}
1+\mu_{t,i}(\bm{q}_c)=0\,,\\
\bm{v}_{t,i}(\bm{q}_c) \cdot \nabla \mu_{t,i}(\bm{q}_c)=0\,,\\
\bm{v}_{t,i}(\bm{q}_c) \cdot \nabla (\bm{v}_{t,i}(\bm{q}_c) \cdot \nabla \mu_{t,i}(\bm{q}_c)) = 0\,.
\end{align}
The swallowtail caustic only exists for an instance in time and marks the location of a cluster, highlighting the location where different cusp curves join (see the second row of figure \ref{fig:caustics_Examples_Big}). 

\bigskip
In addition to the swallowtail caustic, the Lagrangian fluid also develops point-like caustics when both eigenvalue fields lead to a spike in the density field. The points for which
\begin{align}
1+\mu_{t,1}(\bm{q}_c) =0\,,\\
1+\mu_{t,2}(\bm{q}_c) = 0\,,
\end{align}
are known as the umbilic caustics $D_4^\pm$. The umbilic caustics consist of the elliptic $D_4^+$ and hyperbolic caustics $D_4^-$, both related to the clusters of the cosmic web. The umbilic caustics mark the locations where the cusp curves corresponding to the first and the second eigenvalue fields join (see the third and fourth rows of figure \ref{fig:caustics_Examples_Big}). 

\begin{figure}
\centering
\begin{subfigure}[b]{0.24\textwidth}
\includegraphics[width=\textwidth]{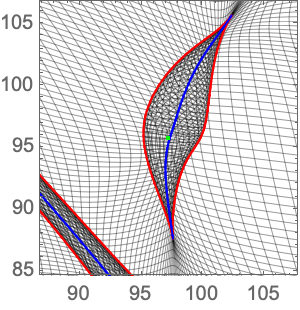}
\end{subfigure}~
\begin{subfigure}[b]{0.24\textwidth}
\includegraphics[width=\textwidth]{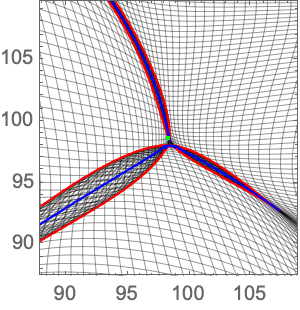}
\end{subfigure}~
\begin{subfigure}[b]{0.24\textwidth}
\includegraphics[width=\textwidth]{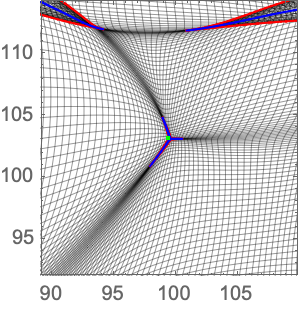}
\end{subfigure}~
\begin{subfigure}[b]{0.24\textwidth}
\includegraphics[width=\textwidth]{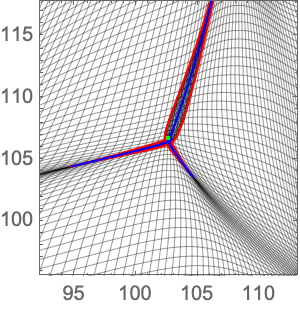}
\end{subfigure}\\
\begin{subfigure}[b]{0.24\textwidth}
\includegraphics[width=\textwidth]{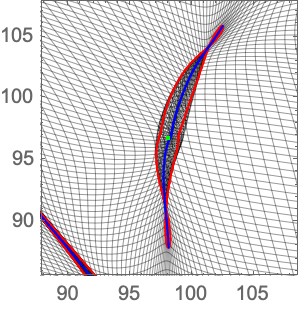}
\end{subfigure}~
\begin{subfigure}[b]{0.24\textwidth}
\includegraphics[width=\textwidth]{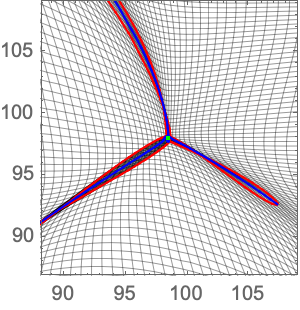}
\end{subfigure}~
\begin{subfigure}[b]{0.24\textwidth}
\includegraphics[width=\textwidth]{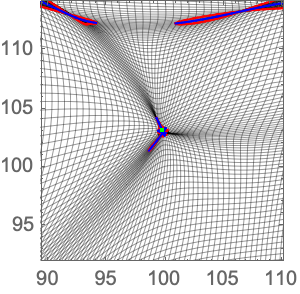}
\end{subfigure}~
\begin{subfigure}[b]{0.24\textwidth}
\includegraphics[width=\textwidth]{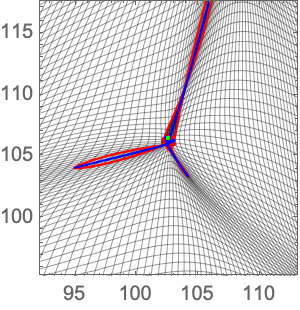}
\end{subfigure}
\caption{Zoom in with a $(20 \text{ Mpc})^2$ box centered around the caustic, with the fold curve (red) and the cusp curve (blue). The swallowtail caustic and the umbilic caustics (green) mark the knots where the filaments merge. From left to right, we see the cusp, the swallowtail, the elliptic umbilic, and the hyperbolic umbilic caustics. The upper and lower panels display the Zel'dovich approximation and the $N$-body simulation \cite{Hidding:2020}.}\label{fig:caustics_Examples_Small}
\end{figure}

\begin{table}
\centering
{\scriptsize
\begin{tabular}{ |l | l | l | l | l|}
\hline
  \ &&&& \\
\textbf{Name} & \textbf{Symbol} & \textbf{2D cosmic web} && \textbf{Caustic conditions}\\
  \ &&&& \\
\hline
  \ &&&& \\
Fold & $A_2$ & shall-crossing && $1+ \mu_{t,i} = 0$ \\
Cusp & $A_3$ & filament && $1+ \mu_{t,i} = 0$, $\bm{v}_i \cdot \nabla \mu_{t,i} = 0$\\
Swallowtail &$A_4$ &  cluster && $1+ \mu_{t,i} = 0$, $\bm{v}_i \cdot \nabla \mu_{t,i} = 0,$ $\bm{v}_i \cdot \nabla(\bm{v}_i \cdot \nabla \mu_{i,t}) = 0$\\
Elliptic/hyperbolic & $D_4^{\pm}$ & cluster && $1+ \mu_{t,1} = 1+ \mu_{t,2} = 0$\\
Morse point & $A_3^+$ & creation/annihilation point&& maximum/minimum of the eigenvalue field $\lambda_i$\\
Morse point & $A_3^-$ & merger point && saddle point of the eigenvalue field $\lambda_i$\\
  \ &&&& \\
\hline
\end{tabular}
}
\caption{Elements of the two-dimensional caustic skeleton and their caustic conditions.}
\label{table:caustics}
\end{table}

\bigskip
A zoom-in of the various caustics in the Zel'dovich approximation and the $N$-body simulation, from the same simulation as in figure~\ref{fig:caustics_Examples_Big}, is shown in figure \ref{fig:caustics_Examples_Small}. The eight panels show the Eulerian space caustic skeleton in different regions of $20 \text{ Mpc} \times 20 \text{ Mpc}$. With the caustic conditions being evaluated on the basis of the Zel'dovich approximation (see sect.~\ref{sec:Zeldovich}), the depicted skeleton is obtained by mapping from Lagrangian to Eulerian space by the ballistic first-order Zeldovich displacement expression. The resulting skeleton structure depicts various caustic features. 

The (red) fold lines delineate the multi-stream regions, while the (blue) cusp curves that trace the filaments bisect these multi-stream regions. The swallowtail and umbilic caustics (green dots) mark the knots where the filaments merge.

%%%%%%%%%%%%%%%%%%%%%%%%%%%%%%%%%%%%%%%%%%%%%%%%%%%%%%%%%%%%%%%%
\subsection{Morse points}
A final important element of the caustic skeleton is that of the Morse points. These are the critical points at which the multi-stream regions emerge, vanish, and merge, changing the topology and connectivity of the cosmic web. The Morse points are defined as the critical points of the eigenvalue fields, satisfying the condition
\begin{align}
\nabla \mu_{t,i}(\bm{q}_c)=\bm{0}\,.
\end{align}
A critical point that undergoes shell-crossing always lies on a cusp curve, as a point with a vanishing gradient automatically satisfies equation \eqref{eq:cuspCondition}. The maxima and minima of the eigenvalue field $\mu_i$, known as $A_3^+$ points, mark the locations and times at which a multi-stream region forms or vanishes. The saddle point of the eigenvalue field $\mu_{t,i}$, known as $A_3^-$ points, mark the location and time at which two multi-stream regions merge to form a larger structure. The Morse points $A_3^-$ determine the connectivity of the cosmic web.

\bigskip
The study of the Morse points of the caustic skeleton is intimately coupled to the topological character of the matter distribution. Morse theory \citep{Milnor:1963} establishes the relation between the spatial distribution of the field's singularities and topological transitions: the topology of manifold changes at its singularities. It is at the field level of a singularity that we see the emergence of a new topological feature or its disappearance as it merges with neighboring features. This property has been exploited in a range of cosmic web classification schemes, such as that of DisPerSE \cite{Sousbie:2011a, Sousbie:2011b}, Spineweb \cite{Aragon:2010b} and Felix \citep{Shivashankar:2016}. A key -- and often tacit -- presumption is that the structure and connectivity of the cosmic web are determined by the singularity structure of the density field. The maxima in the density field are connected via integral lines, and the assumption is that the filaments of the cosmic web are associated with these integral lines.

Of instrumental significance in our caustic skeleton model is that the physically relevant fields for the connections in the cosmic web are that of the deformation field, specifically that of its eigenvalues. In this paper, we, therefore, focus on the deformation tensor field instead of the density field. There are several arguments for assessing the evolution of the cosmic web on the basis of the deformation tensor. First, it follows the Zel'dovich approximation (see sect.~\ref{sec:zeldovich} \citep{Zeldovich:1970,Shandarin:1989,Shandarin:2009}), and implicitly its successful description of the first stages of the buildup of the cosmic web.  Secondly, it also provides a natural means of following the (hierarchical) buildup of the cosmic web. Assessing the eigenvalue field through a sequence of eigenvalue thresholds corresponds to following the evolution of the cosmic web in time, through the latter's specification in terms of the (linear) growing mode and the corresponding expression for the density evolution in the Zeldovich approximation (see eqn.~\ref{eq:denszeld}). Finally, and perhaps most outstanding, is the realization that rather than connecting saddle points and maxima via integral lines satisfying a differential equation, the caustic skeleton connects the critical points using the simpler level sets \eqref{eq:cuspCondition} corresponding to the cusp caustic. 

%%%%%%%%%%%%%%%%%%%%%%%%%%%%%%%%%%%%%%%%%%%%%%%%%%%%%%%%%%%%%%%%
\subsection{The Zel'dovich approximation}\label{sec:Zeldovich}
\label{sec:zeldovich}
In this paper, we define the caustic skeleton using the Zel'dovich approximation \cite{Zeldovich:1970} and study the resulting structures using a dark matter $N$-body simulation \cite{Hidding:2020}. The Zel'dovich approximation is the first-order approximation in Lagrangian fluid dynamics, describing a ballistic motion of the fluid elements in which their displacement $\bm{s}_t(\bm{q})$ factorizes into a spatial and a temporal part
\begin{align}
\bm{s}_t(\bm{q}) = - b_+(t) \nabla_{\bm{q}} \Psi(\bm{q})\,.
\end{align}
The growing mode $b_+(t)$ is a natural time parameter, increasing monotonically, satisfying the differential equation
\begin{align}
\frac{\mathrm{d}^2 b_+(t)}{\mathrm{d}t^2} + 2 \frac{\dot{a}(t)}{a(t)} \frac{\mathrm{d} b_+(t)}{\mathrm{d}t} = 4 \pi G \rho_u(t) b_+(t)\,,
\end{align}
in terms of the scale factor $a$, the mean density $\rho_u = \langle \rho \rangle$, and Newton's gravitational constant $G$. The displacement potential $\Psi$ captures the geometry of the cosmic web
\begin{align}
\Psi(\bm{q}) &=\frac{1}{4 \pi G a^2 \rho_0}\phi_0(\bm{q})%\\
%&= \frac{2}{3\Omega_0 H_0^2}\phi_0(\bm{q})
\,,
\end{align}
expressed in terms of the current total energy density %$\Omega_0$ and
$\rho_0$,
%the current Hubble parameter $H_0$,
and the linearly extrapolated gravitational potential $\phi_0$.

The Zel'dovich approximation describes the evolution of mass elements in terms of their ballistic motion. The mass elements follow linear trajectories determined by the primordial gravitational potential. The approximation accurately describes single-stream regions at early times but fails at late times in multi-stream regions when gravitational backreaction of the mass elements becomes important (see the lower left panel of \ref{fig:Eulerian}).

\bigskip
While the motion of the mass elements starts to deviate from their linear trajectorties as the mass distribution evolves, the Zeldovich approximation does allow us to accurately follow the density field evolution up to advanced stages. This is established through the 1-1 relation between the displacement potential $\Psi(\bm{q})$ and the initial gravitational potential $\phi_0$, whereby the latter establishes a relation between the primordial density perturbations through the Poisson equation,
\begin{align}
\nabla ^2 \phi_0 &= 4 \pi G a^2 \rho_u \delta_0\,,%\\
%&= \frac{3}{2} \Omega H^2 a^2 \delta\,,
\end{align}
with the dimensionless density contrast
\begin{align}
\delta(\bm{q}) = \frac{\rho(\bm{q})}{\rho_u} -1\,.
\end{align}
When working with the Zel'dovich approximation, it is convenient to work in terms of the Hessian $\bm{\psi}$ of the displacement potential,
\begin{align}
  \bm{\psi}=\mathcal{H}\Psi\,,\qquad \psi_{ij}=\frac{\partial^2 \Psi}{\partial q_i \partial q_j}\,,
  \label{eq:deformatiozeld}
\end{align}
and the corresponding eigenvalue $\lambda_i$ and $\bm{v}_i$
\begin{align}
\bm{\psi}\bm{v}_i = \lambda_i \bm{v}_i\,,
\end{align} 
at early times satisfying the relation $\lambda_1 =-\mu_2/b_+,\lambda_2 =-\mu_1/b_+$. For convenience, we will assume the ordering $\lambda_1(\bm{q}) \geq \lambda_2(\bm{q})$ such that the first collapse corresponds to the first eigenvalue field $\lambda_1$. In terms of the eigenvalue fields $\lambda_i$, the density takes the form 
\begin{align}
\rho_t(\bm{x})
&= \sum_{\bm{q} \in A_t(\bm{x})} \frac{\bar{\rho}}{|1-b_+(t) \lambda_1(\bm{q})||1-b_+(t) \lambda_2(\bm{q})|}\,.
\label{eq:denszeld}
\end{align}

\bigskip
\noindent Initially, the (linear) growing mode $b_+(t)$ vanishes and the universe consists of a single-stream region. As the fluid contracts under self-gravity, the growing mode $b_+$ increases. At a time $t$, with corresponding growing mode $b_+(t)$, we can identify the fluid elements that undergo full gravitational collapse. They are the fluid elements, identified by their Lagrangian location $\bm{q}$, for whom the
eigenvalue $\lambda_i(\bm{q})$ is inversely proportional to the growing mode $b_+(t)$, 
\begin{equation}
  \{\bm{q}\,|\,\lambda_i(\bm{q})=1/b_+(t)\}\,.
  \end{equation}
Through this direct relation between gravitational collapse and eigenvalue levels, Morse theory establishes a direct connection between the critical points of the eigenvalue field, the emergence of structural components of the cosmic web and their connectivity. 

Even though the distribution of the mass elements is not accurately described at late times, we find that the approximation does reliably describe the formation of caustics in the non-linear evolution of the cosmic web. In this paper, we will always work with the caustic skeleton of the Zeld'dovich approximation in Lagrangian space. This means we will use the first-order displacement expression and corresponding deformation tensor expression of the Zeldovich approximation, \textit{i.e.}, expression \eqref{eq:deformatiozeld} to evaluate the caustic conditions for the emergence of caustics in the cosmic mass distribution.

We subsequently move the skeleton to Eulerian space with either the Zel'dovich approximation or an $N$-body simulation. For this reason, the skeleton does not absolutely match the shell-crossing regions in the $N$-body simulations. However, it turns out to be a good first approximation (as shown by the example figure \ref{fig:caustics_Examples_Small}).

%%%%%%%%%%%%%%%%%%%%%%%%%%%%%%%%%%%%%%%%%%%%%%%%%%%%%%%%%%%%%%%%
\section{Constrained random field theory}\label{sec:GRF}
One of the most startling realizations in modern cosmology is that the intricate cosmic web, observed in several cosmological redshift surveys, originated from extremely simple initial conditions. At the moment of recombination, when the baryons and electrons form neutral elements and decouple from the photons, the energy density in our universe, was extremely homogeneous with only tiny density fluctuations. The imprints of these fluctuations have over the last decades been measured with increasing accuracy in the temperature maps of the Cosmic Microwave Background field (CMB) \cite{WMAP:2003, Planck:2016}. Several contemporary paradigms for the big bang, such as inflation theory, envisage these fluctuations to have a quantum origin and interpret them as a realization of a Gaussian random field with only small non-Gaussian deviations. This is in accordance with the most recent statistical analyses of the CMB \cite {Creminelli:2006, Planck:2020} and forms the underpinning of the study of the cosmic web. In this section, we will give a concise definition of Gaussian random field theory. We discuss the implementation of linear constrained Gaussian random field theory, following the notation presented in \cite{Weygaert:1996}, and extend these conditions to non-linear constraints. This will allow us to efficiently implement the caustic conditions and study the properties of the emerging structures.

%%%%%%%%%%%%%%%%%%%%%%%%%%%%%%%%%%%%%%%%%%%%%%%%%%%%%%%%%%%%%%%%
\subsection{Gaussian random fields}
A two-dimensional Gaussian random field $f:\mathbb{R}^2\to \mathbb{R}$ is a generalization of a multi-dimensional normal distribution to the continuum, defined by the probability density
\begin{align}
p(f) = \mathcal{N} e^{- S[f]}\,, \label{eq:functional_Distribution}
\end{align}
with the normalization constant $\mathcal{N}$ and the `action' (in analogy with the Euclidean path integral \cite{Feynman:1965})
\begin{align}
S[f]\equiv \frac{1}{2} \iint [f(\bm{q}_1) - \bar{f}(\bm{q}_1)] K(\bm{q}_1,\bm{q}_2) [f(\bm{q}_2) -\bar{f}(\bm{q}_2)]\mathrm{d}\bm{q}_1 \mathrm{d}\bm{q}_2,\label{eq:action}
\end{align}
defined in terms of the mean field $\bar{f}(\bm{q})$ and the kernel $K(\bm{q}_1,\bm{q}_2)$ \cite{Longuet-Higgins:1957, Adler:1981, bbks:1986}. The probability that the random field $f$ is included in a set of functions $\mathcal{S}$ is defined by the path integral
\begin{align}
P[f \in \mathcal{S}] = \mathcal{N} \int \bm{1}_\mathcal{S}(f) e^{-S[f]}\,\mathcal{D}f\,,
\end{align}
with $\bm{1}_\mathcal{S}$ the identity function\footnote{Defined by $\bm{1}_\mathcal{S}(x)=1$ when $x \in \mathcal{S}$ and $\bm{1}_\mathcal{S}(x)=0$ when $x \notin \mathcal{S}$.} and $\mathcal{D}f$ the path integral measure. The expectation value of a functional $Q[f]$ is given by
\begin{align}
\left\langle Q[f] \right\rangle = \mathcal{N}\int Q[f]\, e^{-S[f]}\,\mathcal{D}f\,,
\end{align}
analogous to the Euclidean path integrals in statistical field theory. It can be shown that the expectation value of the Gaussian random field is given by the mean field
\begin{align}
\langle f(\bm{q})\rangle &= \bar{f}(\bm{q})\,,
\end{align}
and that the two-point correlation function
\begin{align}
\xi(\bm{q}_1, \bm{q}_2) &= \langle (f(\bm{q}_1) - \bar{f}(\bm{q}_1)) (f(\bm{q}_2) - \bar{f}(\bm{q}_2))\rangle\nonumber\\
&= \int (f(\bm{q}_1) - \bar{f}(\bm{q}_1)) (f(\bm{q}_2) - \bar{f}(\bm{q}_2)) e^{-S[f]}\mathcal{D}f
\end{align}
is the inverse of the kernel $K$, \textit{i.e.,}
\begin{align}
\int K(\bm{q}_1,\bm{q}) \xi(\bm{q},\bm{q}_2) \mathrm{d}\bm{q}= \delta_D^{(2)}(\bm{q}_1-\bm{q}_2)\,,\label{eq:defK}
\end{align}
with the two-dimensional Dirac delta function $\delta_D^{(2)}$. The Gaussian random field is thus fully determined by the mean field $\bar{f}$ and the two-point correlation function $\xi$. 

In cosmology, the cosmological principle often leads to the study of statistically homogeneous and isotropic random fields for which the mean field is constant $\bar{f}(\bm{q})=\bar{f}=0$ and the two-point correlation function only depends on the magnitude of the difference of the inserted points, \textit{i.e.}, $\xi(\bm{q}_1,\bm{q}_2)=\xi(\|\bm{q}_1-\bm{q}_2\|)$, and consequently $K(\bm{q}_1,\bm{q}_2)=K(\|\bm{q}_1-\bm{q}_2\|)$. In the present paper, we work with statistically homogeneous and isotropic Gaussian random fields. However, the theory generalizes to generic Gaussian random fields.

The statistical properties of homogeneous and isotropic random fields are most transparently expressed in terms of the Fourier transform of the random field
\begin{align}
\hat{f}(\bm{k}) = \int f(\bm{q})e^{i\bm{k}\cdot \bm{q}}\mathrm{d}\bm{q}\,,
\end{align}
and the inverse transform
\begin{align}
f(\bm{q}) = \int \hat{f}(\bm{k})e^{-i \bm{k} \cdot \bm{q}}\frac{\mathrm{d}\bm{k}}{(2\pi)^2}\,.
\end{align}
The Fourier modes of real-valued Gaussian random fields satisfy the reality condition $\hat{f}(\bm{k}) = \hat{f}^*(-\bm{k})$. Using the double convolution theorem, we express the action \eqref{eq:action} as
\begin{align}
S[f] = \frac{1}{2} \int |\hat{f}(\bm{k})|^2 \hat{K}(\bm{k}) \frac{\mathrm{d}\bm{k}}{(2\pi)^2}\,.
\end{align}
In Fourier space, equation \eqref{eq:defK} takes the form
\begin{align}
\int \hat{K}(\bm{k})\, P(\bm{k})\, e^{i\bm{k}(\bm{q}_1-\bm{q}_2)} \frac{\mathrm{d}\bm{k}}{(2\pi)^2} = \delta_D^{(2)}(\bm{q}_1 - \bm{q}_2)\,,
\end{align}
with the power spectrum defined as the Fourier transform of the two-point correlation function,
\begin{align}
P(\bm{k}) = \int \xi(\bm{q}) \, e^{i\bm{k}\cdot \bm{q}}\mathrm{d}\bm{q}\,,
\end{align}
implying the relation $\hat{K}(\bm{k}) = 1/P(\bm{k})$. For a statistically homogeneous and isotropic random field $P(\bm{k})=P(\|\bm{k}\|)$. The resulting probability density of the Fourier modes is diagonal
\begin{align}
p(\hat{f}) \propto \exp\left[ -\frac{1}{2} \int \frac{|\hat{f}(\bm{k})|^2}{P(\bm{k})} \frac{\mathrm{d}\bm{k}}{(2\pi)^2}\right]\,,
\end{align}
implying the covariance
\begin{align}
\langle \hat{f}(\bm{k}_1)\hat{f}^*(\bm{k}_2) \rangle = (2\pi)^2 \delta_D^{(2)}(\bm{k}_1-\bm{k}_2) P(\bm{k}_1)\,.
\end{align}

% \begin{framed}
% In this paper, we for simplicity always assume the initial gravitational potential to be a realization of a homogeneous and isotropic Gaussian random field with a power-law power spectrum $P(k) \propto k^{n_s}$ smoothed with a Gaussian filter $W(k)=e^{-R_s^2 k^2/2}$ with the spectral index $n_s$ and the smoothing scale $R_s$. The effective power spectrum takes the form 
% \begin{align}
% P_{eff}(k)\propto k^{ns}e^{-R_s^2 k^2}\,.
% \end{align}
% Unless otherwise specified, we use a spectral index $n_s=-1$ and a smoothing length scale $R_s = 5\text{ Mpc}$. 
% \end{framed}

In practice, we often consider realizations of Gaussian random fields on a lattice, or more generally a finite set of linear statistics $\bm{Y}=(Y_1,Y_2,\dots,Y_M)$. In this setting, the functional distribution \eqref{eq:functional_Distribution} reduces to the multi-dimensional Gaussian distribution,
\begin{align}
p(\bm{Y}) = \frac{\exp\left[-\frac{1}{2}  \Delta \bm{Y}^T M^{-1} \Delta \bm{Y}\right]}{[(2\pi)^n \det M]^{1/2}}\,,
\end{align}
with the deviation from the mean $\Delta \bm{Y} = \bm{Y} - \langle \bm{Y}\rangle$ and the covariance matrix
\begin{align}
M = \text{cov}(\bm{Y},\bm{Y}) = \langle \Delta \bm{Y}^T \Delta \bm{Y}\rangle\,.
\end{align}
The distribution of the discrete Fourier modes $\bm{Y}=(\hat{f}(\bm{k}_1),\hat{f}(\bm{k}_2),\dots)$ takes the form 
\begin{align}
p\left(\hat{f}(\bm{k}_1), \hat{f}(\bm{k}_2), \dots\right) = \prod_{i} \frac{1}{\sqrt{2\pi P( \bm{k}_i)}} \exp\left[-\frac{|\hat{f}(\bm{k}_i)|^2}{2P(\bm{k}_i)}\right],
\end{align}
yielding an efficient method to generate realizations (see appendix \ref{ap:GRF}).

The statistical properties of random fields are often conveniently expressed in terms of the moments
\begin{align}
\sigma_i^2 &= \frac{1}{(2\pi)^2} \int \|\bm{k}\|^{2i}P(\bm{k})\mathrm{d}\bm{k}\nonumber\\
&= \frac{1}{2\pi} \int_0^\infty k^{2i+1}P(k)\mathrm{d}k\,,
\end{align}
with the magnitude $k = \|\bm{k}\|$. The power spectrum determines the covariance matrix $M$ through these moments. Whereas the variance of the random field is given by the moment $\sigma_0^2$, \textit{i.e.},
\begin{align}
\langle f(\bm{q})^2 \rangle &= \iint e^{i (\bm{p} - \bm{k})\cdot \bm{q}} \langle \hat{f}(\bm{k})\hat{f}^*(\bm{p})\rangle \frac{\mathrm{d}\bm{k}\mathrm{d}\bm{p}}{(2\pi)^4}\nonumber\\
&=\frac{1}{(2\pi)^2}\int P(\bm{k})\mathrm{d}\bm{k}=\sigma_0^2\,,
\end{align}
the expectation value of the square of the norm of the gradient takes the form
\begin{align}
\langle \|\nabla f(\bm{q})\|^2\rangle &= \langle \partial_1f(\bm{q})^2 + \partial_2 f(\bm{q})^2 \rangle \nonumber \\
&= \iint [(ik_1)(-ip_1)+(ik_2)(-ip_2)] e^{i (\bm{p} - \bm{k})\cdot \bm{q}} \langle \hat{f}(\bm{k})\hat{f}^*(\bm{p})\rangle \frac{\mathrm{d}\bm{k}\mathrm{d}\bm{p}}{(2\pi)^4}\nonumber \\
&=\frac{1}{(2\pi)^2}\int \|\bm{k}\|^2 P(\bm{k})\mathrm{d}\bm{k}=\sigma_1^2\,,
\end{align}
with $\bm{k}=(k_1,k_2)$ and $\bm{p}=(p_1,p_2)$. By statistical isotropy, we obtain the variance of the first-order partial derivatives
\begin{align}
\langle \partial_1 f(\bm{q})^2\rangle =\langle \partial_2 f(\bm{q})^2\rangle =\frac{\sigma_1^2}{2}\,.
\end{align}

%%%%%%%%%%%%%%%%%%%%%%%%%%%%%%%%%%%%%%%%%%%%%%%%%%%%%%%%%%%%%%%%
\bigskip

\begin{figure}
\centering
\begin{subfigure}[b]{0.45\textwidth}
\includegraphics[width=\textwidth]{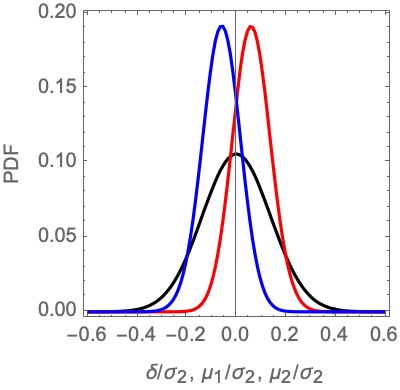}
\end{subfigure}~
\begin{subfigure}[b]{0.45\textwidth}
\includegraphics[width=\textwidth]{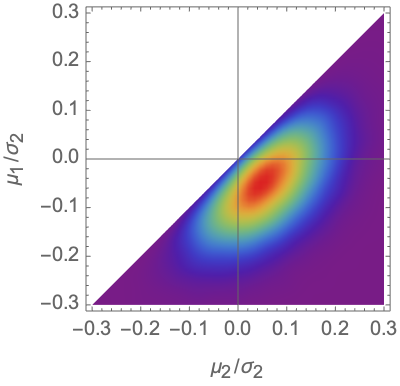}
\end{subfigure}
\caption{The distribution of the eigenvalue fields. \textit{Left:} the PDF of the initial density perturbation $\delta$ (black) and the initial eigenvalue fields $\mu_{i,t}$ (red and blue). \textit{Right:} the joint distribution of the eigenvalue fields.}\label{fig:eigen_stats}
\end{figure}

%%%%%%%%%%%%%%%%%%%%%%%%%%%%%%%%%%%%%%%%%%%%%%%%%%%%%%%%%%%%%%%%
\subsection{Non-Gaussian random fields: deformation tensor eigenvalues}
While the primordial density field has a distinctly Gaussian character, we are primarily interested in the deformation tensor eigenvalues. The eigenvalue fields of the deformation tensor are non-Gaussian. The joint probability distribution of the two deformation eigenvalues $\lambda_1$ and $\lambda_2$ is given by the two-dimensional Doroshkevich formula \cite{Doroshkevich:1970, Feldbrugge:2014}
\begin{align}
  p(\mu_1,\mu_2) = \frac{2\sqrt{2}}{\sqrt{\pi} \sigma_2^3} e^{-\frac{3(\mu_1+\mu_2)^2-8\mu_1 \mu_2}{2 \sigma_2^2}}|\mu_1-\mu_2|\,,
  \label{eq:doroshkevich}
\end{align}
with $\sigma_2$ the generalized moment defined in section \ref{sec:GRF}. The sum of the eigenvalues, according to the Poisson equation, is equal to the initial density contrast $\delta$,
\begin{equation}
\mu_1+\mu_2 \propto \delta_0\,.
\end{equation}
Figure~\ref{fig:eigen_stats} shows the pdfs for both $\delta$ and the eigenvalues $\mu_1$ and $\mu_2$ (lefthand panel). While that for $\delta$ is a clear bell-shaped Gaussian distribution, the non-Gaussian nature of the pdfs for the eigenvalues is directly visible. Their non-Gaussian nature becomes even more clear when looking at the joint probability function, as specified by the 2D Doroshkevich formula above (eqn.\ \eqref{eq:doroshkevich}), shown in the righthand panel of figure~\ref{fig:eigen_stats}.

\begin{figure}
\centering
\begin{subfigure}[b]{0.32\textwidth}
\includegraphics[width=\textwidth]{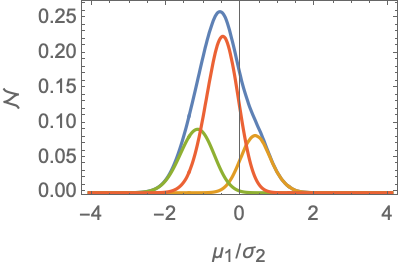}
\caption{The eigenvalue field $\mu_1$}
\end{subfigure}~
\begin{subfigure}[b]{0.32\textwidth}
\includegraphics[width=\textwidth]{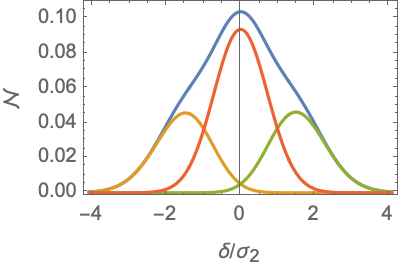}
\caption{The density perturbation $\delta$}
\end{subfigure}~
\begin{subfigure}[b]{0.32\textwidth}
\includegraphics[width=\textwidth]{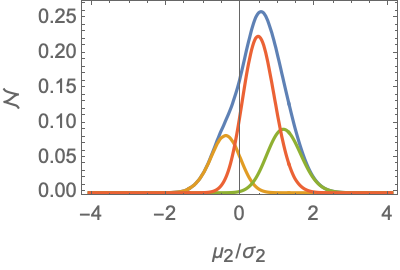}
\caption{The eigenvalue field $\mu_2$}
\end{subfigure}
\caption{The critical point densities of the density perturbation and the eigenvalue fields $\mu_1$ and $\mu_2$. The total critical point density (blue), the density of maxima (green), saddle points (red), and minima (orange).}\label{fig:critical}
\end{figure}

Of particular significance for the gravitational evolution of these fields are the critical points (maxima, minima, saddles). These are the sites where we see the emergence of structural features and key transitions in the hierarchical buildup of the cosmic web. Figure~\ref{fig:critical} provides a comparison between the critical point densities in the primordial Gaussian density field (central panel), and that of the non-Gaussian eigenvalue fields $\mu_1$ and $\mu_2$ (lefthand and righthand panel). The green solid curve depicts the distribution of the maxima in these fields, the orange curve that for the minima, and the red curve that for the saddle points. The combined distribution for all singularities is the blue curve. While the curves for the minima and maxima are symmetric for the Gaussian density field, we find an asymmetry in the case of the eigenvalues $\mu_1$ and $\mu_2$. Even more outstanding is the difference in the saddle point distribution between the density field ones and the eigenvalue ones. 

%%%%%%%%%%%%%%%%%%%%%%%%%%%%%%%%%%%%%%%%%%%%%%%%%%%%%%%%%%%%%%%%
\bigskip
\subsection{Linear constraints: constrained Gaussian random fields}
In this paper, we study how the different geometric features of the cosmic web emerge from different initial conditions. To systematically study these initial conditions we use constrained Gaussian random field theory \cite{Bertschinger:1987, Hoffman:1991, Sheth:1995, Weygaert:1996}. In this section, we develop the theory of linear constraints following the analysis of \cite{Weygaert:1996}. In the next section, we generalize this theory to a large class of non-linear constraints. %We restrict both discussions to statistically homogeneous and isotropic fields with vanishing mean.

\bigskip
\noindent For a random field $f$, consider a set of linear constraints
\begin{align}
\Gamma =\{ C_i[f;\bm{q}_i] = c_i,\,\ i=1,\dots,M \}\,,
\end{align}
with the linear functional $C_i[f;\bm{q}]$ assuming the value $c_i$ in $\bm{q}_i$. A linear functional can take the form of the function value at a point
\begin{align}
C[f;\bm{q}'] &= f(\bm{q}')\,,
\end{align}
it's derivative at a point
\begin{align}
C[f;\bm{q}'] &= \frac{\partial}{\partial q_i}f(\bm{q}')\,,
\end{align}
or more generally a convolution
\begin{align}
C[f;\bm{q}'] &= \int g(\bm{q}' - \bm{q})f(\bm{q})\mathrm{d}\bm{q}\,,
\end{align}
with the convolution kernel $g$. 

\bigskip
\noindent The constrained random field follows the infinite-dimensional distribution
\begin{align}
p(f|\Gamma) = \frac{p(f,\Gamma)}{p(\Gamma)}\,,
\end{align}
where the constraints follow the finite-dimensional Gaussian marginal distribution
\begin{align}
p(\Gamma) = \frac{\exp\left[-\frac{1}{2} \Delta\bm{C}^T Q^{-1} \Delta\bm{C} \right]}{[(2\pi)^M \det Q]^{1/2}}\,,
\end{align}
with the vector $\bm{C}=(C_1[f;\bm{q}_1], \dots, C_M[f;\bm{q}_M])$ and the covariance matrix $Q = \text{cov}(\bm{C}, \bm{C})$. We can write the constrained distribution as the Gaussian probability density
\begin{align}
p(f|\Gamma) \propto  \exp\left[-\frac{1}{2} \left[\iint f(\bm{q}_1) K(\|\bm{q}_1 - \bm{q}_2\|) f(\bm{q}_2)\mathrm{d}\bm{q}_1 \mathrm{d}\bm{q}_2 -\Delta \bm{C}^TQ^{-1}\Delta \bm{C}\right]\right],\label{eq:constraint1}
\end{align}
in the space of functions satisfying the constraints $\Gamma$.

\bigskip
Following the formalism of \cite{Bertschinger:1987}, we may write a field realization $f(\bm{q})$ as the sum of a \textit{mean field} $\bar{f}_{\mathcal{\tilde C}}(\bm{q})$ and a \textit{residual field} $\delta f_{\bm{C}}(\bm{q})$,
 \begin{framed}
  \begin{align}
    f(\bm{q})\,=\,\bar{f}_{\mathcal{\tilde C}}(\bm{q})\,+\,\delta f_{\mathcal{\tilde C}}(\bm{q})\,,
  \end{align}
  \end{framed}
\noindent in which the mean field encapsulates the principal signature of the constraints, being the average field satisfying the constraints. The residual field encapsulates the corresponding spectral fluctuation signal, augmented by the influence of the constraints. 
  
\bigskip
\noindent In appendix \ref{ap:constraintDensity}, the expression for the mean field $\bar{f}_{\bm{c}}(\bm{q})$ is derived, 
\begin{align}
\bar{f}_{\bm{c}}(\bm{q})&=\langle f(\bm{q})|\Gamma\rangle \nonumber \\
&= \bar{f}(\bm{q}) + \sum_{i,j=1}^M\xi_i (\bm{q})\xi_{ij}^{-1}(c_j-\bar{C}_j)\,,
\end{align}
and the covariance
\begin{align}
\text{cov}(f(\bm{q}_1),f(\bm{q}_2)|\Gamma)  = \xi(\bm{q}_1,\bm{q}_2) - \sum_{i,j=1}^M\xi_i(\bm{q}_1)\xi_{ij}^{-1}\xi_j(\bm{q}_2)\,,
\end{align}
with the covariance of the random field and the constraints $\xi_{i}(\bm{q}) = \text{cov}( f(\bm{q}), C_i)$ and the covariance matrix of the constraints $\xi_{ij} = \text{cov}( C_i ,C_j)$. Remarkably, the covariance $\text{cov}(f(\bm{q}_1),f(\bm{q}_2)|\Gamma)$ is independent of the value $\bm{c}$ the constraint $\bm{C}$ assumes. This is a special property of Gaussian distributions and linear constraints. Moreover, with respect to the mean field the residual field is normally distributed (see appendix \ref{ap:constraintDensity}). Consequently, the probability density takes the form
\begin{align}
p( f|\Gamma) \propto  \exp\left[-\frac{1}{2} \iint \delta{f}(\bm{q}_1) \tilde{K}(\bm{q}_1,\bm{q}_2) \delta f(\bm{q}_2)\mathrm{d}\bm{q}_1 \mathrm{d}\bm{q}_2 \right]\,,\label{eq:constraint2}
\end{align}
with the residual field $\delta f = f-\langle f(\bm{q})|\Gamma\rangle$ and the constrained kernel $\tilde{K}$ defined as the inverse of the constrained two-point correlation function, \textit{i.e.},
\begin{align}
\int \tilde{K}(\bm{q}_1,\bm{q}) \left[\xi(\bm{q},\bm{q}_2) - \sum_{i,j=1}^M\xi_i(\bm{q})\xi_{ij}^{-1}\xi_j(\bm{q}_2)\right]\mathrm{d}\bm{q}= \delta_D^{(2)}(\bm{q}_1-\bm{q}_2)\,.
\end{align}

Note that earlier papers \cite{Bertschinger:1987, Hoffman:1991, Sheth:1995, Weygaert:1996} worked in the space of functions satisfying the constraints $\Gamma$, neglecting the correction $\sum_{i,j=1}^M\xi_i(\bm{q})\xi_{ij}^{-1}\xi_j(\bm{q}_2)$, and using the original kernel $K$ for the residual field. In this study, we prefer to work in the space of unrestricted functions, as the correction makes the inhomogeneity and anisotropy of the residual field manifest. These properties are most clearly observed in the variance of the residual field,
\begin{align}
  \langle \delta f(\bm{q})^2|\Gamma \rangle = \sigma_0^2- \sum_{i,j=1}^M\xi_i(\bm{q})\xi_{ij}^{-1}\xi_j(\bm{q})\,,
  \label{eq:variance_Linear}
\end{align}
which vanishes on the constraints. Away from the constraints, the variance of the residual field approaches the variance of the unconstrained field $\text{var}(f)=\sigma_0^2$. Note that this relation implies an analogous relation for the Laplacian of the random field $\nabla^2 f$, \textit{i.e.}, we find the mean
\begin{align}
\langle \nabla^2 f(\bm{q})\,|\,\Gamma\rangle = \nabla^2 f_{\bm{c}}(\bm{q})
\end{align}
and the variance
\begin{align}
\langle \delta (\nabla^2f(\bm{q}_1))^2\,|\,\Gamma \rangle = \sigma_2^2- \sum_{i,j=1}^M\nabla^2\xi_i(\bm{q})\xi_{ij}^{-1}\nabla^2\xi_j(\bm{q})\,.\label{eq:variance_Linear_Laplace}
\end{align}
\noindent The Hoffman-Ribak method cleverly uses the property that the statistics of the residual field $\delta f$ are independent of $\bm{c}$, to generate realizations of the constrained Gaussian random field \cite{Hoffman:1991, Weygaert:1996}. Firstly, generate a realization $g$ of the unconstrained Gaussian random field with the required power spectrum. Secondly, evaluate the constraints for this realization $C_i[g;\bm{q}'_i]=d_i$ and the corresponding mean field $\bar{g}$. The statistical properties of the residual field with respect to this mean field $\delta g = g-\bar{g}$ are independent of the value the constraints assume and thus identical to the properties of the residual field $\delta f$. We can thus identify the residual field $\delta g$ of the field $g$ with the residual field $\delta f$ of the constrained field $f$. By adding the residual field of the unconstrained field to the mean field with the required constraints, we obtain the realization of the constrained Gaussian random field. For more details see appendix \ref{ap:GRF}.

To judge the relevance of a particular set of constraints, it is useful to evaluate the $\chi^2$ of the set of values $c_i$,
\begin{align}
\chi^2 = \sum_{i,j=1}^M (c_i-\bar{C}_i)\, \xi_{ij}^{-1}(c_j-\bar{C}_j)\,,
\end{align}
which follows the chi-squared distribution with $M$ degrees of freedom. This is the Mahalanobis distance. The probability that one obtains a $\chi^2$ higher than this value in an unconstrained Gaussian random field, with the same power spectrum, is given by $\Gamma\left(\frac{M}{2}, \frac{\chi^2}{2}\right)/\Gamma\left(\frac{M}{2}\right)$, with the gamma function $\Gamma(x)$ and the upper incomplete gamma function $\Gamma(s,x)$.

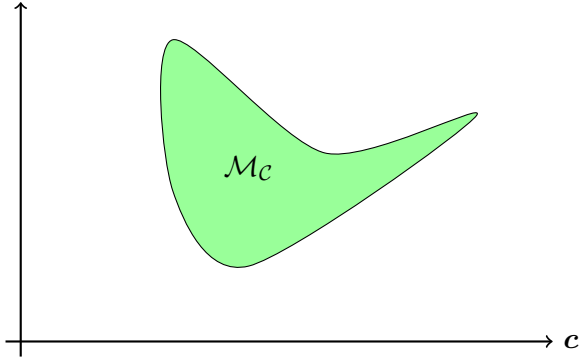
\begin{figure}
\centering
\begin{subfigure}[b]{0.49\textwidth}
\begin{tikzpicture}
%\draw[draw=black] (0,0) rectangle ++(7,5);
\draw[->, thick] (-0.2,0)--(7,0) node[right]{$\bm{c}$};
\draw[->, thick] (0,-0.2)--(0,4.5);
%\draw [fill=blue!80] plot [mark=none, smooth cycle] coordinates {(2,2) (3,1) (6,3) (4,2.5) (2,4)};
\draw [fill=green!40] plot [mark=none, smooth cycle] coordinates {(2,2) (3,1) (6,3) (4,2.5) (2,4)};
%\draw [fill=lime!70] plot [mark=none, smooth cycle] coordinates {(2,2) (3,1) (6,3) (4,2.5) (2,4)};
\node[above] at (3,2) {$\mathcal{M}_\mathcal{C}$};
\end{tikzpicture}
\end{subfigure}
\caption{The constrained manifold $\mathcal{M}_\mathcal{\cal C}$ in the space of linear constraint values.}\label{fig:constraintManifold}
\end{figure}

%%%%%%%%%%%%%%%%%%%%%%%%%%%%%%%%%%%%%%%%%%%%%%%%%%%%%%%%%%%%%%%%
\bigskip
\subsection{Non-linear constrained random fields: formalism}
For the systematic investigation of the caustic skeleton, we extend the theory for linear constraints on Gaussian random fields to that for the generic situation of \textit{non-linear constraints} on random fields. In this section, we present the formalism for {\textit non-linear constrained random fields}. Specifically, it pertains to the large class of non-linear constraints that are (non-linear) functions of a finite number of linear functionals of the random field \citep{Bertschinger:1987,Hoffman:1991,Rybicki:1992,Weygaert:1996}.  

\bigskip
Consider the set of linear functionals $C_i[f,\bm{q}_i]$ with $i=1,\dots,M$, and the corresponding space of the  values $\bm{c}=(c_1,\dots,c_M) \in \mathbb{R}^M$ the linear functionals may assume:
\begin{align}
  C_i[f,\bm{q}_i]=c_i\,,\quad i=1,\dots,M\,.
\end{align}
On the space of constraint values $\bm{c}$, we define the non-linear constraint set

\begin{align}
  \Gamma\, = \,\{\mathcal{\tilde C}_i(\bm{c}) = 0\}_{i=1}^N
\end{align}
consisting of the $N$ non-linear functions $\mathcal{\tilde C}_k:\mathbb{R}^M \to \mathbb{R}$ ($k=1,\dots,N$ and $N\leq M$). Geometrically, we can visualize this by imagining the space of $M$ constraint values  $\bm{c}\in \mathbb{R}^M$. In this space, the (non-linear) constraint set $\Gamma$ outline a $(M-N)$-dimensional manifold $\mathcal{M}_\mathcal{\tilde C}$
(see figure~\ref{fig:constraintManifold}),
\begin{align}
\mathcal{M}_{\mathcal{\tilde C}} = 
\{\bm{c}\, |\, \mathcal{\tilde C}_i(\bm{c}) = 0\,,\quad \text{ for all } i=1,\dots,N\}.
\end{align}
\bigskip

Of key significance for this observation is that it allows us to assign straightforwardly a probability to the non-linear constraints. It simply is the probability for the linear constraint values $\bm{c}=(c_1,\dots,c_M) \in \mathbb{R}^M$ to be located on the non-linear constraint manifold $\mathcal{M}_\mathcal{\tilde C}$. Given the Gaussian distribution for the linear constraint values, we arrive at the non-linear constraint probability density,

\begin{framed}
\begin{align}
p(\bm{c}\,|\,\bm{c}\in \mathcal{M}_{\mathcal{\tilde C}})  = \frac{p(\bm{c})}{\int_{\mathcal{M}_\mathcal{\tilde C}} p(\bm{c})\, \mathrm{d}\bm{c}}\,.
\end{align}
\end{framed}
From the geometric point of perspective, we may observe that for the generic situation of the constraint manifold $\mathcal{M}_{\mathcal{\tilde C}}$ being a non-flat curved multidimensional manifold the probability distribution density $p(\bm{c}\,|\,\bm{c}\in \mathcal{M}_{\mathcal{\tilde C}})$ is non-Gaussian.

On the constraint manifold $\mathcal{M}_\mathcal{\tilde C}$, the constraint probability density is simply proportional to the original Gaussian distribution. Hence, the problem of generating realizations of the more generic non-linear constrained random fields is reduced to sampling realizations of the distribution $p(\bm{c}|\bm{c}\in \mathcal{M}_\mathcal{\tilde C})$. Once the values ${\bm c}$ have been successfully sampled, we have simplified the problem towards one of the regular sampling Gaussian random fields. To this end, we may use the Hoffman-Ribak method \citep{Bertschinger:1987,Hoffman:1991,Weygaert:1996} to generate Gaussian realizations obeying the constraint values $\bm{c}$. This results in realizations of fields $f({\bm q})$ obeying the imposed non-linear constraints.

%%%%%%%%%%%%%%%%%%%%%%%%%%%%%%%%%%%%%%%%%%%%%%%%%%%%%%%%%%%%%%%%
\bigskip
\subsection{Non-linear constrained random fields: procedure}
The non-linear constrained random field formalism outlined above translates into the following procedure for generating constrained realizations. By formulating the non-linear constraint conditions in this form, we have come to the realization that the finite-dimensional non-Gaussian probability density $p(\bm{c}\,|\,\bm{c}\in \mathcal{M}_{\mathcal{\tilde C}})$ can be evaluated in terms of Gaussian distribution functions by exploring the geometry of the corresponding manifold $\mathcal{M}_\mathcal{\tilde C}$. It results in the following procedure:

\begin{itemize}
\item As for the situation with linear constraints. we follow \cite{Bertschinger:1987} in writing a field realization $f(\bm{q})$ as the sum of a \textit{mean field}, the average field satisfying the constraints and expressing the principal imprint of the constraints, and a \textit{residual field} $\delta f_{\mathcal{C}}(\bm{q})$,

\begin{framed}
\begin{align}
    f(\bm{q})\,=\,\bar{f}_{\mathcal{\tilde C}}(\bm{q})\,+\,\delta f_{\mathcal{\tilde C}}(\bm{q})
\end{align}
\end{framed}
in which the mean field encapsulates the principal signature of the constraints, being the average field satisfying the constraints. The residual field encapsulates the corresponding spectral fluctuation signal, augmented by the influence of the constraints. 
  
\item
The mean field of the constrained field assumes the surprisingly suggestive form 
\begin{framed}
\begin{align}
\bar{f}_{\mathcal{\tilde C}}(\bm{q}) =\bar{f}_{\bar{\bm{c}}}(\bm{q})\,,
\end{align}
with the mean constraint
\begin{align}
\bar{\bm{c}} &\equiv \langle \bm{c} \,|\, \Gamma\rangle = \int_{\bm{c} \in \mathcal{M}_{\mathcal{\tilde C}}}  \bm{c}\, p(\bm{c}\,|\,\bm{c}\in \mathcal{M}_{\mathcal{\tilde C}}) \mathrm{d}\bm{c}\,.
\end{align}
\end{framed}
This mean field expression follows directly from the realization  
\begin{align}
\bar{f}_{\mathcal{\tilde C}}(\bm{q}) 
&=\left\langle f(\bm{q})\,|\,\Gamma\right \rangle \nonumber\\
&= \int_{\bm{c} \in \mathcal{M}_{\mathcal{\tilde C}}} \bar{f}_{\bm{c}}(\bm{q})\, p(\bm{c}|\bm{c}\in \mathcal{M}_{\mathcal{\tilde C}}) \mathrm{d}\bm{c}\nonumber\\
&= \sum_{i,j=1}^M\xi_i(\bm{q}) \xi_{ij}^{-1}\int_{\bm{c} \in \mathcal{M}_{\mathcal{\tilde C}}}  (c_j-\bar{C}_j)\, p(\bm{c}\,|\,\bm{c}\in \mathcal{M}_{\mathcal{\tilde C}}) \mathrm{d}\bm{c}\nonumber\\
&= \sum_{i,j=1}^M \xi_i(\bm{q}) \xi_{ij}^{-1} (\bar{c}_j-\bar{C}_j)\nonumber\\
&= \bar{f}_{\bar{\bm{c}}}(\bm{q})\,.
\end{align}

\item
The residual field  $\delta f = f - \bar{f}_{\mathcal{\tilde C}}$ is generically no longer a Gaussian random field. The variance of the residual field yields
\begin{framed}
\begin{align}
\left \langle \delta f (\bm{q})^2\,|\,\Gamma\right\rangle =
\sigma_0^2 - \sum_{i,j=1}^M\xi_i(\bm{q}) \zeta_{ij}^{-1} \xi_j(\bm{q}) 
\,,\label{eq:variance_Nonlinear}
\end{align}
with the coefficients
\begin{align}
\zeta_{ij}^{-1}= \xi_{ij}^{-1}-\sum_{k,l=1}^M\xi_{ik}^{-1}\,\text{cov}(c_k, c_l\,|\,\bm{c}\in\mathcal{M}_\mathcal{\tilde C})\,\xi_{lj}^{-1}\,,
\label{eqn:zetacoeff}
\end{align}
\end{framed}
where the constraint covariance $\text{cov}(c_j, c_l\,|\,\bm{c}\in\mathcal{M}_\mathcal{\tilde C})$ captures the geometry of the constraint manifold $\mathcal{M}_\mathcal{\tilde C}$. The expression for the residual field $\delta f(\bm{q})$ follows directly from
\begin{align}
\left \langle \delta f (\bm{q})^2\,|\,\Gamma\right\rangle 
&= \int_{\mathcal{M}_\mathcal{\tilde C}} \langle (f(\bm{q})-\bar{f}_{\bar{\bm{c}}}(\bm{q}))^2\,|\,C_i[f,\bm{q}_i]=c_i\rangle p(\bm{c}\,|\,\bm{c}\in\mathcal{M}_{\mathcal{\tilde C}})\mathrm{d}\bm{c}\nonumber \\
&= \int_{\mathcal{M}_\mathcal{\tilde C}} \langle f^2(\bm{q})\,|\,C_i[f,\bm{q}_i]=c_i\rangle p(\bm{c}\,|\,\bm{c}\in\mathcal{M}_{\mathcal{\tilde C}})\mathrm{d}\bm{c} - \bar{f}_{\bar{\bm{c}}}^2(\bm{q})\nonumber \\
&= \int_{\mathcal{M}_\mathcal{\tilde C}} \langle (f(\bm{q})-\bar{f}_{\bm{c}}(\bm{q}))^2\,|\,C_i[f,\bm{q}_i]=c_i\rangle p(\bm{c}\,|\,\bm{c}\in\mathcal{M}_{\mathcal{\tilde C}})\mathrm{d}\bm{c} \nonumber \\
&\ \ +\int_{\mathcal{M}_\mathcal{\tilde C}} (\bar{f}_{\bm{c}}(\bm{q})- \bar{f}_{\bar{\bm{c}}}(\bm{q}))^2p(\bm{c}\,|\,\bm{c}\in\mathcal{M}_{\mathcal{\tilde C}})\mathrm{d}\bm{c} \nonumber \\
&=  \sigma_0^2 - \sum_{i,j=1}^M\xi_i(\bm{q}) \xi_{ij}^{-1} \xi_j(\bm{q}) + \sum_{i,j,k,l=1}^M \xi_{i}(\bm{q})\xi_{ij}^{-1}\,\text{cov}(c_j, c_l\,|\,\bm{c}\in\mathcal{M}_\mathcal{\tilde C})\,\xi_{lk}^{-1}\xi_{k}(\bm{q})\nonumber\\
&=
\sigma_0^2 - \sum_{i,j=1}^M\xi_i(\bm{q}) \zeta_{ij}^{-1} \xi_j(\bm{q})\,,
\end{align}
where
\begin{align}
\int_{\mathcal{M}_\mathcal{\tilde C}} (\bar{f}_{\bm{c}}(\bm{q})- \bar{f}_{\bar{\bm{c}}}(\bm{q}))^2p(\bm{c}\,|\,\bm{c}\in\mathcal{M}_{\mathcal{\tilde C}})\mathrm{d}\bm{c} 
&= \text{var}\left(\sum_{i,j=1}^M\xi_i(\bm{q})\xi_{ij}^{-1}c_j\,|\,\bm{c}\in \mathcal{M}_\mathcal{\tilde C}\right)\\
&=\sum_{i,j,k,l=1}^M \xi_i(\bm{q})\xi_{ij}^{-1}\,\text{cov}(c_j,c_k\,|\,\bm{c}\in\mathcal{M}_\mathcal{\tilde C})\,\xi_{kl}^{-1}\xi_l(\bm{q})\,.\nonumber
\end{align}
A telling difference between the residual for the non-linear constraints $\left \langle \delta f (\bm{q})^2\,|\,\Gamma\right\rangle$ and that in the case of linear Gaussian random field constraints (eqn.~\ref{eq:variance_Linear}) is that they are no longer independent of the values $\bm{c}$ the constraint(s) assume. Only when the manifold $\mathcal{M}_\mathcal{\tilde C}$ consists of a point, the residual expression for the linear constraints is recovered.

\item
A similar relation as for the residual field applies to the Laplacian $\nabla^2 f(\bm{q})$ of the random field. In the context of the caustic skeleton the field $f({\bm q})=\phi({\bm q})$ is the gravitational potential

\begin{align}
f({\bm q})=\phi({\bm q})\,.
\end{align}

Hence, according to the Poisson equation, the Laplacian of the field represents the density perturbation. Following an analogous derivation, we obtain 

\begin{framed}
the mean constrained field Laplacian  
\begin{align}
\langle \nabla^2 f(\bm{q})\,|\,\Gamma\rangle = \sum_{i,j=1}^M \nabla^2\xi_i(\bm{q})\xi_{ij}^{-1}(\bar{c}_j-\bar{C}_j)
\end{align}
and the variance

\begin{align}
\langle \delta (\nabla^2 f(\bm{q}))\,|\,\Gamma\rangle = \sigma_2^2 - \sum_{i,j=1}^M \nabla^2 \xi_i(\bm{q})\zeta_{ij}^{-1}\nabla^2\xi_j(\bm{q})\,,
\end{align}
\end{framed}
in which $\bar{\bm{c}}$ are the main constraint values and $\zeta_{ij}$ the coefficient defined in expression~\ref{eqn:zetacoeff}.
\end{itemize}

%%%%%%%%%%%%%%%%%%%%%%%%%%%%%%%%%%%%%%%%%%%%%%%%%%%%%%%%%%%%%%%%
\bigskip
\subsection{Non-linear constraints: case study}
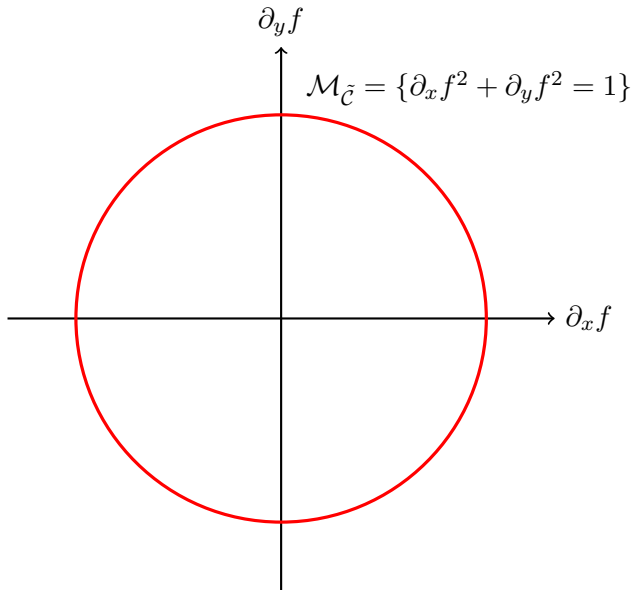
\begin{figure}
\centering
\begin{subfigure}[b]{0.49\textwidth}
\begin{tikzpicture}[scale=0.9]
%\draw[help lines, color=gray!30, dashed] (-0.9,-0.9) grid (4.9,4.9);
\draw[->, thick] (-4,0)--(4,0) node[right]{$\partial_x f$};
\draw[->, thick] (0,-4)--(0,4) node[above]{$\partial_y f$};
\draw[color=red, very thick](0,0) circle (3);
\node[above] at (2.75,3) {$\mathcal{M}_\mathcal{\tilde C}=\{\partial_xf^2+\partial_yf^2=1\}$};
\end{tikzpicture}
\end{subfigure}
\caption{The constraint manifold $\mathcal{M}_\mathcal{\tilde C}=\{\partial_xf^2+\partial_yf^2=1\}$ in the space of first-order derivatives $(\partial_x f,\partial_yf)$ for the non-linear constraint $\|\nabla f\|=1$.}\label{fig:constraintManifold_example}
\end{figure}

A good example of a non-linear constraint is the requirement that the gradient of the Gaussian random field has a unit norm $\|\nabla f(\bm{q}_c)\|=1$. For this condition, we consider the space of first-order derivatives
\begin{align}
\bm{C}=(\partial_x f, \partial_y f)\,,
\end{align}
and the non-linear constraint 
\begin{align}
\mathcal{\tilde C}(\bm{C})=\partial_x f^2 +  \partial_y f^2 - 1=0\,.
\end{align}
The constraint manifold $\mathcal{M}_\mathcal{\tilde C}=\{(\partial_xf,\partial_yf)\, |\, \partial_xf^2+\partial_yf^2=1\}$ is the unit circle in the space of first order derivatives (see figure \ref{fig:constraintManifold_example}).

For a statistically isotropic Gaussian random field, the constrained probability density is uniform $p(\bm{c}\, |\, \bm{c}\in\mathcal{M}_\mathcal{\tilde C}) = 1/(2\pi)$. After sampling from the constraint manifold, we construct a corresponding realization using linear constrained Gaussian random field theory. Both the mean and the covariance of the constraint value with respect to the constraint manifold vanish.

When the Gaussian random field is not statistically isotropic, the constraint density $p(\bm{c}\, |\, \bm{c}\in\mathcal{M}_\mathcal{\tilde C})$ is no longer uniformly distributed. We can construct realizations of the constrained Gaussian random field using rejection sampling on the constraint manifold. First, determine the maximum $max$ of $p(\bm{c})$ restricted to $\mathcal{M}_\mathcal{\tilde C}$. Secondly, generate a uniformly sampled point $\bm{c}$ on the unit circle $\mathcal{M}_\mathcal{\tilde C}$. Finally, we accept this sample $\bm{c}$ with probability $p(\bm{c})/max$. Once we have sampled the points on the constraint manifold $\mathcal{M}_\mathcal{\tilde C}$, we again construct a corresponding realization using linear constrained Gaussian random field theory. It is also straightforward to sample the mean and covariance of the constraint value $\bm{c}$ on the constraint manifold $\mathcal{M}_\mathcal{\tilde C}$.

%%%%%%%%%%%%%%%%%%%%%%%%%%%%%%%%%%%%%%%%%%%%%%%%%%%%%%%%%%%%%%%%
\section{The caustic skeleton: constrained initial conditions}\label{sec:caustic_skeleton_constraints}
Having established the procedure for non-linear constrained random fields, in this section we apply the procedure to explore the mass distribution in and around caustic singularities and hence that in and the corresponding structural components of the cosmic web. To this end, we here derive the expressions for the non-linear constraints pertaining to the caustic singularities. We will concentrate on the formalism for the cusp constraints, swallowtail constraints and umbilic constraints and present the corresponding (mean) field realizations. On the basis of this, we set up a \textit{laboratory} for exploring the progenitors of filament and clusters, \textit{i.e.}, the primordial mass distribution in and around filaments, via the cusp constraints, and in around clusters, via the swallowtail and umbilic constraints. This provides us with a powerful instrument for a systematic investigation and inventory of the (hierarchical) buildup, dynamics and (multistream) velocity flows in and around these structural features of the cosmic web.  

%%%%%%%%%%%%%%%%%%%%%%%%%%%%%%%%%%%%%%%%%%%%%%%%%%%%%%%%%%%%%%%%
\subsection{Deformation tensor: definitions}
Central within the non-linear constraint formalism is the specification of the displacement field and corresponding deformation tensor. Within the present study, we follow the most practical approach, that of using the analytical first-order expression for the displacement in the Zeldovich approximation (see sect.~\ref{sec:zeldovich}),
\begin{align}
\bm{s}_t(\bm{q}) = -b_+(t) \nabla_{\bm{q}} \Psi(\bm{q}),
\end{align}
For practical purposes, we will follow the following expression for the deformation tensor in terms of the Hessian of displacement potential $\Psi$, 
\begin{align}
  T_{ij} \equiv \frac{\partial^2 \Psi}{\partial q_i \partial q_j}\,,\\
  \,\\
\bm{\psi}(\bm{q}) = \begin{pmatrix} T_{11}(\bm{q}) & T_{12}(\bm{q}) \\ T_{12}(\bm{q}) & T_{22}(\bm{q})\end{pmatrix}\,,
    \end{align}
Extending this notation, we introduce the following expression for any higher-order partial derivative of the
displacement potential
\begin{equation}
  T_{ij\dots k}=\frac{\partial}{\partial q_i}\frac{\partial}{\partial q_j}\dots \frac{\partial}{\partial q_k}\Psi\,.
  \end{equation}

\bigskip
The eigenvalues and eigenvectors of the deformation tensor $\bm{\psi}$ follow from the eigenequation,
\begin{align}
\bm{\psi}(\bm{q}) \bm{v}_i(\bm{q}) = \lambda_i(\bm{q}) \bm{v}_i(\bm{q}),
\end{align}
with the ordering $\lambda_1\geq \lambda_2$. For the eigenvectors $\bm{v}_i(\bm{q})$ we use the normalized versions, \textit{i.e.}, $\|\bm{v}_{i}\|=1$. It is straightforward to obtain the explicit analytical expressions for the eigenvalue and eigenvector fields in terms of the second-order derivatives of the deformation potential
\begin{align}
\lambda_{1,2} &= \frac{1}{2}\left[T_{11}+T_{22} \pm \sqrt{4 T_{12}^2+(T_{11}-T_{22})^2}\right]\,,\\
\bm{v}_{1,2} &\propto  \left(T_{11}-T_{22} \pm \sqrt{4 T_{12}^2+(T_{11}-T_{22})^2}, 2 T_{12}\right)\,.
\end{align}
These expressions for the eigenvalues and eigenvectors immediately reveal their non-linear dependence on the deformation potential. With these explicit expressions for the eigenvalues and eigenvectors, we may subsequently infer the constraint expressions for the different classes of caustics by inserting these in the expression for the caustic conditions (see table~\ref{table:caustics}). 

\bigskip
Of key importance for the practical implementation of the non-linear constraint formalism for the various caustic constraints is appendix~\ref{ap:eigenvalueRel}. It contains the derivation of the expressions of the eigenvalues $\lambda_1$ and $\lambda_2$, the eigenvectors $\bm{v}_1$ and $\bm{v}_2$, the first order and second order gradients of the eigenvalue fields in terms of the partial derivatives $T_{ij \dots k}$ of the displacement potential $\Psi$. 

%%%%%%%%%%%%%%%%%%%%%%%%%%%%%%%%%%%%%%%%%%%%%%%%%%%%%%%%%%%%%%%%
\subsection{Cusp \& filament constrained fields}
The cusp caustic forms at a time $t$ corresponding to the inverse of the largest eigenvalue $\lambda_1$, and satisfies the following caustic conditions (see table~\ref{table:caustics}),
\begin{align}
b_+(t) \lambda_1(\bm{q}_c) &= 1\,, \\
 \bm{v}_1(\bm{q}_c) \cdot \nabla\lambda_1(\bm{q}_c) &= 0\,.
\end{align}
The orientation of the cusp is determined by the normal vector
\begin{align}
\bm{n} =  \nabla(\bm{v}_1(\bm{q}_c) \cdot \nabla\lambda_1(\bm{q}_c))\,,
\end{align}
which makes an angle 
\begin{align}
\alpha = \text{sign}(\bm{v}_2\cdot \bm{n}) \arccos\left(\frac{|\bm{v}_1\cdot \bm{n}|}{\|\bm{n}\|}\right)
\end{align}
with respect the corresponding eigenvector $\bm{v}_1$. As both the cusp curve and the eigenvector field consist of lines, the angle $\alpha$ assumes a value in the interval $[-\pi/2,\pi/2]$. 

\begin{figure}
\centering
\begin{subfigure}[b]{0.32\textwidth}
\includegraphics[width=\textwidth]{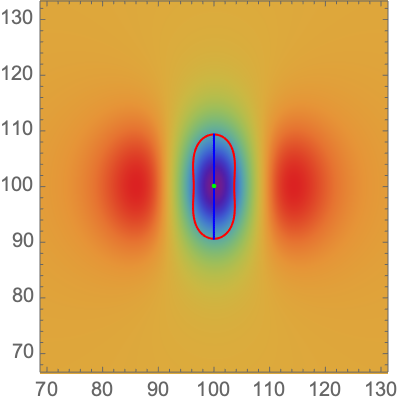}
\end{subfigure}~
\begin{subfigure}[b]{0.32\textwidth}
\includegraphics[width=\textwidth]{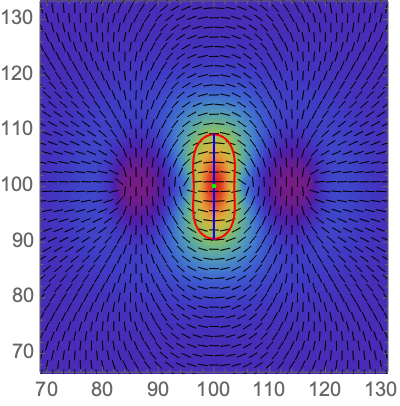}
\end{subfigure}~
\begin{subfigure}[b]{0.32\textwidth}
\includegraphics[width=\textwidth]{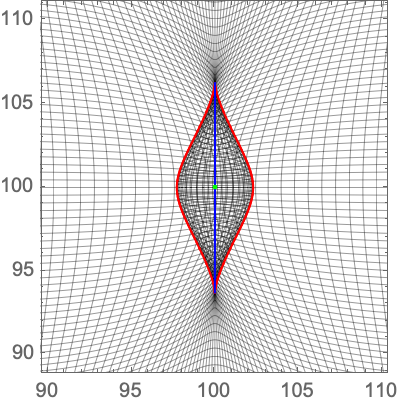}
\end{subfigure}
\caption{The mean cusp caustic. \textit{Left:} the gravitational potential. \textit{Centre:} the first eigenvalue and eigenvector fields. \textit{Right:} the mean field mass distribution around the cusp curve, evolved by the Zel'dovich approximation}\label{fig:meanCusp}
\end{figure}

\begin{figure}
\centering
\includegraphics[width=0.5\textwidth]{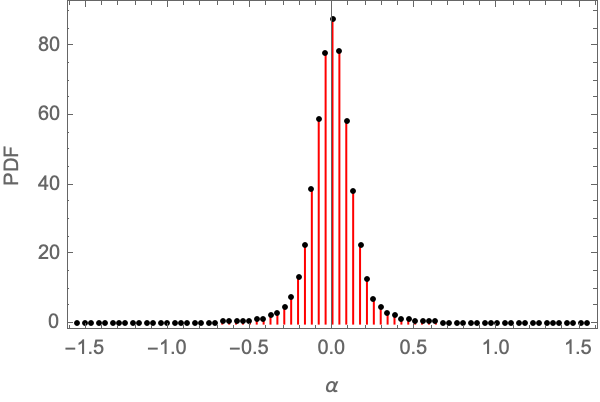}
\caption{The distribution of the angle $\alpha$ between the cusp curve and the eigenvector field.} 
\label{fig:alpha}
\end{figure}

\subsubsection{{\it Cusp mean field}}
The mean field of the cusp filament is shown in figure~\ref{fig:meanCusp}. The mean field has been set such that the cusp is at the center of the box, and oriented along the vertical direction. The lefthand panel of the figure shows the gravitational potential of the cusp, the central panel the corresponding (first) eigenvalue and eigenvector fields, while the righthand panel shows the corresponding evolved mass distribution according to the Zeldovich approximation. 

The gravitational potential around a cusp is close to a quadrupole. This is in line with the expectation for the constrained mass distribution around protofilaments, as has been pointed out by the study of \cite{Weygaert:1996}. The blue central structure is a potential trough that induces an inflow of matter, whereas the two red structures are two potential hills related to the implied low-density protovoids surrounding the embryonic cups filament. This leads to the outflow of mass from these low-density regions towards the higher-density filamentary cusp region.

The critical curve in the eigenvalue field consists of an elongated oval which forms the traditional Zeldovich pancake in the Zeldovich approximation. The cusp curve, the blue solid line, bisects the oval in Lagrangian space. Turning to the structure evolving out of this configuration, we see the emergence of the well-known Zeldovich pancake configuration in Eulerian space. The pancake is bisected by the critical curve in Eulerian space (blue solid line).

On average, the orientation of the cusp curve is normal to the eigenvector field $\bm{v}_1$. The distribution of the angle $\alpha$ in the mean field realizations is plotted in figure\ref{fig:alpha}~. The distribution is a steep -- near bell-shaped -- curve centered around $\alpha=0$, with an exponential fall-off towards the angles $\alpha =\pm \pi/2$.

%%%%%%%%%%%%%%%%%%%%%%%%%%%%%%%%%%%%%%%%%%%%%%%%%%%%%%%%%%%%%%%%
\subsubsection{{\it Cusp residual field}}
To obtain full realizations of the caustic-constrained fields, we have to sample the residual fields $\delta f$ that comprise the spectrally induced field fluctuations augmented by the imposed caustic constraint. The combination of mean field and residual field yields a field realization representative for the applied caustic constraint.

To obtain a representatively sampled residual field, we need to infer the non-linear probability distribution $p({\bm c})$ for the corresponding linear constraint set ${\bm c}$. This involves first
determining which linear Gaussian field functionals $c_k$ are incorporated in the non-linear caustic constraint. Subsequently, we need to specify the (non-linear) probability distribution $p({\bm c})$ and properly sample this distribution in the linear functional space ${\bm c}$. Once this has been accomplished, the Gaussian field constraint set ${\bm c}$ is applied to the Hoffman-Ribak method for finding a representative residual field realization.  

\bigskip
Without significantly affecting the result, it is insightful to analyze the constrained random field of the cusp caustic in the eigenframe of the deformation tensor. In this frame, the eigenvectors are $\bm{v}_1=(1,0)$ and $\bm{v}_2=(0,1)$.
The resulting expression for the caustic conditions is a set of linear conditions on the second and third-order derivatives of the deformation potential
\begin{align}
T_{11}=1/b_+(t)\,, \quad T_{12}=0\,,\quad T_{22}\leq T_{11}\,,\quad T_{111}=0\,,\label{eq:cusp_cond_1}
\end{align}
while the normal vector $\bm{n}$ has the following non-linear form, 
\begin{align}
\bm{n}=\left(T_{1111} + \frac{3T_{112}^2}{T_{11}-T_{22}}, T_{1112} + \frac{3T_{112}T_{122}}{T_{11}-T_{22}}\right)\,.\label{eq:cusp_cond_2}
\end{align}
For a derivation of these identities, we refer to appendix~\ref{ap:eigenvalueRel}.

\bigskip
For simplicity, we ignore the orientation of the cusp curve and consider a cusp curve passing through the center of the box. The relevant linear statistics is
\begin{equation}
Y=(T_{11},T_{12},T_{22},T_{111})\,,
\end{equation}
whose probability density factorizes into the density of the joint distribution of $T_{11}$ and $T_{22}$ and the one-dimensional Gaussian distributions for $T_{12}$ and $T_{111}$, \textit{i.e.},
\begin{align}
p(T_{11},T_{12},T_{22},T_{111}) = p(T_{11},T_{22})p(T_{12})p(T_{111})\,.
\end{align}
Generally, two derivatives of a Gaussian random field at a point are independent if and only if they together consist of an odd number of derivatives in either the $q_1$ or the $q_2$ direction. The conditional probability density takes the form
\begin{align}
p&(T_{11},T_{12},T_{22},T_{111}|,T_{22}\leq T_{11}=1/b_+,T_{12}=0,T_{111}=0) \nonumber\\
&= p(T_{22}| T_{22} \leq T_{11}=1/b_+)\nonumber\\
&= \mathcal{N} e^{-\frac{3(T_{11} + T_{22})^2 - 8 T_{11} T_{22}}{2 \sigma_2^2}}\big|_{T_{11}=1/b_+}\,,
\end{align}
with the normalization constant $\mathcal{N}$. After sampling $T_{22}$ from this distribution, we use linear constrained random field theory to obtain the corresponding realization. We may note that this procedure is equivalent to generating realizations for the linear constraints $T_{11}=1/b_+,T_{12}=T_{111}=0$ and rejecting realizations for which $T_{22} > 1/b_+$. Given these realizations, it is straightforward to compute the angle $\alpha$ and sample the distribution. For the resulting distribution, we refer to figure \ref{fig:alpha}.

\bigskip
Subsequently addressing the orientation of the cusp curve, we note that the orientation of the cusp curve is fully characterized by the derivatives  
\begin{align}
Y=(T_{11},T_{12},T_{22},T_{111},T_{112},T_{122},T_{1111},T_{1112})\,,
\end{align}
with the conditional distribution
\begin{align}
&p(T_{11},T_{12},T_{22},T_{111},T_{112},T_{122},T_{1111},T_{1112}|T_{22} \leq T_{11}=1/b_+, T_{111}=0)\\
&=
p(T_{11},T_{22},T_{1111}|T_{22}\leq T_{11}=1/b_+)p(T_{12}T_{1112}|T_{12}=0)p(T_{111},T_{122}|T_{111}=0)p(T_{112})\,.\nonumber
\end{align}
By sampling this distribution with the additional constraint that $\bm{n} \propto (1,0)$, realizations may be generated for which the cusp curve is vertical in the center of the box.  However, in the present study, we use a slightly different method based on the sampling method described above, in combination with the procedure for rotating the (derivatives of) the deformation tensor outlined in appendix~\ref{ap:rotations}. Given a realization, in this method, we first evaluate all second, third, and fourth-order derivatives
and evaluate the angle $\alpha$. Subsequently, these derivatives are rotated. The resulting derivatives form a representative sample of a cusp curve running vertically through the center of the box. By generating the linear constrained realizations with these rotated derivatives, we obtain a realization of the Gaussian random field required cusp curve.

%%%%%%%%%%%%%%%%%%%%%%%%%%%%%%%%%%%%%%%%%%%%%%%%%%%%%%%%%%%%%%%%
\subsection{Swallowtail \& cluster constrained fields}
For the constrained realizations in and around the swallowtail cluster caustic at Lagrangian location $\bm{q}_c$ we invoke the swallowtail caustic conditions, with respect to the field of the largest eigenvalue $\lambda_1$, and that of the corresponding eigenvector $\bm{v}_1$ field,
\begin{align}
b_+(t) \lambda_1(\bm{q}_c) = 1\,,\\
\quad \bm{v}_1(\bm{q}_c) \cdot \nabla\lambda_1(\bm{q}_c) = 0\,,\\
 \quad\bm{v}_1(\bm{q}_c)\cdot \nabla( \bm{v}_1(\bm{q}_c) \cdot \nabla\lambda_1(\bm{q}_c)) = 0\,.
\end{align}
The swallowtail caustic in the cusp curve is parallel to the eigenvector field $\bm{v}_1$, meaning that $\alpha = \pm \pi/2$, since
\begin{equation}
\bm{v}_1 \cdot \bm{n}= \bm{v}_1 \cdot \nabla(\bm{v}_1(\bm{q}_c) \cdot \nabla\lambda_1(\bm{q}_c))  =0\,.
\end{equation}
While the orientation of the swallowtail caustic does not need to be fixed, we do need to specify the direction. The swallowtail feature is normal to the fold curve, in which it forms. In the present study, we use the condition
\begin{equation}
  \bm{v}_2 \cdot \nabla \lambda_1(\bm{q}) <0\,.
  \end{equation}

\subsubsection{{\it Swallowtail mean field}}
\begin{figure}
\centering
\begin{subfigure}[b]{0.32\textwidth}
\includegraphics[width=\textwidth]{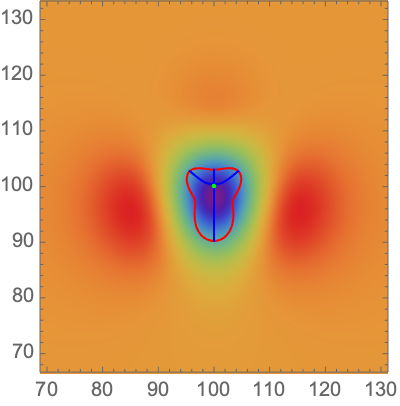}
\end{subfigure}~
\begin{subfigure}[b]{0.32\textwidth}
\includegraphics[width=\textwidth]{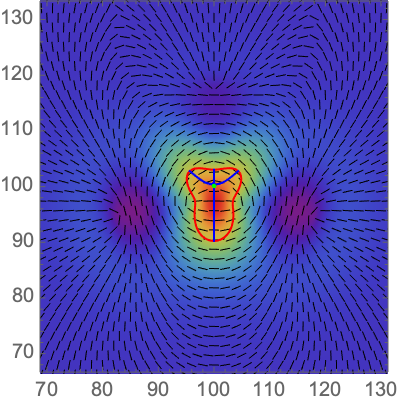}
\end{subfigure}~
\begin{subfigure}[b]{0.32\textwidth}
\includegraphics[width=\textwidth]{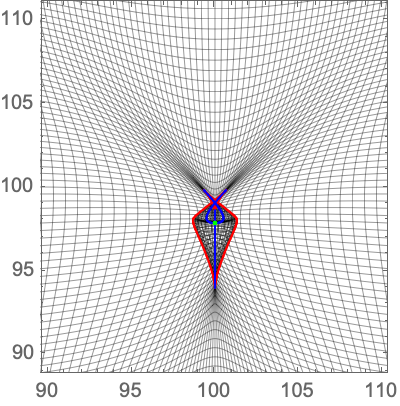}
\end{subfigure}
\caption{The mean swallowtail caustic. \text{Left:} the gravitational potential. \textit{Centre:} the first eigenvalue and eigenvector fields. \textit{Right:} the mean field mass distribution around the swallowtail caustic, evolved by the Zel'dovich approximation}\label{fig:meanSwallowtail}
\end{figure}

In figure \ref{fig:meanSwallowtail}, we plot the mean field of the swallowtail cluster. The lefthand panel of the figure shows the gravitational potential of the cusp, the central panel the corresponding (first) eigenvalue and eigenvector fields, while the righthand panel shows the corresponding evolved mass distribution according to the Zeldovich approximation.

The gravitational field is more faceted than the quadrupolar pattern of the cusp caustic. Also here we find a central gravitational trough. It is more concentrated and has a weakly triangular shape with a somewhat pointed lower side and a more flattened upper edge. It marks the central region of matter inflow. The surrounding region reveals three potential hills, corresponding to low-density outflow regions. Two dominant protovoids are seen at the lefthand and righthand side, reminiscent of the quadrupolar pattern of the cusp caustic, while a weaker outflow region is seen at the upper side. The mean eigenvalue field (central panel) reveals more details of the implied pattern. The red fold curve is near oval with two bifurcating shoulders to the upper right and left. Within the fold curve, we find the blue trident shape cusp curves, which appear as the medial axis of the fold curve.

The resulting matter distribution, evolved through the Zeldovich approximation, in and around the swallowtail reveals an interesting and highly significant aspect of the connections in the cosmic web. The core of the swallowtail evolves into a cluster node of the weblike network. Three filamentary features emanate out of the central region, one tail-like ridge pointing downwards and two whisker-like features on the upper side. It indicates how the caustic structure of the mass distribution leads to the natural pattern of cluster-filament connections in the cosmic web. 

%%%%%%%%%%%%%%%%%%%%%%%%%%%%%%%%%%%%%%%%%%%%%%%%%%%%%%%%%%%%%%%%
\subsubsection{{\it Swallowtail residual field}}
To allow a representative sampling of the residual field, via the probability distribution for the involved linear constraints, we proceed in a similar fashion as in the case of the cusp caustic. To this end, we consider the caustic conditions for the swallowtail in the eigenframe,
\begin{align}
T_{11}=1/b_+(t)\,, \quad T_{12}=0\,,\quad T_{22}\leq T_{11}\,,\quad T_{111}=0\,, \quad T_{1111}+\frac{3T_{112}^2}{T_{11}-T_{22}} =0\,.
\end{align}
The condition that specifies the direction in which the swallowtail develops yields the simple form 
\begin{align}
\bm{v}_2 \cdot \nabla \lambda_1  <0: \qquad T_{112} <0\,.
\end{align}
For notational aspects and detailed derivations consult appendix~\ref{ap:eigenvalueRel}.

\bigskip
\noindent The swallowtail caustic is determined by the linear statistics
\begin{equation}
  Y=(T_{11},T_{12},T_{22},T_{111},T_{112},T_{1111})
\end{equation}
with the probability density
\begin{align}
p(T_{11},T_{12},T_{22},T_{111},T_{112},T_{1111})=p(T_{11},T_{22},T_{1111})p(T_{12})p(T_{111})p(T_{112})\,,
\end{align}
and constraint distribution 
\begin{align}
&p\left(T_{22},T_{112},T_{1111}|T_{22}\leq T_{11}=1/b_+, T_{12}=0,T_{111}=0,T_{1111}+\frac{3T_{112}^2}{T_{11}-T_{22}}=0\right)\nonumber\\
=&\, p\left(T_{22},T_{112}, T_{1111}|T_{22}\leq T_{11}=1/b_+, T_{1111}+\frac{3T_{112}^2}{T_{11}-T_{22}}=0\right)\,.
\end{align}

\bigskip
\noindent For the cusp caustic, it sufficed to construct realizations with the linear constraints $T_{11}=1/b_+$, $T_{12}=0$, $T_{111}=0$ and reject realizations for which $T_{22} > T_{11}$. For the swallowtail caustic, we implement a more advanced rejection scheme to efficiently sample the constrained manifold.

In the joint distribution of the statistic $(T_{11},T_{22},T_{112},T_{1111})$, the variable $T_{112}$ is independent of the statistics $T_{11}, T_{22}$ and $T_{1111}$. We also take along the fact that $T_{112}$ is fully specified in terms of the variables $T_{11},T_{22}$ and $T_{1111}$ as a result of the conditions
\begin{align}
  &T_{1111}+\frac{3T_{112}^2}{T_{11}-T_{22}}=0\,,\\
  \ \\
  &T_{112}<0\,.
\end{align}
We use these properties to construct a rejection sampling scheme in which we sample the Gaussian distribution for $(T_{22},T_{1111})$ and accept a sample with a probability determined by the joint distribution for $(T_{22},T_{112},T_{1111})$. To this end, consider the Gaussian joint
distribution $p(T_{11},T_{22},T_{1111})$ with vanishing mean $\bm{\mu}=\bm{0}$ and covariance matrix
\begin{align}
  \begin{pmatrix} a & \bm{b} \\ \bm{b}^T & \Sigma \end{pmatrix},
\end{align}
with
\begin{align}
a=\frac{3 \sigma_2^2}{8}\,, \quad
\bm{b}=\left(\frac{\sigma_2^2}{8}, -\frac{5 \sigma_3^2}{16}\right)\,,\quad
\Sigma = \begin{pmatrix} \frac{3 \sigma_2^2}{8} & -\frac{\sigma_3^2}{16} \\ -\frac{\sigma_3^2}{16} & \frac{35 \sigma_4^2}{128}\end{pmatrix}.
  \end{align}
In addition, we have the independent statistic $T_{112}$, which is normally distributed with zero mean $\langle T_{112}\rangle = 0$ and variance $\langle T_{112}^2\rangle = \sigma_3^2/16$.

\bigskip
The conditional distribution $p(T_{22},T_{1111}|T_{11}=1/b_+)$ is a bivariate Gaussian distribution with non-zero mean $\bar{\bm{\mu}}=\frac{\bm{b}}{a b_+}$ and the covariance matrix $\bar{\Sigma}=\Sigma -\frac{1}{a} \bm{b}\bm{b}^T$. After sampling from this distribution, we first reject samples for which $T_{22}>1/b_+$ or $T_{1111}>0$ to obtain samples from the auxiliary probability density 
\begin{align}
p(T_{22},T_{1111}| T_{22} < T_{11}=1/b_+,T_{1111}\leq 0)\,.
\end{align}

\noindent The variables $T_{22},T_{1111}$ fully determine the variable $T_{112}$,
\begin{align}
T_{112}= -\sqrt{-\frac{T_{1111}(T_{11}-T_{22})}{3}}\,,
\end{align}
when $T_{22} < T_{11}$ and $T_{1111}\leq 0$.  The unnormalized conditional density is the product of the conditional density $p(T_{22},T_{1111}|T_{22}\leq T_{11}=1/b_+,T_{1111}\leq 0)$ and a function assuming values between $0$ and $1$,
\begin{align}
&p(T_{22},T_{1111}|T_{22}\leq T_{11}=1/b_+, T_{1111}+3T_{112}^2/(T_{11}-T_{22}) =0)\nonumber\\
& \propto e^{-\frac{1}{2}((T_{22},T_{1111})-\bar{\bm{\mu}})\bar{\Sigma}^{-1}((T_{22},T_{1111})-\bar{\bm{\mu}}) }\Theta(1/b_+-T_{22})\Theta(-T_{1111})e^{ \frac{8T_{1111} (1/b_+ - T_{22})}{3 \sigma_3^2}}\,,
\end{align}
with the Heaviside step function $\Theta$. Hence, we obtain samples of the target distribution by sampling from the conditional density $p(T_{22},T_{1111}|T_{22}\leq T_{11}=1/b_+,T_{1111}\leq 0)$ and accepting samples with probability 
\begin{align}
\exp\left[ \frac{8T_{1111} (1/b_+ - T_{22})}{3 \sigma_3^2}\right]\,.
\end{align}

\begin{figure}
\centering
\begin{subfigure}[b]{0.32\textwidth}
\includegraphics[width=\textwidth]{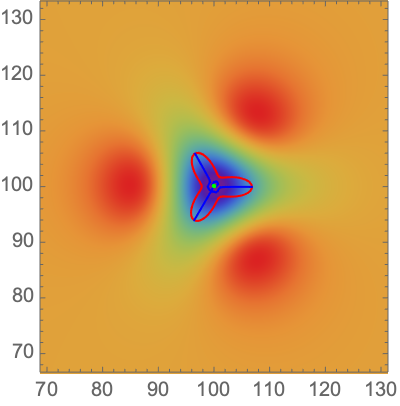}
\end{subfigure}~
\begin{subfigure}[b]{0.32\textwidth}
\includegraphics[width=\textwidth]{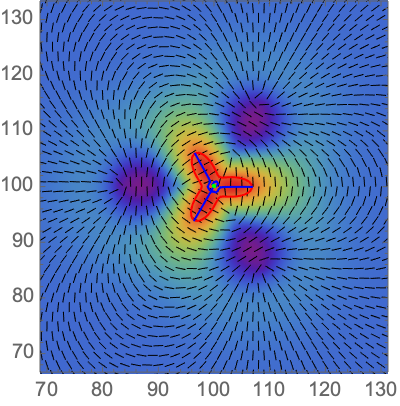}
\end{subfigure}~
\begin{subfigure}[b]{0.32\textwidth}
\includegraphics[width=\textwidth]{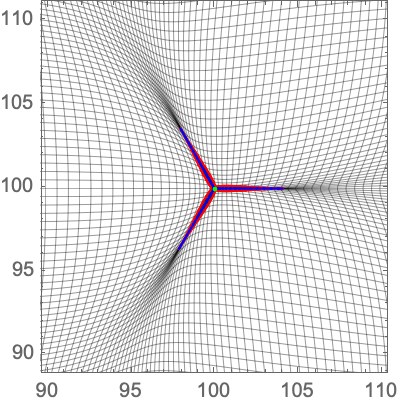}
\end{subfigure}\\
\begin{subfigure}[b]{0.32\textwidth}
\includegraphics[width=\textwidth]{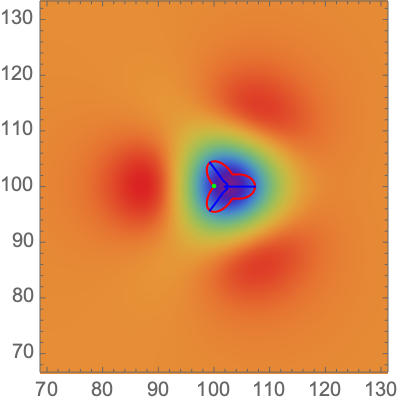}
\end{subfigure}~
\begin{subfigure}[b]{0.32\textwidth}
\includegraphics[width=\textwidth]{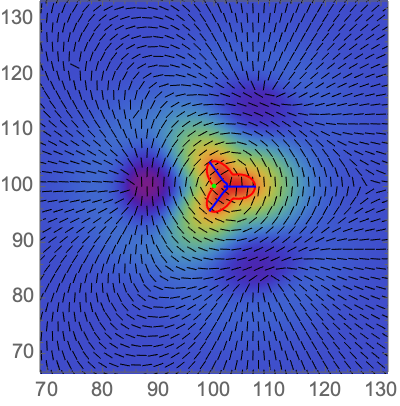}
\end{subfigure}~
\begin{subfigure}[b]{0.32\textwidth}
\includegraphics[width=\textwidth]{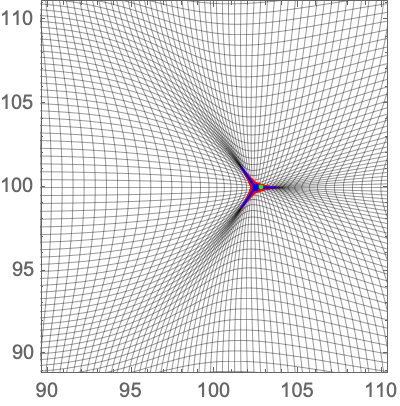}
\end{subfigure}
\caption{The mean umbilic caustic. \textit{Upper:} the  elliptic caustic. \textit{Lower:} the hyperbolic caustic. \textit{Left:} the gravitational potential. \textit{Centre:} the first eigenvalue and eigenvector fields. \textit{Right:} the mean field mass distribution around the elliptic and hyperbolic caustics, evolved by the Zel'dovich approximation}\label{fig:meanHyperbolic}
\end{figure}

%%%%%%%%%%%%%%%%%%%%%%%%%%%%%%%%%%%%%%%%%%%%%%%%%%%%%%%%%%%%%%%%
\subsection{Umbilic \& cluster constrained fields}
The umbilic caustic forming at time $t$ satisfies the simultaneous conditions for the first and second eigenvalues,  
\begin{align}
b_+(t) \lambda_1(\bm{q}_c) = b_+(t) \lambda_2(\bm{q}_c) = 1\,.
\end{align}
In terms of the second-order derivatives of the displacement potential, these eigenvalue conditions can be expressed in terms of three linear conditions
\begin{align}
T_{11}&=T_{22}=1/b_+(t)\,,\\
T_{12}&=0\,.
\end{align}
At the umbilic points, the determinant 
\begin{align}
\det(\nabla \bm{x}_t) = \det (I- b_+ \bm{\psi})
\end{align}
assumes a critical point, whose nature is determined by the Hessian (see \cite{Rozhanskii:1984})
\begin{align}
\mathcal{H}\left[\det (I- b_+(t) \bm{\psi})\right] =
b_+^2(t)\begin{pmatrix} 
2(T_{111}T_{122} -T_{112}^2) & T_{111}T_{222}-T_{112}T_{122} \\
T_{111}T_{222}-T_{112}T_{122} & 2(T_{112}T_{222}-T_{122}^2)
\end{pmatrix}\,,
\end{align}
and it's determinant
\begin{align}
&\det\left( \mathcal{H}\left[\det (I- b_+(t) \bm{\psi})\right]\right) \nonumber\\
&=b_+^4(t)\left[
 6 T_{111} T_{112} T_{122} T_{222} + 3 T_{112}^2 T_{122}^2  - 4 (T_{111} T_{122}^3 + T_{112}^3 T_{222})  - T_{111}^2 T_{222}^2\right]\,.
\end{align}
From this, we note that the Hessian at the critical point only depends on the third-order derivatives of the deformation potential. The fourth order derivatives cancel when $T_{11}=T_{22}=1/b_+$ and $T_{12}=0$.

\bigskip
There are two different classes of umbilic points, the elliptic caustic $D_4^+$ and the hyperbolic caustic $D_4^-$. The umbilic point is an elliptic caustic $D_4^+$ when the critical point is a maximum or a minimum, \textit{i.e.}, for which 
\begin{align}
\det (\mathcal{H}\left[\det (I- b_+ \bm{\psi})\right]) >0\,.
\end{align}
The umbilic point is a hyperbolic caustic when the critical point is a saddle point, \textit{i.e.},
\begin{align}
\det (\mathcal{H}\left[\det (I- b_+ \bm{\psi})\right]) <0\,.
\end{align}
This condition should be contrasted to the condition presented in \cite{Delmarcelle:1995, Lavin:1997, Hidding:2016, Feldbrugge:2018} based on the sign of 
\begin{align}
\delta = \frac{1}{2}(T_{111} T_{122} + T_{112} T_{222} - T_{112}^2  - T_{122}^2 ).
\end{align} 
In fact, this condition describes the behavior of the eigenvector field in the vicinity of the umbilic point and does not differentiate between the elliptic and hyperbolic caustics. In this context, we should note that for the elliptic caustic we always have $\delta <0$.

\bigskip
To describe the structure in and around the umbilics, we may follow a more general consideration of the corresponding displacement field. To this end, we expand the displacement field $\bm{s}_t$ to quadratic order in $q_1$ and $q_2$ around a umbilic point at the origin $(q_1,q_2)=(0,0)$. The determinant of the gradient $\nabla \bm{x}_t$ has a critical point at the origin, as it is the product of two eigenvalue fields that vanish at the origin. The determinant takes the general form
\begin{align}
\det \nabla \bm{x}_t&=\det(I + \nabla \bm{s}_t)\nonumber\\
&= 
A q_1^2 + B q_1 q_2 + C q_2^2 + O(q_1^3,q_1^2 q_2,q_1 q_2^2,q_2^3)\,.
\end{align}
for some constants $A,B,$ and $C$.
%with
%\begin{align}
%A&= s_{1,11} s_{2,12}-s_{1,12} s_{2,11}\,,\\
%B&= s_{1,11} s_{2,22}-s_{1,22} s_{2,11}\,,\\
%C&= s_{1,12} s_{2,22}-s_{1,22} s_{2,12}\,, 
%\end{align}
%where we write the displacement field as $\bm{s}_t=(s_1,s_2)$ and the partial derivatives as $s_{i,jk}=\partial_j \partial_k s_{i}$. 
For small $q_1$ and $q_2$, the level set $\det \nabla \bm{x}_t =0$ approaches a conic section:
\begin{itemize}
\item an ellipse when the discriminant $B^2 - 4AC$ is negative corresponding to the $D_4^+$ caustic,
\item a hyperbole when the discriminant $B^2 - 4AC$ is positive corresponding to the $D_4^-$ caustic,
\item a parabola when the discriminant $B^2 - 4AC$ vanishes, corresponding to the $D_5$ caustic (which does not feature in the 2D cosmic web).
\end{itemize}
Note that the discriminant coincides with the negative of the determinant of the Hessian, \textit{i.e.},
\begin{align}
B^2-4AC = -\det \left(\mathcal{H} \left[\det \nabla \bm{x}_t \right]\right)\,.
\end{align}
When the displacement field is a gradient field, the constants $D$ and $E$ vanish and the condition reduces to the constant discussed above.

The orientation of the elliptic umbilic caustic is determined by the behavior of the eigenvector fields of the deformation tensor in the vicinity of the caustic. The fold curve in the vicinity of the elliptic umbilic caustic is a small ellipse. This forms three cusp caustics when the eigenvector field is parallel to the eigenvector field, ie. $\bm{v}_i \cdot \bm{m} = 0$ with $\bm{m}$ the vector normal to this circle.

The orientation and configuration of the hyperbolic umbilic caustic is determined by the eigenvalues $\nu_i$ and eigenvectors $\bm{w}_i$ of the Hessian $\mathcal{H}\left[\det (I- b_+ \bm{\psi})\right]$. The cusp curve of the hyperbolic umbilic is directed towards the eigenvector corresponding to the positive eigenvalue. The angle $\beta = 2 \arctan(\sqrt{|\nu_1/\nu_2|})$ describes the angle between the two wedges of the fold curves. 

\subsubsection{{\it Umbilic mean field}}
The mean field of the elliptic and the swallowtail caustic are plotted in figure \ref{fig:meanHyperbolic}. The most outstanding observation is that both the elliptic and hyperbolic umbilic cluster are connected to three cusp curves, revealing the natural association of the clusters with three filaments in the cosmic web.

\begin{figure}
\centering
\begin{subfigure}[b]{0.24\textwidth}
\includegraphics[width=\textwidth]{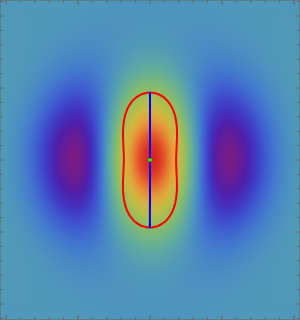}
\end{subfigure}
\begin{subfigure}[b]{0.24\textwidth}
\includegraphics[width=\textwidth]{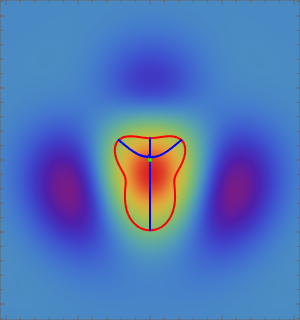}
\end{subfigure}
\begin{subfigure}[b]{0.24\textwidth}
\includegraphics[width=\textwidth]{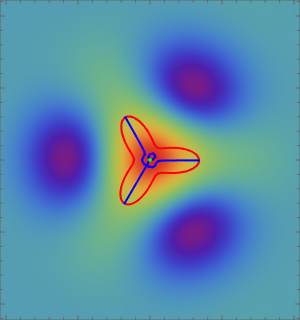}
\end{subfigure}
\begin{subfigure}[b]{0.24\textwidth}
\includegraphics[width=\textwidth]{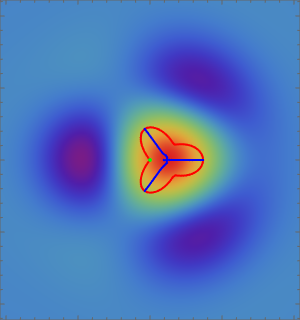}
\end{subfigure}\\ \vspace{0.2\baselineskip}
\begin{subfigure}[b]{0.24\textwidth}
\includegraphics[width=\textwidth]{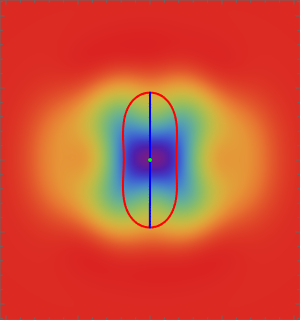}
\end{subfigure}
\begin{subfigure}[b]{0.24\textwidth}
\includegraphics[width=\textwidth]{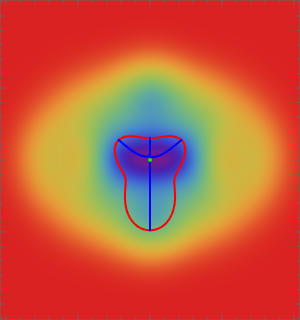}
\end{subfigure}
\begin{subfigure}[b]{0.24\textwidth}
\includegraphics[width=\textwidth]{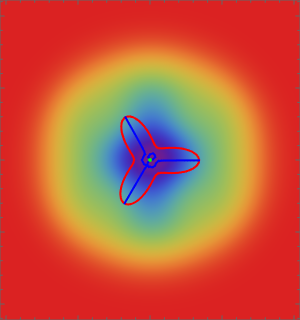}
\end{subfigure}
\begin{subfigure}[b]{0.24\textwidth}
\includegraphics[width=\textwidth]{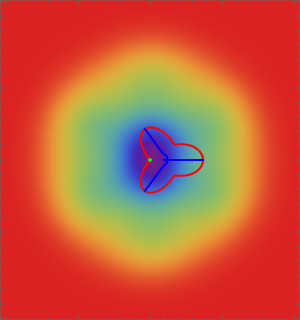}
\end{subfigure}
\caption{Mean and the variance of the density field around the caustics in a $[80\text{ Mpc},120 \text{ Mpc}]^2$ box. \textit{Left to right:} the cusp, the swallowtail, the elliptic umbilic, and the hyperbolic umbilic caustic. \textit{Top:} the mean density field. \textit{Bottom:} the variance of the density field.}\label{fig:mean_var_dens}
\end{figure}

%%%%%%%%%%%%%%%%%%%%%%%%%%%%%%%%%%%%%%%%%%%%%%%%%%%%%%%%%%%%%%%%
\subsubsection{{\it Umbilic residual field}}
We implement the umbilic caustics by sampling random fields with the linear constraints
\begin{align}
T_{11}&=T_{22}=1/b_+\,,\\
T_{12}&=0\,.
\end{align}
We subsequently determine the sign of the determinant $\det\left( \mathcal{H}\left[\det (I- b_+ \bm{\psi})\right]\right)$ to identify whether it is an elliptic or a hyperbolic caustic. Given the identification, we compute its orientation. By rotating the second-order and third-order derivatives (using appendix \ref{ap:rotations}), we generate linear constrained conditions corresponding to elliptic and hyperbolic caustics with a specified orientation. Given the appropriate sampling of the constraint manifold $\mathcal{M}_\mathcal{\tilde C}$, we can evaluate the mean field, the variance of the residual field, and representative realizations.

%%%%%%%%%%%%%%%%%%%%%%%%%%%%%%%%%%%%%%%%%%%%%%%%%%%%%%%%%%%%%%%%
\subsection{Constrained Field Variance}
By definition, the residual field $\delta f$ vanishes at the point where we impose the caustic condition. However, the caustic condition not only fixes the random field at a point but influences the direct environment of the condition. This behavior is formally described by the variance $\left \langle \delta f (\bm{q})^2\,|\,\Gamma\right\rangle$ in the density perturbation near the constraint (equation \eqref{eq:variance_Linear} for linear and equation \eqref{eq:variance_Nonlinear} for non-linear constraints).

To this end, figure \ref{fig:mean_var_dens} presents the variance (density) fields of the cusp, swallowtail and umbilic caustics.
For reference, the residual fields are depicted together with the corresponding mean density fields of the caustics. An important observation is that the statistical properties of the residual field near the cusp and swallowtail caustic are not isotropic or homogeneous. On the other hand, the residual field is (nearly) isotropic near the umbilic caustics.

Finally, for reference note that for the power-law power spectrum with the Gaussian smoothing at the length scale $R_s=5\text{ Mpc}$ considered in this paper, the characteristic correlation length for the random field near the caustics is about $20\text{ Mpc}$.

%%%%%%%%%%%%%%%%%%%%%%%%%%%%%%%%%%%%%%%%%%%%%%%%%%%%%%%%%%%%%%%%
\section{Composite constrained realizations}\label{sec:composite_constraints}
To study the interplay between the voids, walls, filaments, and clusters in the cosmic web and study the connectivity, it is important to extend the constrained random field theory of the caustic skeleton to include multiple constraints with specified orientations. In this section, we show how to construct realizations with composite constraints.

%%%%%%%%%%%%%%%%%%%%%%%%%%%%%%%%%%%%%%%%%%%%%%%%%%%%%%%%%%%%%%%%
\subsection{Orienting constraint conditions}\label{sec:orientation}
In section \ref{sec:caustic_skeleton_constraints}, we demonstrated a method to construct realizations of Gaussian random fields satisfying a caustic constraint at a single point with a fixed orientation. For example, we constructed realizations that included a cusp filament running vertically through the origin of the initial conditions. From the formalism, it follows that the caustic constraint can be translated to any point in the box by shifting the underlying linear functionals $\bm{C}$ (the Fourier transform of the constraints receives a phase shift). Alternatively, we can generate a realization with a cusp at the origin and translate the origin to the required location using the symmetry of the torus on which the realization lives. Unfortunately, the latter method does not work when considering multiple caustic constraints at different locations. Moreover, the second method yields difficulties for even a single caustic constraint when attempting to change the orientation with a rotation due to the symmetries of the boundary conditions of the box.

Fortunately, there exists an efficient method to generate realizations with any preferred orientation at any location:
\begin{enumerate}
\item 
Generate a realization satisfying the caustic conditions at a given point in the box with a fixed orientation as described in section \ref{sec:caustic_skeleton_constraints}. 
\item 
Analyze the derivatives in the caustic conditions. For example, from equations \eqref{eq:cusp_cond_1} and \eqref{eq:cusp_cond_2}, we note that the cusp caustic is fully determined by a number of second, third, and fourth-order derivatives.
\item
Evaluate all derivatives of the realization to the relevant order. Generally, some derivatives were obtained while sampling the constraint manifold $\mathcal{M}_\mathcal{\tilde C}$. Others do not directly influence the caustic and are sampled from the realization. For the cusp, we evaluate the partial derivatives 
\begin{align}
T_{11},T_{12},T_{22},T_{111},T_{112},T_{122},T_{222},T_{1111},T_{1112},T_{1122},T_{1222},T_{2222}\,.
\end{align}
Alternatively, we can extend the constraint manifold to include all these partial derivatives and sample them directly.
\item 
While rotating the coordinate system by an angle $\theta$, the $n$-th order derivatives transform into each other (see appendix \ref{ap:rotations}). We use these transformation rules to obtain a representative sample of the required partial derivatives for the caustic with the desired orientation.
\item 
Finally, we apply the Hoffman Ribak algorithm, with respect to the linear constraints consisting of the relevant partial derivatives, to obtain a realization satisfying the caustic condition with the desired orientation.
\end{enumerate}

\begin{figure}
\centering
\begin{subfigure}[b]{0.4\textwidth}
\includegraphics[width=\textwidth]{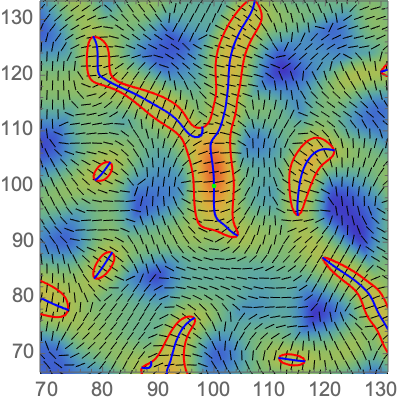}
\end{subfigure}~
\begin{subfigure}[b]{0.4\textwidth}
\includegraphics[width=\textwidth]{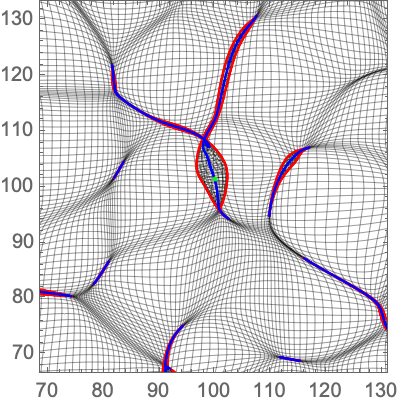}
\end{subfigure}\\
\begin{subfigure}[b]{0.4\textwidth}
\includegraphics[width=\textwidth]{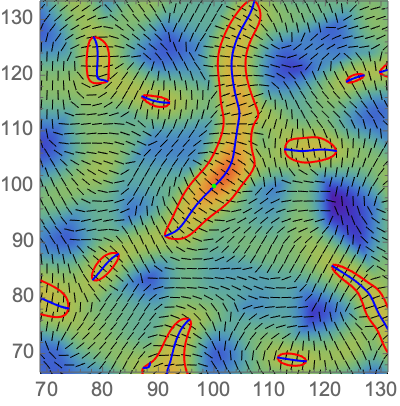}
\end{subfigure}~
\begin{subfigure}[b]{0.4\textwidth}
\includegraphics[width=\textwidth]{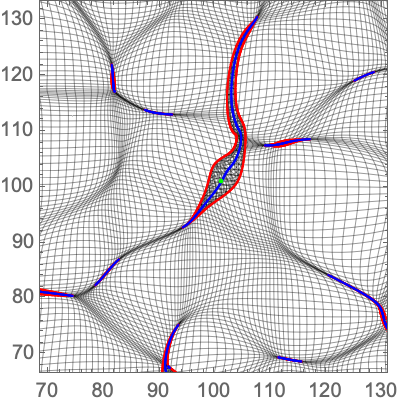}
\end{subfigure}\\
\begin{subfigure}[b]{0.4\textwidth}
\includegraphics[width=\textwidth]{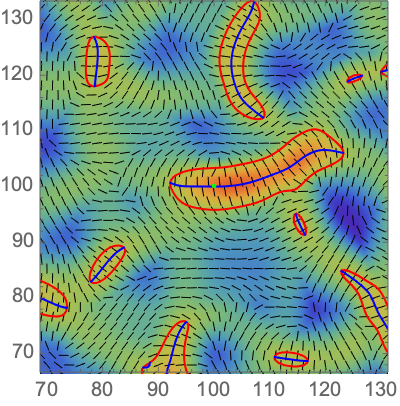}
\end{subfigure}~
\begin{subfigure}[b]{0.4\textwidth}
\includegraphics[width=\textwidth]{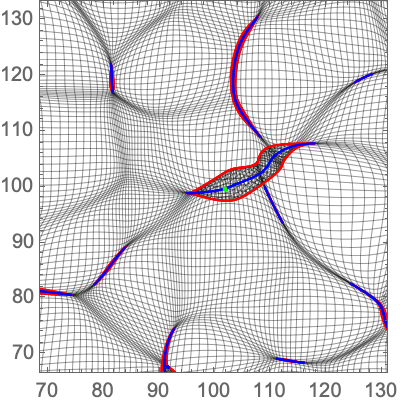}
\end{subfigure}
\caption{The caustic skeleton consisting of the fold curves (red) and cusp curve (blue) in Lagrangian and Eulerian space with a rotation of the cusp caustic at the center of the box. \textit{Upper:} the eigenvalue and eigenvector fields in Lagrangian space. \textit{Lower:} the Zel'dovich approximation in Eulerian space. \textit{Left:} with the vertical orientation. \textit{Center:} with a diagonal orientation. \textit{Right:} with the horizontal orientation.}\label{fig:rotation}
\end{figure}

See figure \ref{fig:rotation} for an illustration of the cusp caustic with three orientations. In the left panel, the cusp caustic passes vertically through the center of the box. In the central and right panels, we generate realizations for which the cusp curve passes diagonally and horizontally through the center of the box. These three realizations share a single unconstrained random field. As a consequence, we see that the structure of the initial conditions away from the center $(100,100)$ remains unchanged. Using constrained Gaussian random field theory, we can locally tweak the random field to satisfy the desired non-linear constraints.

%%%%%%%%%%%%%%%%%%%%%%%%%%%%%%%%%%%%%%%%%%%%%%%%%%%%%%%%%%%%%%%%
\subsection{Multiple constraints}
In section \ref{sec:orientation}, we showed a method to orient a single non-linear constraint. In this section, we extend the treatment to multiple caustic constraints at different locations with specified orientations. 

Given the locations $\bm{q}_i$ and orientations, $\theta_i$ of the $M$ caustics constraints with $i=1,\dots,M$, one might naively repeat the analysis presented in section \ref{sec:orientation} for each constrained point and construct a realization using the Hoffman Ribak algorithm on the sampled partial derivatives. While this is a good approximation when the pairwise separation of the points $\bm{q}_i$ is larger than the typical correlation length of the random field, it fails when partial derivatives at a single point $\bm{q}_i$ become significantly correlated to the partial derivatives of a neighboring point $\bm{q}_j$. Ideally, we would sample the partial derivatives of the deformation potential at the constrained points $\bm{q}_i$ simultaneously. However, such a procedure cannot be straightforwardly implemented when sampling constraints in the eigenframe and orienting the constraints with the procedure described in the previous section.

To correct for these correlations we propose a rejection sampling method. For each caustic constraint corresponding to the point $\bm{q}_i$, we determine the $n$-th order derivatives required to specify the caustic and its orientation. Construct the conditional (unnormalized) joint probability density $p_i$ for the partial derivatives of the constraint $\bm{q}_i$ with respect to the corresponding constrained manifold $\mathcal{M}_{\mathcal{\tilde C},i}$, and the Gaussian joint probability density $p$ including all the partial derivatives at the constrained points. Next, determine the maximum $max$ of the ratio of the joint distribution of all constraints and the individual constraints $p/\prod p_i$. Finally, generate a sample of the partial derivatives satisfying the oriented caustics constraints as described above. Accept this sample with probability $p/(max\, \prod p_i)$. Alternatively, we construct the big constraint manifold $\mathcal{M}_\mathcal{\tilde C}$ by taking the union over the small constrained manifold $\mathcal{M}_{\mathcal{\tilde C},i}$, and directly sample from the conditional probability density $p(\cdot |\mathcal{M}_\mathcal{\tilde C})$ using the rejection sampling method.

\begin{figure}
\centering
\begin{subfigure}[b]{0.49\textwidth}
\includegraphics[width=\textwidth]{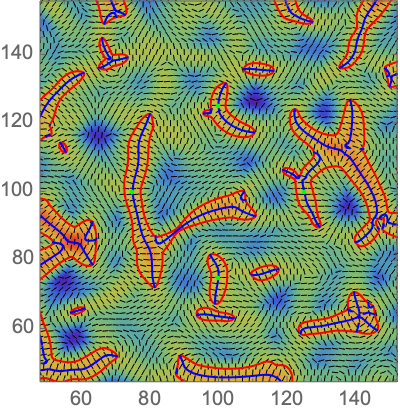}
\caption{$\lambda_1$, $\bm{v}_1$}
\end{subfigure}~
\begin{subfigure}[b]{0.49\textwidth}
\includegraphics[width=\textwidth]{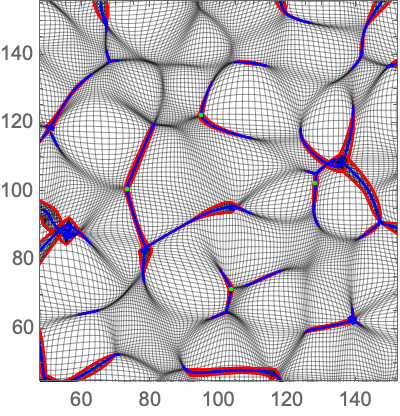}
\caption{Zel'dovich approximation}
\end{subfigure}
\caption{The caustic skeleton consists of the fold curves (red) and cusp curves (blue) in Lagrangian and Eulerian space, with four critical point constraints on the first eigenvalue field. \textit{Left:} the first eigenvalue and eigenvector fields in Lagrangian space. \textit{Right:} the Zel'dovich approximation in Eulerian space.}\label{fig:composite}
\end{figure}

We illustrate this procedure in figure \ref{fig:composite}, which includes four Morse point constraints on the first eigenvalue field. A critical point $\bm{q}_c$ appearing at time $t$ satisfies the condition
\begin{align}
\nabla \lambda_1(\bm{q}_c) = 0\,,\\
\lambda_1(\bm{q}_c)=1/b_+(t)\,.
\end{align}
In the eigenvector frame, this condition reduces to the linear constraints
\begin{align}
1/b_+ = T_{11} \geq T_{22}\,, \quad  T_{12}=0\,,\quad T_{111}=T_{112}=0\,.
\end{align}
See appendix \ref{ap:eigenvalueRel} for a derivation. The critical point is either a maximum, minimum -- corresponding to the Morse point $A_3^+$ -- or a saddle point -- corresponding to the Morse point $A_3^-$. We can differentiate between the two cases by evaluating the sign of the determinant of the Hessian $\mathcal{H}\lambda_1$. In the eigenframe, the Hessian takes the form
\begin{align}
\mathcal{H}\lambda_1 = 
\begin{pmatrix} 
T_{1111}+\frac{2T_{112}^2}{T_{11}-T_{22}} & T_{1112} + \frac{2 T_{112}T_{122}}{T_{11}-T_{22}} \\
T_{1112}+\frac{2 T_{112}T_{122}}{T_{11}-T_{22}} & T_{1122} + \frac{2 T_{122}^2}{T_{11}-T_{22}}
\end{pmatrix}\,.
\end{align}
See appendix \ref{ap:eigenvalueRel} for a derivation of this result. In figure \ref{fig:composite}, the lower, the lower, and the right critical points are $A_3^+$ points. The left and the upper critical points are $A_3^-$ points. In the right panel, we can clearly distinguish the character of the $A_3^+$ and $A_3^-$ points. The $A_3^+$ points correspond to the creation of a multi-stream region forming a Zel'dovich pancake. The $A_3^-$ points correspond to the merger of two multi-stream regions.

%%%%%%%%%%%%%%%%%%%%%%%%%%%%%%%%%%%%%%%%%%%%%%%%%%%%%%%%%%%%%%%%
\section{Conclusion}\label{sec:conclusion}
The study of caustics in the cosmic web can be traced back to early work by Arnol'd and collaborators \cite{Arnold:1982a,Arnold:1982b,Shandarin:2019,Shandarin:2021}. Caustic skeleton theory provides a set of conditions for the emergence of various caustics classified by Lagrangian catastrophe theory in fluids dynamics \cite{Arnold:1982a,Hidding:2014,Feldbrugge:2018}. In large-scale structure formation, these different caustics can be associated with the walls, filaments, and clusters of the present-day cosmic web. When combined with an analytic Lagrangian model for structure formation, such as the Zel'dovich approximation \cite{Zeldovich:1970}, we obtain a direct mapping between a set of points, curves, and sheets in Lagrangian space and the spine of the current cosmic web \cite{Feldbrugge:2018}. These manifolds are defined as non-linear conditions on the eigenvalue and eigenvector fields of the primordial deformation potential. The conditions signal distinct environments in the initial conditions forming the different features of the cosmic web.

In the present paper, we extend linear-constrained Gaussian random field theory \cite{Hoffman:1991,Weygaert:1996} to non-linear constraints, to go beyond the caustics conditions and develop a framework to study the properties of the primordial gravitational potential and density perturbations in the vicinity of these caustics. This way, we aim to put flesh on the bones of the caustic skeleton. By running $N$-body simulations on these initial conditions, this framework paves the way for a systematic investigation of the different elements of the cosmic web. Note that this non-linear method allows for efficient experimentation with the primordial density perturbations, as the numerical implementation comes down to but a small number of Fast Fourier transforms. This should be compared to the splice method \cite{Cadiou:2021} which relies on a large number of linear constraints.

In this paper, we limited the exploration of the cosmic web to two-dimensional toy models. We find that the cusp caustic, associated with the filaments, indeed influences its local environment. Moreover, the mean field displays the typical Zel'dovich pancake. The swallowtail and umbilic caustics, in the two-dimensional cosmic web models linked to the clusters, indeed link the filaments and form either as a singularity on the cusp curves or as a point where gravitational collapse occurs along two directions. In particular, striking is the emergence of the three-fold symmetry of the mean fields of the umbilic caustics from the caustic conditions.

In an upcoming paper, we will use non-linear constrained Gaussian random field theory to investigate the properties of caustics in the three-dimensional cosmic web. Using a set of $N$-body simulations, we will study the density and velocity distributions in the vicinity of different parts of the three-dimensional caustic skeleton of the cosmic web in both Lagrangian and Eulerian space. Note that we can investigate these properties directly as a function of the formation time and smoothing scale of the caustic in the Zel'dovich approximation. 

This will in particular be interesting when it comes to filaments, as caustic skeleton theory -- as yet not fully appreciated -- predicts the existence of two filament classes \cite{Feldbrugge:2018}. The major share of the filamentary network is represented by $A_4$ swallowtail caustics, representing the sectional interfaces between two wall-like membranes. These have not yet fully collapsed along two directions. Umbilic $D_4$ caustics are also filaments, corresponding to the more prominent and dense filamentary extensions of the cluster nodes in the cosmic web and representing filaments that have fully collapsed along two directions. These classes are expected to form at different times and have different characteristic geometric properties. Recently, some numerical evidence on different filament classes has been found \cite{Galarraga-Espinosa:2020}. However, it is currently not yet clear whether these theoretical and numerical populations are related to each other.

Another fruitful field where this method might be very valuable is the systematic investigation of the influence of the cosmic web on the embedded galaxies. Different caustics mark different environments of the cosmic web, with distinct formation histories. It is expected that the formation history of a patch of the cosmic web correlates with the embedded galaxies. In a future paper, we for example plan to use these techniques to systematically investigate the recently discovered galaxy alignment at different locations of the cosmic web \cite{Aragon:2007, Ganeshaiah:2018, Ganeshaiah:2019, Hellwing:2021,Lopez:2021,Ganeshaiah:2021}.

%%%%%%%%%%%%%%%%%%%%%%%%%%%%%%%%%%%%%%%%%%%%%%%%%%%%%%%%%%%%%%%%
\section*{Acknowledgement}
We thank Sergei Shandarin for having raised our interest in caustics as a key to the dynamical understanding of the cosmic web. We thank Bernard Jones for the long-standing discussions on constrained random field theory and Johan Hidding for his comments on caustic skeleton theory and its relation to constraint theory. Finally, JF would like to thank Nynke Niezink for the enlightening discussions on probability theory and statistics. JF is supported in part by the Higgs Fellowship.

%%%%%%%%%%%%%%%%%%%%%%%%%%%%%%%%%%%%%%%%%%%%%%%%%%%%%%%%%%%%%%%%

\bibliographystyle{JHEP}

\bibliography{mybibliography}

\providecommand{\href}[2]{#2}\begingroup\raggedright\begin{thebibliography}{100}

\bibitem{Zeldovich:1970}
Y.B.~{Zel'dovich}, \emph{{Gravitational instability: An approximate theory for
  large density perturbations.}}, {\emph{A\&A} {\bfseries 5} (1970) 84}.

\bibitem{Bond:1996}
J.R.~{Bond}, L.~{Kofman} and D.~{Pogosyan}, \emph{{How filaments of galaxies
  are woven into the cosmic web}},
  \href{https://doi.org/10.1038/380603a0}{\emph{Nature} {\bfseries 380} (1996)
  603} [\href{https://arxiv.org/abs/arXiv:astro-ph/9512141}{{\ttfamily
  arXiv:astro-ph/9512141}}].

\bibitem{Weygaert:2008}
R.~{van de Weygaert} and J.R.~{Bond}, \emph{{Clusters and the Theory of the
  Cosmic Web}},  in \emph{A Pan-Chromatic View of Clusters of Galaxies and the
  Large-Scale Structure}, M.~{Plionis}, O.~{L{\'o}pez-Cruz} and D.~{Hughes},
  eds., vol.~740 of \emph{Lecture Notes in Physics, Berlin Springer Verlag},
  p.~335, 2008.

\bibitem{Feldbrugge:2018}
J.~{Feldbrugge}, R.~{van de Weygaert}, J.~{Hidding} and J.~{Feldbrugge},
  \emph{{Caustic Skeleton \& Cosmic Web}},
  \href{https://doi.org/10.1088/1475-7516/2018/05/027}{\emph{\jcap} {\bfseries
  2018} (2018) 027} [\href{https://arxiv.org/abs/1703.09598}{{\ttfamily
  1703.09598}}].

\bibitem{Arnold:1982a}
V.I.~{Arnol'd}, S.F.~{Shandarin} and I.B.~{Zel'dovich}, \emph{{The large scale
  structure of the universe. I - General properties One- and two-dimensional
  models}}, \href{https://doi.org/10.1080/03091928208209001}{\emph{Geophysical
  and Astrophysical Fluid Dynamics} {\bfseries 20} (1982) 111}.

\bibitem{Arnold:1982b}
V.I.~{Arnol'd}, \emph{{Evolution of singularities of potential flows in
  collisionless media and transformations of caustics in three-dimensional
  space}}, {\emph{Trudy Seminar imeni G Petrovskogo} {\bfseries 8} (1982) 21}.

\bibitem{Bertschinger:1987}
E.~{Bertschinger}, \emph{{Path Integral Methods for Primordial Density
  Perturbations: Sampling of Constrained Gaussian Random Fields}},
  \href{https://doi.org/10.1086/185066}{\emph{\apjl} {\bfseries 323} (1987)
  L103}.

\bibitem{Hoffman:1991}
Y.~{Hoffman} and E.~{Ribak}, \emph{{Constrained Realizations of Gaussian
  Fields: A Simple Algorithm}},
  \href{https://doi.org/10.1086/186160}{\emph{\apjl} {\bfseries 380} (1991)
  L5}.

\bibitem{Weygaert:1996}
R.~{van de Weygaert} and E.~{Bertschinger}, \emph{{Peak and gravity constraints
  in Gaussian primordial density fields: An application of the Hoffman-Ribak
  method}}, \href{https://doi.org/10.1093/mnras/281.1.84}{\emph{\mnras}
  {\bfseries 281} (1996) 84}
  [\href{https://arxiv.org/abs/astro-ph/9507024}{{\ttfamily
  astro-ph/9507024}}].

\bibitem{Shandarin:1989}
S.F.~{Shandarin} and Y.B.~{Zel'dovich}, \emph{{The large-scale structure of the
  universe: Turbulence, intermittency, structures in a self-gravitating
  medium}}, \href{https://doi.org/10.1103/RevModPhys.61.185}{\emph{Reviews of
  Modern Physics} {\bfseries 61} (1989) 185}.

\bibitem{Cautun:2014}
M.~{Cautun}, R.~{van de Weygaert}, B.J.T.~{Jones} and C.S.~{Frenk},
  \emph{{Evolution of the cosmic web}},
  \href{https://doi.org/10.1093/mnras/stu768}{\emph{\mnras} {\bfseries 441}
  (2014) 2923} [\href{https://arxiv.org/abs/1401.7866}{{\ttfamily 1401.7866}}].

\bibitem{Ganeshaiah:2019}
P.~{Ganeshaiah Veena}, M.~{Cautun}, E.~{Tempel}, R.~{van de Weygaert} and
  C.S.~{Frenk}, \emph{{The Cosmic Ballet II: spin alignment of galaxies and
  haloes with large-scale filaments in the EAGLE simulation}},
  \href{https://doi.org/10.1093/mnras/stz1343}{\emph{\mnras} {\bfseries 487}
  (2019) 1607} [\href{https://arxiv.org/abs/1903.06716}{{\ttfamily
  1903.06716}}].

\bibitem{Hossen:2022}
M.R.~{Hossen}, S.A.~{Ema}, K.~{Bolejko} and G.F.~{Lewis}, \emph{{Mapping the
  cosmic mass distribution with stacked weak gravitational lensing and Doppler
  lensing}}, \href{https://doi.org/10.1093/mnras/stab3292}{\emph{\mnras}
  {\bfseries 509} (2022) 5142}
  [\href{https://arxiv.org/abs/2111.05439}{{\ttfamily 2111.05439}}].

\bibitem{Kovacs:2022}
A.~{Kov{\'a}cs}, P.~{Vielzeuf}, I.~{Ferrero}, P.~{Fosalba}, U.~{Demirbozan},
  R.~{Miquel} et~al., \emph{{Dark Energy Survey Year 3 results: Imprints of
  cosmic voids and superclusters in the Planck CMB lensing map}},
  \href{https://doi.org/10.1093/mnras/stac2011}{\emph{\mnras} {\bfseries 515}
  (2022) 4417} [\href{https://arxiv.org/abs/2203.11306}{{\ttfamily
  2203.11306}}].

\bibitem{ABACUSSUMMIT:2021}
N.A.~{Maksimova}, L.H.~{Garrison}, D.J.~{Eisenstein}, B.~{Hadzhiyska},
  S.~{Bose} and T.P.~{Satterthwaite}, \emph{{ABACUSSUMMIT: a massive set of
  high-accuracy, high-resolution N-body simulations}},
  \href{https://doi.org/10.1093/mnras/stab2484}{\emph{\mnras} {\bfseries 508}
  (2021) 4017} [\href{https://arxiv.org/abs/2110.11398}{{\ttfamily
  2110.11398}}].

\bibitem{Quijote:2020}
F.~{Villaescusa-Navarro}, C.~{Hahn}, E.~{Massara}, A.~{Banerjee},
  A.M.~{Delgado}, D.K.~{Ramanah} et~al., \emph{{The Quijote Simulations}},
  \href{https://doi.org/10.3847/1538-4365/ab9d82}{\emph{\apjs} {\bfseries 250}
  (2020) 2} [\href{https://arxiv.org/abs/1909.05273}{{\ttfamily 1909.05273}}].

\bibitem{Shandarin:2010}
S.~{Shandarin}, S.~{Habib} and K.~{Heitmann}, \emph{{Origin of the cosmic
  network in {\ensuremath{\Lambda}}CDM: Nature vs nurture}},
  \href{https://doi.org/10.1103/PhysRevD.81.103006}{\emph{\prd} {\bfseries 81}
  (2010) 103006} [\href{https://arxiv.org/abs/0912.4471}{{\ttfamily
  0912.4471}}].

\bibitem{Shandarin:2011}
S.F.~{Shandarin}, \emph{{The multi-stream flows and the dynamics of the cosmic
  web}}, \href{https://doi.org/10.1088/1475-7516/2011/05/015}{\emph{\jcap}
  {\bfseries 2011} (2011) 015}
  [\href{https://arxiv.org/abs/1011.1924}{{\ttfamily 1011.1924}}].

\bibitem{Shandarin:2012}
S.~{Shandarin}, S.~{Habib} and K.~{Heitmann}, \emph{{Cosmic web, multistream
  flows, and tessellations}},
  \href{https://doi.org/10.1103/PhysRevD.85.083005}{\emph{Phys. Rev. D}
  {\bfseries 85} (2012) 083005}
  [\href{https://arxiv.org/abs/1111.2366}{{\ttfamily 1111.2366}}].

\bibitem{Abel:2012}
T.~{Abel}, O.~{Hahn} and R.~{Kaehler}, \emph{{Tracing the dark matter sheet in
  phase space}},
  \href{https://doi.org/10.1111/j.1365-2966.2012.21754.x}{\emph{\mnras}
  {\bfseries 427} (2012) 61} [\href{https://arxiv.org/abs/1111.3944}{{\ttfamily
  1111.3944}}].

\bibitem{Falck:2012}
B.L.~{Falck}, M.C.~{Neyrinck} and A.S.~{Szalay}, \emph{{ORIGAMI: Delineating
  Halos Using Phase-space Folds}},
  \href{https://doi.org/10.1088/0004-637X/754/2/126}{\emph{\apj} {\bfseries
  754} (2012) 126} [\href{https://arxiv.org/abs/1201.2353}{{\ttfamily
  1201.2353}}].

\bibitem{Einasto:1978}
M.~{Joeveer} and J.~{Einasto}, \emph{{Has the Universe the Cell Structure?}},
  in \emph{Large Scale Structures in the Universe}, M.S.~{Longair} and
  J.~{Einasto}, eds., vol.~79, p.~241, Jan., 1978.

\bibitem{Lapparent:1986}
V.~{de Lapparent}, M.J.~{Geller} and J.P.~{Huchra}, \emph{{A Slice of the
  Universe}}, \href{https://doi.org/10.1086/184625}{\emph{\apjl} {\bfseries
  302} (1986) L1}.

\bibitem{Colless:2003}
M.~{Colless} and et. al., \emph{{The 2dF Galaxy Redshift Survey: Final Data
  Release}}, {\emph{ArXiv Astrophysics e-prints} (2003) }
  [\href{https://arxiv.org/abs/arXiv:astro-ph/0306581}{{\ttfamily
  arXiv:astro-ph/0306581}}].

\bibitem{Huchra:2012}
J.P.~{Huchra}, L.M.~{Macri}, K.L.~{Masters}, T.H.~{Jarrett}, P.~{Berlind},
  M.~{Calkins} et~al., \emph{{The 2MASS Redshift Survey-Description and Data
  Release}}, \href{https://doi.org/10.1088/0067-0049/199/2/26}{\emph{ApJS}
  {\bfseries 199} (2012) 26} [\href{https://arxiv.org/abs/1108.0669}{{\ttfamily
  1108.0669}}].

\bibitem{Granett:2012}
B.R.~{Granett}, L.~{Guzzo}, J.~{Coupon}, S.~{Arnouts}, P.~{Hudelot},
  O.~{Ilbert} et~al., \emph{{The power spectrum from the angular distribution
  of galaxies in the CFHTLS-Wide fields at redshift {\ensuremath{\sim}}0.7}},
  \href{https://doi.org/10.1111/j.1365-2966.2011.20297.x}{\emph{\mnras}
  {\bfseries 421} (2012) 251}
  [\href{https://arxiv.org/abs/1112.0008}{{\ttfamily 1112.0008}}].

\bibitem{Thom:1975}
R.~Thom, D.~Fowler and C.~Waddington, \emph{Structural Stability and
  Morphogenesis: An Outline of a General Theory of Models}, Advanced book
  program, W. A. Benjamin (1975).

\bibitem{Zeeman:1972}
C.~{Zeeman}, \emph{A catastrophe machine},  in \emph{Biological Process in
  Living Systems}, C.H.~Waddington, ed., (New York), Routledge (1972).

\bibitem{Zeeman:1976}
E.C.~{Zeeman}, \emph{{Catastrophe Theory}},
  \href{https://doi.org/10.1038/scientificamerican0476-65}{\emph{Scientific
  American} {\bfseries 234} (1976) 65}.

\bibitem{Arnold:1972}
V.I.~{Arnol'd}, \emph{{Normal forms for functions near degenerate critical
  points, the Weyl groups of $A_k,D_k,E_k$ and Lagrangian singularities}},
  {\emph{Functional Anal. Appl} {\bfseries 6} (1972) 1972}.

\bibitem{Arnold:1976}
V.I.~{Arnol'd}, \emph{{Wave front evolution and equivalent Morse lemma}},
  {\emph{Communications in Pure Applied Mathematics} {\bfseries 29} (1976)
  557}.

\bibitem{Arnold:1984}
V.I.~{Arnold}, \emph{{Catastrophe theory}}, Springer Berlin, Heidelberg (1984).

\bibitem{Hidding:2014}
J.~{Hidding}, S.F.~{Shandarin} and R.~{van de Weygaert}, \emph{{The Zel'dovich
  approximation: key to understanding cosmic web complexity}},
  \href{https://doi.org/10.1093/mnras/stt2142}{\emph{\mnras} {\bfseries 437}
  (2014) 3442} [\href{https://arxiv.org/abs/1311.7134}{{\ttfamily 1311.7134}}].

\bibitem{Aragon:2010a}
M.A.~{Arag{\'o}n-Calvo}, R.~{van de Weygaert} and B.J.T.~{Jones},
  \emph{{Multiscale phenomenology of the cosmic web}},
  \href{https://doi.org/10.1111/j.1365-2966.2010.17263.x}{\emph{\mnras}
  {\bfseries 408} (2010) 2163}
  [\href{https://arxiv.org/abs/1007.0742}{{\ttfamily 1007.0742}}].

\bibitem{Aragon:2010b}
M.A.~{Arag{\'o}n-Calvo}, E.~{Platen}, R.~{van de Weygaert} and A.S.~{Szalay},
  \emph{{The Spine of the Cosmic Web}},
  \href{https://doi.org/10.1088/0004-637X/723/1/364}{\emph{\apj} {\bfseries
  723} (2010) 364} [\href{https://arxiv.org/abs/0809.5104}{{\ttfamily
  0809.5104}}].

\bibitem{Weygaert:1993}
R.~{van de Weygaert} and E.~{van Kampen}, \emph{{Voids in Gravitational
  Instability Scenarios - Part One - Global Density and Velocity Fields in an
  Einstein - De-Sitter Universe}},
  \href{https://doi.org/10.1093/mnras/263.2.481}{\emph{\mnras} {\bfseries 263}
  (1993) 481}.

\bibitem{Haarlem:1993}
M.~{van Haarlem} and R.~{van de Weygaert}, \emph{{Velocity Fields and
  Alignments of Clusters in Gravitational Instability Scenarios}},
  \href{https://doi.org/10.1086/173416}{\emph{\apj} {\bfseries 418} (1993)
  544}.

\bibitem{Rybicki:1992}
G.B.~{Rybicki} and W.H.~{Press}, \emph{{Interpolation, Realization, and
  Reconstruction of Noisy, Irregularly Sampled Data}},
  \href{https://doi.org/10.1086/171845}{\emph{\apj} {\bfseries 398} (1992)
  169}.

\bibitem{Bertschinger:1989}
E.~{Bertschinger} and A.~{Dekel}, \emph{{Recovering the Full Velocity and
  Density Fields from Large-Scale Redshift-Distance Samples}},
  \href{https://doi.org/10.1086/185348}{\emph{\apjl} {\bfseries 336} (1989)
  L5}.

\bibitem{Dekel:1999}
A.~{Dekel}, A.~{Eldar}, T.~{Kolatt}, A.~{Yahil}, J.A.~{Willick}, S.M.~{Faber}
  et~al., \emph{{POTENT Reconstruction from Mark III Velocities}},
  \href{https://doi.org/10.1086/307636}{\emph{\apj} {\bfseries 522} (1999) 1}
  [\href{https://arxiv.org/abs/astro-ph/9812197}{{\ttfamily
  astro-ph/9812197}}].

\bibitem{Courtois:2012}
H.M.~{Courtois}, Y.~{Hoffman}, R.B.~{Tully} and S.~{Gottl{\"o}ber},
  \emph{{Three-dimensional Velocity and Density Reconstructions of the Local
  Universe with Cosmicflows-1}},
  \href{https://doi.org/10.1088/0004-637X/744/1/43}{\emph{\apj} {\bfseries 744}
  (2012) 43} [\href{https://arxiv.org/abs/1109.3856}{{\ttfamily 1109.3856}}].

\bibitem{Hoffman:2015}
Y.~{Hoffman}, H.M.~{Courtois} and R.B.~{Tully}, \emph{{Cosmic bulk flow and the
  local motion from Cosmicflows-2}},
  \href{https://doi.org/10.1093/mnras/stv615}{\emph{\mnras} {\bfseries 449}
  (2015) 4494} [\href{https://arxiv.org/abs/1503.05422}{{\ttfamily
  1503.05422}}].

\bibitem{Sorce:2016}
J.G.~{Sorce}, S.~{Gottl{\"o}ber}, G.~{Yepes}, Y.~{Hoffman}, H.M.~{Courtois},
  M.~{Steinmetz} et~al., \emph{{Cosmicflows Constrained Local UniversE
  Simulations}}, \href{https://doi.org/10.1093/mnras/stv2407}{\emph{\mnras}
  {\bfseries 455} (2016) 2078}
  [\href{https://arxiv.org/abs/1510.04900}{{\ttfamily 1510.04900}}].

\bibitem{Zaroubi:1995}
S.~{Zaroubi}, Y.~{Hoffman}, K.B.~{Fisher} and O.~{Lahav}, \emph{{Wiener
  Reconstruction of the Large-Scale Structure}},
  \href{https://doi.org/10.1086/176070}{\emph{\apj} {\bfseries 449} (1995) 446}
  [\href{https://arxiv.org/abs/astro-ph/9410080}{{\ttfamily
  astro-ph/9410080}}].

\bibitem{Erdogdu:2004}
P.~{Erdo{\v{g}}du}, O.~{Lahav}, S.~{Zaroubi}, G.~{Efstathiou}, S.~{Moody},
  J.A.~{Peacock} et~al., \emph{{The 2dF Galaxy Redshift Survey: Wiener
  reconstruction of the cosmic web}},
  \href{https://doi.org/10.1111/j.1365-2966.2004.07984.x}{\emph{\mnras}
  {\bfseries 352} (2004) 939}
  [\href{https://arxiv.org/abs/astro-ph/0312546}{{\ttfamily
  astro-ph/0312546}}].

\bibitem{Cadiou:2021}
C.~{Cadiou}, A.~{Pontzen}, H.V.~{Peiris} and L.~{Lucie-Smith}, \emph{{The
  causal effect of environment on halo mass and concentration}},
  \href{https://doi.org/10.1093/mnras/stab2650}{\emph{\mnras} {\bfseries 508}
  (2021) 1189} [\href{https://arxiv.org/abs/2107.03407}{{\ttfamily
  2107.03407}}].

\bibitem{Platen:2011}
E.~{Platen}, R.~{van de Weygaert}, B.J.T.~{Jones}, G.~{Vegter} and
  M.A.A.~{Calvo}, \emph{{Structural analysis of the SDSS Cosmic Web - I.
  Non-linear density field reconstructions}},
  \href{https://doi.org/10.1111/j.1365-2966.2011.18905.x}{\emph{\mnras}
  {\bfseries 416} (2011) 2494}
  [\href{https://arxiv.org/abs/1107.1488}{{\ttfamily 1107.1488}}].

\bibitem{Doumler:2013}
T.~{Doumler}, S.~{Gottl{\"o}ber}, Y.~{Hoffman} and H.~{Courtois},
  \emph{{Reconstructing cosmological initial conditions from galaxy peculiar
  velocities - III. Constrained simulations}},
  \href{https://doi.org/10.1093/mnras/sts614}{\emph{\mnras} {\bfseries 430}
  (2013) 912} [\href{https://arxiv.org/abs/1212.2810}{{\ttfamily 1212.2810}}].

\bibitem{Peebles:1989}
P.J.E.~{Peebles}, \emph{{Tracing Galaxy Orbits Back in Time}},
  \href{https://doi.org/10.1086/185529}{\emph{\apjl} {\bfseries 344} (1989)
  L53}.

\bibitem{Mohayaee:2006}
R.~{Mohayaee}, H.~{Mathis}, S.~{Colombi} and J.~{Silk}, \emph{{Reconstruction
  of primordial density fields}},
  \href{https://doi.org/10.1111/j.1365-2966.2005.09774.x}{\emph{\mnras}
  {\bfseries 365} (2006) 939}
  [\href{https://arxiv.org/abs/astro-ph/0501217}{{\ttfamily
  astro-ph/0501217}}].

\bibitem{Lavaux:2008}
G.~{Lavaux}, R.~{Mohayaee}, S.~{Colombi}, R.B.~{Tully}, F.~{Bernardeau} and
  J.~{Silk}, \emph{{Observational biases in Lagrangian reconstructions of
  cosmic velocity fields}},
  \href{https://doi.org/10.1111/j.1365-2966.2007.12539.x}{\emph{\mnras}
  {\bfseries 383} (2008) 1292}
  [\href{https://arxiv.org/abs/0707.3483}{{\ttfamily 0707.3483}}].

\bibitem{Hada:2018}
R.~{Hada} and D.J.~{Eisenstein}, \emph{{An iterative reconstruction of
  cosmological initial density fields}},
  \href{https://doi.org/10.1093/mnras/sty1203}{\emph{\mnras} {\bfseries 478}
  (2018) 1866} [\href{https://arxiv.org/abs/1804.04738}{{\ttfamily
  1804.04738}}].

\bibitem{Shi:2018}
Y.~{Shi}, M.~{Cautun} and B.~{Li}, \emph{{New method for initial density
  reconstruction}},
  \href{https://doi.org/10.1103/PhysRevD.97.023505}{\emph{\prd} {\bfseries 97}
  (2018) 023505} [\href{https://arxiv.org/abs/1709.06350}{{\ttfamily
  1709.06350}}].

\bibitem{Zhu:2018}
H.-M.~{Zhu}, Y.~{Yu} and U.-L.~{Pen}, \emph{{Nonlinear reconstruction of
  redshift space distortions}},
  \href{https://doi.org/10.1103/PhysRevD.97.043502}{\emph{\prd} {\bfseries 97}
  (2018) 043502} [\href{https://arxiv.org/abs/1711.03218}{{\ttfamily
  1711.03218}}].

\bibitem{Kitaura:2008}
F.S.~{Kitaura} and T.A.~{En{\ss}lin}, \emph{{Bayesian reconstruction of the
  cosmological large-scale structure: methodology, inverse algorithms and
  numerical optimization}},
  \href{https://doi.org/10.1111/j.1365-2966.2008.13341.x}{\emph{\mnras}
  {\bfseries 389} (2008) 497}
  [\href{https://arxiv.org/abs/0705.0429}{{\ttfamily 0705.0429}}].

\bibitem{Kitaura:2009}
F.S.~{Kitaura}, J.~{Jasche}, C.~{Li}, T.A.~{En{\ss}lin}, R.B.~{Metcalf},
  B.D.~{Wandelt} et~al., \emph{{Cosmic cartography of the large-scale structure
  with Sloan Digital Sky Survey data release 6}},
  \href{https://doi.org/10.1111/j.1365-2966.2009.15470.x}{\emph{\mnras}
  {\bfseries 400} (2009) 183}
  [\href{https://arxiv.org/abs/0906.3978}{{\ttfamily 0906.3978}}].

\bibitem{Jasche:2010}
J.~{Jasche}, F.S.~{Kitaura}, C.~{Li} and T.A.~{En{\ss}lin}, \emph{{Bayesian
  non-linear large-scale structure inference of the Sloan Digital Sky Survey
  Data Release 7}},
  \href{https://doi.org/10.1111/j.1365-2966.2010.17313.x}{\emph{\mnras}
  {\bfseries 409} (2010) 355}
  [\href{https://arxiv.org/abs/0911.2498}{{\ttfamily 0911.2498}}].

\bibitem{Leclercq:2015}
F.~{Leclercq}, J.~{Jasche} and B.~{Wandelt}, \emph{{Bayesian analysis of the
  dynamic cosmic web in the SDSS galaxy survey}},
  \href{https://doi.org/10.1088/1475-7516/2015/06/015}{\emph{\jcap} {\bfseries
  6} (2015) 015} [\href{https://arxiv.org/abs/1502.02690}{{\ttfamily
  1502.02690}}].

\bibitem{Hess:2016}
S.~{He{\ss}} and F.-S.~{Kitaura}, \emph{{Cosmic flows and the expansion of the
  local Universe from non-linear phase-space reconstructions}},
  \href{https://doi.org/10.1093/mnras/stv2928}{\emph{\mnras} {\bfseries 456}
  (2016) 4247} [\href{https://arxiv.org/abs/1412.7310}{{\ttfamily 1412.7310}}].

\bibitem{Bos:2016}
E.G.P.~{Bos}, R.~{van de Weygaert}, F.~{Kitaura} and M.~{Cautun},
  \emph{{Bayesian Cosmic Web Reconstruction: BARCODE for Clusters}},  in
  \emph{The Zeldovich Universe: Genesis and Growth of the Cosmic Web}, R.~{van
  de Weygaert}, S.~{Shandarin}, E.~{Saar} and J.~{Einasto}, eds., vol.~308,
  pp.~271--288, Oct., 2016,
  \href{https://doi.org/10.1017/S1743921316009996}{DOI}
  [\href{https://arxiv.org/abs/1611.01220}{{\ttfamily 1611.01220}}].

\bibitem{Leclercq:2017}
F.~{Leclercq}, J.~{Jasche}, G.~{Lavaux}, B.~{Wandelt} and W.~{Percival},
  \emph{{The phase-space structure of nearby dark matter as constrained by the
  SDSS}}, \href{https://doi.org/10.1088/1475-7516/2017/06/049}{\emph{\jcap}
  {\bfseries 6} (2017) 049} [\href{https://arxiv.org/abs/1601.00093}{{\ttfamily
  1601.00093}}].

\bibitem{McAlpine:2022}
S.~{McAlpine}, J.C.~{Helly}, M.~{Schaller}, T.~{Sawala}, G.~{Lavaux},
  J.~{Jasche} et~al., \emph{{SIBELIUS-DARK: a galaxy catalogue of the local
  volume from a constrained realization simulation}},
  \href{https://doi.org/10.1093/mnras/stac295}{\emph{\mnras} {\bfseries 512}
  (2022) 5823} [\href{https://arxiv.org/abs/2202.04099}{{\ttfamily
  2202.04099}}].

\bibitem{Hess:2013}
S.~{He{\ss}}, F.-S.~{Kitaura} and S.~{Gottl{\"o}ber}, \emph{{Simulating
  structure formation of the Local Universe}},
  \href{https://doi.org/10.1093/mnras/stt1428}{\emph{\mnras} {\bfseries 435}
  (2013) 2065} [\href{https://arxiv.org/abs/1304.6565}{{\ttfamily 1304.6565}}].

\bibitem{Hidding:2016}
J.~{Hidding}, R.~{van de Weygaert} and S.~{Shandarin}, \emph{{The Zeldovich \&
  Adhesion approximations and applications to the local universe}},  in
  \emph{The Zeldovich Universe: Genesis and Growth of the Cosmic Web}, R.~{van
  de Weygaert}, S.~{Shandarin}, E.~{Saar} and J.~{Einasto}, eds., vol.~308 of
  \emph{IAU Symposium}, pp.~69--76, Oct., 2016,
  \href{https://doi.org/10.1017/S1743921316009650}{DOI}
  [\href{https://arxiv.org/abs/1611.01221}{{\ttfamily 1611.01221}}].

\bibitem{Feynman:1965}
R.P.~Feynman and A.R.~Hibbs, \emph{{Quantum mechanics and path integrals}},
  International series in pure and applied physics, McGraw-Hill, New York, NY
  (1965).

\bibitem{Feldbrugge:2022}
J.~{Feldbrugge} and {van de Weygaert}, \emph{{Caustic skeleton of the cosmic
  web: non-linear constraint realizations of structural components; the 3D
  case}}, .

\bibitem{Hidding:2020}
J.~Hidding, \emph{jhidding/nbody2d: 2d pm n-body code},  Oct., 2020.
\newblock 10.5281/zenodo.4158731.

\bibitem{Doroshkevich:1970}
A.G.~{Doroshkevich}, \emph{{Spatial structure of perturbations and origin of
  galactic rotation in fluctuation theory}},
  \href{https://doi.org/10.1007/BF01001625}{\emph{Astrophysics} {\bfseries 6}
  (1970) 320}.

\bibitem{Shandarin:2009}
S.F.~{Shandarin} and R.A.~{Sunyaev}, \emph{{The conjecture of the cosmic web.
  Commentary on: Zel'dovich Ya. B., 1970, A\&A, 5, 84}},
  \href{https://doi.org/10.1051/0004-6361/200912144}{\emph{\aap} {\bfseries
  500} (2009) 19}.

\bibitem{Sheth:2004}
R.K.~{Sheth} and R.~{van de Weygaert}, \emph{{A hierarchy of voids: much ado
  about nothing}},
  \href{https://doi.org/10.1111/j.1365-2966.2004.07661.x}{\emph{\mnras}
  {\bfseries 350} (2004) 517}
  [\href{https://arxiv.org/abs/astro-ph/0311260}{{\ttfamily
  astro-ph/0311260}}].

\bibitem{Buchert:1992}
T.~{Buchert}, \emph{{Lagrangian theory of gravitational instability of
  Friedman-Lemaitre cosmologies and the 'Zel'dovich approximation'}},
  \href{https://doi.org/10.1093/mnras/254.4.729}{\emph{\mnras} {\bfseries 254}
  (1992) 729}.

\bibitem{Buchert:1993a}
T.~{Buchert}, \emph{{Lagrangian perturbation theory - A key-model for
  large-scale structure}}, {\emph{\aap} {\bfseries 267} (1993) L51}.

\bibitem{Buchert:1993b}
T.~{Buchert} and J.~{Ehlers}, \emph{{Lagrangian theory of gravitational
  instability of Friedman-Lemaitre cosmologies -- second-order approach: an
  improved model for non-linear clustering}},
  \href{https://doi.org/10.1093/mnras/264.2.375}{\emph{\mnras} {\bfseries 264}
  (1993) 375}.

\bibitem{Buchert:1994a}
T.~{Buchert}, \emph{{Lagrangian Theory of Gravitational Instability of
  Friedman-Lemaitre Cosmologies - a Generic Third-Order Model for Nonlinear
  Clustering}}, \href{https://doi.org/10.1093/mnras/267.4.811}{\emph{\mnras}
  {\bfseries 267} (1994) 811}
  [\href{https://arxiv.org/abs/astro-ph/9309055}{{\ttfamily
  astro-ph/9309055}}].

\bibitem{Buchert:1994b}
T.~{Buchert}, A.L.~{Melott} and A.G.~{Weiss}, \emph{{Testing higher-order
  Lagrangian perturbation theory against numerical simulations I. Pancake
  models}}, {\emph{\aap} {\bfseries 288} (1994) 349}
  [\href{https://arxiv.org/abs/astro-ph/9309056}{{\ttfamily
  astro-ph/9309056}}].

\bibitem{Bouchet:1995}
F.R.~{Bouchet}, S.~{Colombi}, E.~{Hivon} and R.~{Juszkiewicz},
  \emph{{Perturbative Lagrangian approach to gravitational instability.}},
  {\emph{\aap} {\bfseries 296} (1995) 575}
  [\href{https://arxiv.org/abs/astro-ph/9406013}{{\ttfamily
  astro-ph/9406013}}].

\bibitem{Springel:2005}
V.~{Springel}, S.D.M.~{White}, A.~{Jenkins}, C.S.~{Frenk}, N.~{Yoshida},
  L.~{Gao} et~al., \emph{{Simulations of the formation, evolution and
  clustering of galaxies and quasars}},
  \href{https://doi.org/10.1038/nature03597}{\emph{Nature} {\bfseries 435}
  (2005) 629} [\href{https://arxiv.org/abs/arXiv:astro-ph/0504097}{{\ttfamily
  arXiv:astro-ph/0504097}}].

\bibitem{illustris:2014}
M.~{Vogelsberger}, S.~{Genel}, V.~{Springel}, P.~{Torrey}, D.~{Sijacki},
  D.~{Xu} et~al., \emph{{Introducing the Illustris Project: simulating the
  coevolution of dark and visible matter in the Universe}},
  \href{https://doi.org/10.1093/mnras/stu1536}{\emph{\mnras} {\bfseries 444}
  (2014) 1518} [\href{https://arxiv.org/abs/1405.2921}{{\ttfamily 1405.2921}}].

\bibitem{eagle:2015}
J.~{Schaye}, R.A.~{Crain}, R.G.~{Bower}, M.~{Furlong}, M.~{Schaller},
  T.~{Theuns} et~al., \emph{{The EAGLE project: simulating the evolution and
  assembly of galaxies and their environments}},
  \href{https://doi.org/10.1093/mnras/stu2058}{\emph{\mnras} {\bfseries 446}
  (2015) 521} [\href{https://arxiv.org/abs/1407.7040}{{\ttfamily 1407.7040}}].

\bibitem{Hahn:2007}
O.~{Hahn}, C.~{Porciani}, C.M.~{Carollo} and A.~{Dekel}, \emph{{Properties of
  dark matter haloes in clusters, filaments, sheets and voids}},
  \href{https://doi.org/10.1111/j.1365-2966.2006.11318.x}{\emph{\mnras}
  {\bfseries 375} (2007) 489}
  [\href{https://arxiv.org/abs/astro-ph/0610280}{{\ttfamily
  astro-ph/0610280}}].

\bibitem{Feldbrugge:2014b}
J.~{Feldbrugge}, J.~{Hidding} and R.~{van de Weygaert}, \emph{{Statistics of
  Caustics in Large-Scale Structure Formation}}, {\emph{Proceedings of IAU
  Symposium 308 ``The Zeld'ovich Universe: Genesis and Growth of the Cosmic
  Web''} (2014) } [\href{https://arxiv.org/abs/1412.5121}{{\ttfamily
  1412.5121}}].

\bibitem{Ramachandra:2015}
N.S.~Ramachandra and S.F.~Shandarin, \emph{Multi-stream portrait of the cosmic
  web}, \href{https://doi.org/10.1093/mnras/stv1389}{\emph{Monthly Notices of
  the Royal Astronomical Society} {\bfseries 452} (2015) 1643}
  [\href{https://arxiv.org/abs/http://mnras.oxfordjournals.org/content/452/2/1643.full.pdf+html}{{\ttfamily
  http://mnras.oxfordjournals.org/content/452/2/1643.full.pdf+html}}].

\bibitem{Ramachandra:2017}
N.S.~{Ramachandra} and S.F.~{Shandarin}, \emph{{Topology and geometry of the
  dark matter web: A multi-stream view}},
  \href{https://doi.org/10.1093/mnras/stx183}{\emph{\mnras} {\bfseries 467}
  (2017) 1748} [\href{https://arxiv.org/abs/1608.05469}{{\ttfamily
  1608.05469}}].

\bibitem{Shandarin:2019}
S.F.~{Shandarin} and N.S.~{Ramachandra}, \emph{{The Caustic Design of the Dark
  Matter Web}}, {\emph{arXiv e-prints} (2019) arXiv:1906.05920}
  [\href{https://arxiv.org/abs/1906.05920}{{\ttfamily 1906.05920}}].

\bibitem{Shandarin:2021}
S.F.~{Shandarin}, \emph{{Identifying dark matter haloes by the caustic
  boundary}}, \href{https://doi.org/10.1088/1475-7516/2021/01/044}{\emph{\jcap}
  {\bfseries 2021} (2021) 044}
  [\href{https://arxiv.org/abs/2005.14548}{{\ttfamily 2005.14548}}].

\bibitem{Shandarin:1983}
S.F.~{Shandarin}, A.G.~{Doroshkevich} and I.B.~{Zeldovich}, \emph{{The
  large-scale structure of the universe}}, {\emph{Uspekhi Fizicheskikh Nauk}
  {\bfseries 139} (1983) 83}.

\bibitem{Rozhanskii:1984}
L.~{Rozhanskii} and S.~{Shandarin}, \emph{{Large-scale structure of the
  Universe. Three-dimensional model}}, {\emph{Keldysh Inst. Appl. Math., Akad.
  Nauk SSSR, Moscow, 28 pp.} (1984) }.

\bibitem{Berry:1977}
M.V.~{Berry} and J.F.~{Nye}, \emph{{Fine structure in caustic junctions}},
  \href{https://doi.org/10.1038/267034a0}{\emph{\nat} {\bfseries 267} (1977)
  34}.

\bibitem{Berry:1980}
M.V.~{Berry} and C.~{Upstill}, \emph{{IV Catastrophe Optics: Morphologies of
  Caustics and Their Diffraction Patterns}},
  \href{https://doi.org/10.1016/S0079-6638(08)70215-4}{\emph{Progess in Optics}
  {\bfseries 18} (1980) 257}.

\bibitem{Feldbrugge:2019}
J.~{Feldbrugge}, U.-L.~{Pen} and N.~{Turok}, \emph{{Oscillatory path integrals
  for radio astronomy}}, {\emph{arXiv e-prints} (2019) arXiv:1909.04632}
  [\href{https://arxiv.org/abs/1909.04632}{{\ttfamily 1909.04632}}].

\bibitem{Poston:1978}
T.~Poston and I.~Stewart, \emph{Catastrophe Theory and Its Applications},
  Pitman (1978).

\bibitem{Gilmore:1981}
R.~Gilmore, \emph{Catastrophe theory for scientists and engineers}, Wiley
  (1981).

\bibitem{Kravtsov:1983}
I.A.~{Kravtsov} and I.I.~{Orlov}, \emph{{Caustics, catastrophes, and wave
  fields}}, {\emph{Uspekhi Fizicheskikh Nauk} {\bfseries 141} (1983) 591}.

\bibitem{Arnold:2012a}
V.~Arnol'd, A.~Varchenko and S.~Gusein-Zade, \emph{Singularities of
  Differentiable Maps: Volume I: The Classification of Critical Points Caustics
  and Wave Fronts}, Monographs in Mathematics, Birkhauser Boston (2012).

\bibitem{Arnold:2012b}
V.~Arnol'd, S.~Gusein-Zade and A.~Varchenko, \emph{Singularities of
  Differentiable Maps, Volume 2: Monodromy and Asymptotics of Integrals},
  Modern Birkh{\"a}user Classics, Birkhauser Boston (2012).

\bibitem{Milnor:1963}
J.~Milnor, \emph{Morse theory}, Based on lecture notes by M. Spivak and R.
  Wells. Annals of Mathematics Studies, No. 51, Princeton University Press,
  Princeton, N.J. (1963).

\bibitem{Sousbie:2011a}
T.~{Sousbie}, \emph{{The persistent cosmic web and its filamentary structure -
  I. Theory and implementation}},
  \href{https://doi.org/10.1111/j.1365-2966.2011.18394.x}{\emph{\mnras}
  {\bfseries 414} (2011) 350}
  [\href{https://arxiv.org/abs/1009.4015}{{\ttfamily 1009.4015}}].

\bibitem{Sousbie:2011b}
T.~{Sousbie}, C.~{Pichon} and H.~{Kawahara}, \emph{{The persistent cosmic web
  and its filamentary structure - II. Illustrations}},
  \href{https://doi.org/10.1111/j.1365-2966.2011.18395.x}{\emph{\mnras}
  {\bfseries 414} (2011) 384}
  [\href{https://arxiv.org/abs/1009.4014}{{\ttfamily 1009.4014}}].

\bibitem{Shivashankar:2016}
N.~{Shivashankar}, P.~{Pranav}, V.~{Natarajan}, R.~{van de Weygaert},
  E.G.P.~{Bos} and S.~{Rieder}, \emph{{Felix: A Topology Based Framework for
  Visual Exploration of Cosmic Filaments}},
  \href{https://doi.org/10.1109/TVCG.2015.2452919}{\emph{IEEE Transactions on
  Visualizations and Computer Graphics. 2016. Vol. 22(6} {\bfseries 22} (2016)
  1745}.

\bibitem{WMAP:2003}
E.~{Komatsu}, A.~{Kogut}, M.R.~{Nolta}, C.L.~{Bennett}, M.~{Halpern},
  G.~{Hinshaw} et~al., \emph{{First-Year Wilkinson Microwave Anisotropy Probe
  (WMAP) Observations: Tests of Gaussianity}},
  \href{https://doi.org/10.1086/377220}{\emph{The Astrophysical Journal
  Supplement} {\bfseries 148} (2003) 119}
  [\href{https://arxiv.org/abs/astro-ph/0302223}{{\ttfamily
  astro-ph/0302223}}].

\bibitem{Planck:2016}
{Planck Collaboration}, P.A.R.~{Ade}, N.~{Aghanim}, M.~{Arnaud}, F.~{Arroja},
  M.~{Ashdown} et~al., \emph{{Planck 2015 results. XVII. Constraints on
  primordial non-Gaussianity}},
  \href{https://doi.org/10.1051/0004-6361/201525836}{\emph{\aap} {\bfseries
  594} (2016) A17} [\href{https://arxiv.org/abs/1502.01592}{{\ttfamily
  1502.01592}}].

\bibitem{Creminelli:2006}
P.~{Creminelli}, A.~{Nicolis}, L.~{Senatore}, M.~{Tegmark} and
  M.~{Zaldarriaga}, \emph{{Limits on non-Gaussianities from WMAP data}},
  \href{https://doi.org/10.1088/1475-7516/2006/05/004}{\emph{Journal of
  Cosmology and Astroparticle Physics} {\bfseries 5} (2006) 004}
  [\href{https://arxiv.org/abs/astro-ph/0509029}{{\ttfamily
  astro-ph/0509029}}].

\bibitem{Planck:2020}
{Planck Collaboration}, Y.~{Akrami}, F.~{Arroja}, M.~{Ashdown}, J.~{Aumont},
  C.~{Baccigalupi} et~al., \emph{{Planck 2018 results. IX. Constraints on
  primordial non-Gaussianity}},
  \href{https://doi.org/10.1051/0004-6361/201935891}{\emph{\aap} {\bfseries
  641} (2020) A9} [\href{https://arxiv.org/abs/1905.05697}{{\ttfamily
  1905.05697}}].

\bibitem{Longuet-Higgins:1957}
M.S.~{Longuet-Higgins}, \emph{{Statistical Properties of an Isotropic Random
  Surface}}, \href{https://doi.org/10.1098/rsta.1957.0018}{\emph{Philosophical
  Transactions of the Royal Society of London Series A} {\bfseries 250} (1957)
  157}.

\bibitem{Adler:1981}
R.J.~{Adler}, \emph{{The Geometry of Random Fields}}, Society for Industrial
  and Applied Mathematics (1981).

\bibitem{bbks:1986}
J.M.~{Bardeen}, J.R.~{Bond}, N.~{Kaiser} and A.S.~{Szalay}, \emph{{The
  statistics of peaks of Gaussian random fields}},
  \href{https://doi.org/10.1086/164143}{\emph{The Astrophysical Journal}
  {\bfseries 304} (1986) 15}.

\bibitem{Feldbrugge:2014}
J.~Feldbrugge, \emph{{Statistics of caustics in large-scale structure
  formation}},  Master's thesis, Rijksuniversiteit Groningen, the Netherlands,
  2014.

\bibitem{Sheth:1995}
R.K.~{Sheth}, \emph{{Constrained realizations and minimum variance
  reconstruction of non-Gaussian random fields}},
  \href{https://doi.org/10.1093/mnras/277.3.933}{\emph{\mnras} {\bfseries 277}
  (1995) 933} [\href{https://arxiv.org/abs/astro-ph/9511096}{{\ttfamily
  astro-ph/9511096}}].

\bibitem{Delmarcelle:1995}
T.~{Delmarcelle}, \emph{{The Visualization of Second-Order Tensor Fields}},
  Ph.D. thesis, Stanford University, Jan., 1995.

\bibitem{Lavin:1997}
Y.~Lavin, Y.~Levy and L.~Hesselink, \emph{Singularities in nonuniform tensor
  fields},  in \emph{8th {IEEE} Visualization Conference, {IEEE} Vis 1997,
  Phoenix, AZ, USA, October 19-24, 1997, Proceedings}, pp.~59--66, {IEEE}
  Computer Society and {ACM}, 1997,
  \href{https://doi.org/10.1109/VISUAL.1997.663857}{DOI}.

\bibitem{Galarraga-Espinosa:2020}
D.~{Gal{\'a}rraga-Espinosa}, N.~{Aghanim}, M.~{Langer}, C.~{Gouin} and
  N.~{Malavasi}, \emph{{Populations of filaments from the distribution of
  galaxies in numerical simulations}},
  \href{https://doi.org/10.1051/0004-6361/202037986}{\emph{\aap} {\bfseries
  641} (2020) A173} [\href{https://arxiv.org/abs/2003.09697}{{\ttfamily
  2003.09697}}].

\bibitem{Aragon:2007}
M.A.~{Arag{\'o}n-Calvo}, R.~{van de Weygaert}, B.J.T.~{Jones} and J.M.~{van der
  Hulst}, \emph{{Spin Alignment of Dark Matter Halos in Filaments and Walls}},
  \href{https://doi.org/10.1086/511633}{\emph{\apjl} {\bfseries 655} (2007) L5}
  [\href{https://arxiv.org/abs/astro-ph/0610249}{{\ttfamily
  astro-ph/0610249}}].

\bibitem{Ganeshaiah:2018}
P.~{Ganeshaiah Veena}, M.~{Cautun}, R.~{van de Weygaert}, E.~{Tempel},
  B.J.T.~{Jones}, S.~{Rieder} et~al., \emph{{The Cosmic Ballet: spin and shape
  alignments of haloes in the cosmic web}},
  \href{https://doi.org/10.1093/mnras/sty2270}{\emph{\mnras} {\bfseries 481}
  (2018) 414} [\href{https://arxiv.org/abs/1805.00033}{{\ttfamily
  1805.00033}}].

\bibitem{Hellwing:2021}
W.A.~{Hellwing}, M.~{Cautun}, R.~{van de Weygaert} and B.T.~{Jones},
  \emph{{Caught in the cosmic web: Environmental effect on halo concentrations,
  shape, and spin}},
  \href{https://doi.org/10.1103/PhysRevD.103.063517}{\emph{\prd} {\bfseries
  103} (2021) 063517} [\href{https://arxiv.org/abs/2011.08840}{{\ttfamily
  2011.08840}}].

\bibitem{Lopez:2021}
P.~{L{\'o}pez}, M.~{Cautun}, D.~{Paz}, M.~{Merch{\'a}n} and R.~{van de
  Weygaert}, \emph{{Deviations from tidal torque theory: Evolution of the halo
  spin-filament alignment}},
  \href{https://doi.org/10.1093/mnras/stab451}{\emph{\mnras} {\bfseries 502}
  (2021) 5528} [\href{https://arxiv.org/abs/2012.01638}{{\ttfamily
  2012.01638}}].

\bibitem{Ganeshaiah:2021}
P.~{Ganeshaiah Veena}, M.~{Cautun}, R.~{van de Weygaert}, E.~{Tempel} and
  C.S.~{Frenk}, \emph{{Cosmic Ballet III: Halo spin evolution in the cosmic
  web}}, \href{https://doi.org/10.1093/mnras/stab411}{\emph{\mnras} {\bfseries
  503} (2021) 2280} [\href{https://arxiv.org/abs/2007.10365}{{\ttfamily
  2007.10365}}].

\end{thebibliography}\endgroup

\appendix
%%%%%%%%%%%%%%%%%%%%%%%%%%%%%%%%%%%%%%%%%%%%%%%%%%%%%%%%%%%%%%%%
\section{Gaussian random fields algorithms}\label{ap:GRF}
Gaussian random fields are most naturally expressed in Fourier space, as the Fourier modes of an unconstrained Gaussian random field are normally and independently distributed, 
\begin{align}
p(\hat{f}(\bm{k}_1), \hat{f}(\bm{k}_2), \dots) = \prod_{i} \frac{1}{\sqrt{2\pi P(\bm{k}_i)}} \exp\left[-\frac{|\hat{f}(\bm{k}_i)|^2}{2P(\bm{k}_i)}\right],
\end{align}
up to the reality condition, $\hat{f}(\bm{k})=\hat{f}^*(-\bm{k})$. In this paper, we use Fast Fourier Transform routines to efficiently generate realizations of both unconstrained and linearly constrained Gaussian random.

%%%%%%%%%%%%%%%%%%%%%%%%%%%%%%%%%%%%%%%%%%%%%%%%%%%%%%%%%%%%%%%%
\subsection{Unconstrained Gaussian random fields}
We can efficiently generate realizations of unconstrained Gaussian random fields on a rectangular lattice using Fast Fourier Transform routines. The Fourier modes of an isotropic and homogeneous Gaussian random field are independent distributed Gaussian variables with zero mean, $\mu=0$, and the standard deviation $\sigma = \sqrt{P(\|\bm{k}\|)}$. Using this observation we generate the realization by multiplying the Fourier transform of white noise with the square root of the power spectrum:
\bigskip

\begin{algorithm}[H]
\SetAlgoLined
\begin{enumerate}[itemsep=1ex, leftmargin=0cm, rightmargin=1cm]
\item Generate a realization of white noise $n_w(\bm{q})$, consisting of identically independently distributed normal numbers $\mathcal{N}(\mu=0,\sigma^2=1)$ on a regular rectangular lattice.
\item Fast Fourier transform the white noise $\hat{n}_w(\bm{k}) = \text{FFT}[n_w(\bm{q})](\bm{k})$. The Fourier modes are independently normally distributed satisfying the reality condition $\hat{n}_w(\bm{k}) = \hat{n}_w^*(-\bm{k})$.
\item Rescale the Fourier modes with the power spectrum $\hat{f}(\bm{k}) = \sqrt{P(\bm{k})}\hat{n}_w(\bm{k})$.
\item Evaluate the inverse fast Fourier transform the rescaled modes to obtain the realization of the unconstrained Gaussian random field $f(\bm{q}) = \text{FFT}^{-1}[\hat{f}(\bm{k})](\bm{q})$.
\end{enumerate}
 \caption{Generating a realization of an unconstrained Gaussian random field on a rectangular lattice}
 \label{alg:GRF}
\end{algorithm}
\bigskip

%%%%%%%%%%%%%%%%%%%%%%%%%%%%%%%%%%%%%%%%%%%%%%%%%%%%%%%%%%%%%%%%
\subsection{Linearly constrained Gaussian random fields}
We efficiently generate realizations of constrained Gaussian random fields, satisfying a set of linear constraints $\Gamma=\{\bm{C}[f]=\bm{c}\}$, using the Hoffman Ribak algorithm \cite{Hoffman:1991, Weygaert:1996}. The algorithm is based on the observation that the residue of the random field with respect to the mean field, $\delta f = f - \langle f | \Gamma\rangle$, is a Gaussian random field whose properties are independent of the constraint values $\bm{c}$. Note that this is a special property of Gaussianity of the random field and the linearity of the constraints $\bm{C}$.

Given an unconstrained Gaussian random field $f$ with the required power spectrum, we compute the value of the linear constraint $C_i$ using the convolution theorem
\begin{align}
C_i[f,\bm{q}_i]=\text{FFT}^{-1}[\hat{C}_i^*(\bm{k}) \hat{f}(\bm{k})](\bm{0})
\end{align}
evaluated at the origin, with the Fourier transform of the constraint $\hat{C}_i$. Assuming that the unconstrained mean field vanishes, $\langle f \rangle =0$, we evaluate the constrained mean field using the formula
\begin{align}
\langle f(\bm{q}) | \Gamma\rangle = \sum_{i,j} \xi_i(\bm{q})\xi_{ij}^{-1} c_j\,,
\end{align}
with the covariance between the random field and the constraint
\begin{align}
\xi_i(\bm{q}) = \text{FFT}^{-1}[\hat{C}_i(\bm{k}) P(\bm{k})](\bm{q})\,,
\end{align}
and the covariance matrix of the constraints
\begin{align}
\xi_{ij}=\text{FFT}^{-1}[\hat{C}_i^*(\bm{k}) \hat{C}_i(\bm{k}) P(\bm{k})](\bm{0})\,,
\end{align}
evaluated at the origin. In these equations, we assume that $\bar{f}=\bar{C}_i=0$. Note that the vector $\bm{\xi}(\bm{q})=[\sum_i \xi_i(\bm{q}) \xi_{ij}^{-1}]$ is a normal basis of the constraints. The mean field is the inner product of $\bm{x}$ with the constraint values, 
\begin{align}
\langle f(\bm{q}) | \Gamma\rangle = \bm{\xi}(\bm{q})\cdot \bm{c}\,.
\end{align}
We use these routines to implement the Hoffman-Ribak method:
\bigskip

\begin{algorithm}[H]
\SetAlgoLined
\begin{enumerate}[itemsep=1ex, leftmargin=0cm, rightmargin=1cm]
\item Generate a realization of an unconstrained Gaussian random field $g$ with the required power spectrum using algorithm \ref{alg:GRF}.
\item Evaluate the linear constraints $C_i$ of the unconstrained field $g$,
\begin{align}
d_i = C_i[g;\bm{q}_i]\,.
\end{align} 
\item Evaluate the mean field corresponding to the constraint values $d_i$,
\begin{align}
\bar{g}(\bm{q}) = \sum_{ij}\xi_i(\bm{q}) \xi^{-1}_{ij} d_j\,.
\end{align}
\item Evaluate the residue of the unconstrained field 
\begin{align}
\delta g=g - \bar{g}\,.
\end{align}
\item Since the residual field $\delta g$ is statistically independent of the values $d_j$, we can identify $\delta g$ with the residue $\delta f$ of the constrained random field. The realization of the constrained Gaussian random field takes the form
\begin{align}
f(\bm{q}) = \sum_{ij}\xi_i(\bm{q}) \xi^{-1}_{ij} c_j + \delta g(\bm{q})\,.
\end{align}
\end{enumerate}
 \caption{The Hoffman-Ribak method for Gaussian random fields with linear constraints}
 \label{alg:HoffmanRibak}
\end{algorithm}

%%%%%%%%%%%%%%%%%%%%%%%%%%%%%%%%%%%%%%%%%%%%%%%%%%%%%%%%%%%%%%%%
\section{Constraint probability density}\label{ap:constraintDensity}
The conditional density 
\begin{align}
p(f|\Gamma) =\frac{p(f, \Gamma)}{p(\Gamma)}
\end{align}
with $\Gamma = \{\bm{C} = \bm{c}\}$ is a Gaussian density as both $p(f,\Gamma)$ and $p(\Gamma)$ are normally distributed. It thus suffices to evaluate the mean and covariance matrix. For completeness, we write the mean of the function and the constraints as $\bar{f}(\bm{q})=\langle f(\bm{q}) \rangle$ and $\bar{\bm{C}}=\langle \bm{C}\rangle$, and the covariance as
\begin{align}
\xi(\bm{q}_1,\bm{q}_2)& = \text{cov}(f(\bm{q}_1),f(\bm{q}_2))= \langle (f(\bm{q}_1) - \bar{f}(\bm{q}_1))(f(\bm{q}_2) - \bar{f}(\bm{q}_2))\rangle\,,\\
\xi_i(\bm{q}) &= \text{cov}(f(\bm{q}),C_i) = \langle (f(\bm{q})-\bar{f}(\bm{q}))(C_i - \bar{C}_i)\rangle\,,\\
\xi_{ij} &= \text{cov}(C_i,C_j) = \langle (C_i - \bar{C}_i)(C_j-\bar{C}_j)\rangle\,.
\end{align}
For conciseness, we use the Einstein summation convention over the repeated dummy indices.

Define an auxiliary function $g(\bm{q}) = f(\bm{q}) + \bm{A}(\bm{q}) \cdot \bm{C}$ with $A_i(\bm{q}) = -\xi_j(\bm{q}) \xi_{ji}^{-1}$. The function $g$ is by construction independent of the variable $\bm{C}$, as they are jointly normally distributed and their covariance matrix vanishes, \textit{i.e.}, 
\begin{align}
\text{cov}(g(\bm{q}),\bm{C})
&=\text{cov}(f(\bm{q}),\bm{C}) +  \text{cov}(\bm{C},\bm{C}) \bm{A}(\bm{q})\nonumber \\
&=[\xi_k(\bm{q}) - \xi_i(\bm{q}) \xi_{ij}^{-1}\xi_{jk}]_{k=1}^N\nonumber\\
&=\bm{0}\,.
\end{align}
The expectation value of $g$ is $\langle g(\bm{q})\rangle = \bar{f}(\bm{q}) + \bm{A}(\bm{q}) \cdot \bar{\bm{C}}$, which we can use to evaluate the expectation value of the constrained field
\begin{align}
\langle f(\bm{q}) | \Gamma \rangle 
&=\langle g(\bm{q}) - \bm{A}(\bm{q})\cdot \bm{C}|\Gamma\rangle\nonumber\\
&=\langle g(\bm{q})\rangle - \bm{A}(\bm{q}) \cdot \bm{c}\nonumber\\
&=\bar{f}(\bm{q}) + \bm{A}(\bm{q})\cdot (\bar{\bm{C}}-\bm{c})\nonumber\\
&= \bar{f}(\bm{q}) + \xi_i(\bm{q})\xi_{ij}^{-1}(c_j -\bar{C}_j)\,.
\end{align}
When $\bar{f} = 0$ and $\bar{\bm{C}}=\bm{0}$ the expectation value reduces to the identity $\langle f(\bm{q}) | \Gamma \rangle =\xi_i(\bm{q})\xi_{ij}^{-1}c_j$. The covariance of the constraint field follows along similar lines
%\begin{align}
%\text{var}(f(\bm{q})|\Gamma) 
%&= \text{var}(g(\bm{q}) - \bm{A}(\bm{q})\cdot \bm{C}|\Gamma) \\
%&= \text{var}(g(\bm{q})|\Gamma) + \text{var}(\bm{A}(\bm{q})\cdot \bm{C}|\Gamma) -2 \bm{A}(\bm{q})\cdot \text{cov}(g(\bm{q}), \bm{C}|\Gamma)\\
%&= \text{var}(g(\bm{q})) \\
%&=\text{var}(f(\bm{q}) + \bm{A}(\bm{q})\cdot \bm{C})\\
%&=\text{var}(f(\bm{q})) + \bm{A}^T(\bm{q}) \text{var}(\bm{C}) \bm{A}(\bm{q}) + 2\bm{A}(\bm{q}) \cdot \text{cov}(f(\bm{q}),\bm{C}) \\
%&=\text{var}(f(\bm{q})) + \xi_i(\bm{q})\xi_{ij}^{-1}\xi_{jk}\xi_{kl}^{-1}\xi_{l}(\bm{q}) - 2\xi_i(\bm{q})\xi_{ij}^{-1}\xi_{j}(\bm{q})\\
%&=\xi(\bm{0}) - \xi_i(\bm{q}) \xi_{ij}^{-1} \xi_j(\bm{q})\,.
%\end{align}
\begin{align}
\text{cov}(f(\bm{q}_1), f(\bm{q}_2)|\Gamma) 
&= \text{cov}(g(\bm{q}_1) - \bm{A}(\bm{q}_1)\cdot \bm{C}, g(\bm{q}_2) - \bm{A}(\bm{q}_2)\cdot \bm{C}|\Gamma) \nonumber\\
&= \text{cov}(g(\bm{q}_1), g(\bm{q}_2)|\Gamma) + \text{cov}(\bm{A}(\bm{q}_1)\cdot \bm{C}, \bm{A}(\bm{q}_2)\cdot \bm{C}|\Gamma)\nonumber\\
&\ \ \  - \bm{A}(\bm{q}_1)\cdot \text{cov}(\bm{C}, g(\bm{q}_2)|\Gamma) -\text{cov}(g(\bm{q}_1) , \bm{C}|\Gamma)\cdot \bm{A}(\bm{q}_2)\nonumber\\
&= \text{cov}(g(\bm{q}_1), g(\bm{q}_2)) \nonumber\\
&=\text{cov}(f(\bm{q}_1) + \bm{A}(\bm{q}_1)\cdot \bm{C}, f(\bm{q}_2) + \bm{A}(\bm{q}_2)\cdot \bm{C})\nonumber\\
&=\text{cov}(f(\bm{q}_1), f(\bm{q}_2)) + \bm{A}(\bm{q}_1) \text{cov}(\bm{C},\bm{C}) \bm{A}^T(\bm{q}_2) \nonumber\\
&\ \ \ + \bm{A}(\bm{q}_1) \cdot \text{cov}(\bm{C},f(\bm{q}_2))+ \text{cov}(f(\bm{q}_1), \bm{C}) \cdot \bm{A}(\bm{q}_2) \nonumber\\
&=\text{cov}(f(\bm{q}_1), f(\bm{q}_2)) + \xi_i(\bm{q}_1)\xi_{ij}^{-1}\xi_{jk}\xi_{kl}^{-1}\xi_{l}(\bm{q}_2) - 2\xi_i(\bm{q}_1)\xi_{ij}^{-1}\xi_{j}(\bm{q}_2)\nonumber\\
&=\xi(\bm{q}_1,\bm{q}_2) - \xi_i(\bm{q}_1) \xi_{ij}^{-1} \xi_j(\bm{q}_2)\,.
\end{align}
We conclude that the residual field, \textit{i.e.}, the residue $\delta f$ of $f$ with respect to the mean field $\langle f(\bm{q})|\Gamma\rangle =\bar{f}(\bm{q})+ \xi_i(\bm{q})\xi_{ij}^{-1}(c_j-\bar{C}_j)$ is a Gaussian random field with zero mean $\langle \delta f|\Gamma\rangle=0$ and variance $\langle \delta f(\bm{q})^2|\Gamma\rangle = \sigma_0^2 - \xi_i(\bm{q}) \xi_{ij}^{-1} \xi_j(\bm{q})$. Note that
for a Gaussian random field with linear constraints the residual field $\delta f$ is indeed independent of $\bm{c}$. The covariance of the constraint field yields the constrained probability density
\begin{align}
p(f|\Gamma) \propto  \exp\left[-\frac{1}{2} \iint \delta{f}(\bm{q}_1) \tilde{K}(\bm{q}_1,\bm{q}_2) \delta f(\bm{q}_2)\mathrm{d}\bm{q}_1 \mathrm{d}\bm{q}_2 \right]
\end{align}
with the residual field $\delta f = f-\bar{f}(\bm{q})+ \xi_i(\bm{q})\xi_{ij}^{-1}(c_j-\bar{C}_j)$ and the kernel $\tilde{K}$ defined as the inverse of the modified two-point correlation function, \textit{i.e.},
\begin{align}
\int \tilde{K}(\bm{q}_1,\bm{q}) \left[\xi(\bm{q},\bm{q}_2) - \xi_i(\bm{q})\xi_{ij}^{-1}\xi_j(\bm{q}_2)\right]\mathrm{d}\bm{q}= \delta_D^{(2)}(\bm{q}_1-\bm{q}_2)\,.
\end{align}
The correction $- \xi_i(\bm{q})\xi_{ij}^{-1}\xi_j(\bm{q}_2)$ implements the statistical anisotropy and inhomogeneity of the residual field.

%%%%%%%%%%%%%%%%%%%%%%%%%%%%%%%%%%%%%%%%%%%%%%%%%%%%%%%%%%%%%%%%
\section{Non-linear eigenvalue relations: a direct approach}\label{ap:eigenvalueRel}
The eigenvalue and eigenvector fields are non-linear functions of the deformation tensor
\begin{align}
\bm{\psi} = \begin{pmatrix} T_{11} & T_{12} \\ T_{12} & T_{22}\end{pmatrix},
\end{align}
with the partial derivatives of the displacement potential $T_{ijk\dots} = \frac{\partial}{\partial q_i}\frac{\partial}{\partial q_j}\frac{\partial}{\partial q_k}\dots \Psi$. In two dimensions, the eigenvalue and eigenvector fields are the quadratic roots of the eigenequation $\bm{\psi}\bm{v}_i=\lambda_i \bm{v}_i$, with the explicit identities for the eigenvalue fields
\begin{align}
\lambda_1(T_{11},T_{22},T_{12}) &= \frac{1}{2}\left(T_{11}+T_{22} + \sqrt{4 T_{12}^2 +(T_{11}-T_{22})^2}\right)\,,\label{eq:lambda_1}\\
\lambda_2(T_{11},T_{22},T_{12}) &= \frac{1}{2}\left(T_{11}+T_{22} - \sqrt{4 T_{12}^2 +(T_{11}-T_{22})^2}\right)\,.\label{eq:lambda_2}
\end{align}
The study of Gaussian random fields satisfying non-linear caustic constraints is most concisely performed in the eigenframe, for which $T_{12}=0$ and $T_{11} \geq T_{22}$ leading to the eigenvalues $\lambda_1=T_{11}, \lambda_2=T_{22}$ and and eigenvectors
\begin{align}
\lambda_{1} &= T_{11}\,,\quad \bm{v}_1=(1,0)\,,\\
\lambda_{2} &= T_{22}\,,\quad \bm{v}_2=(0,1)\,.
\end{align}
In this section, we give an explicit derivation of several identities relating to the partial derivatives of the eigenvalue and eigenvector fields and the partial derivatives of the displacement potential in the eigenframe, using an Euler angle. In appendix \ref{ap:eigenvalueRel_alt}, we present a more efficient algebraic derivation, which more easily generalizes to the three-dimensional setting.

Using an eigenvalue decomposition we can write the deformation tensor as $\bm{\psi} = R^T \Lambda R,$ with the diagonal matrix
\begin{align}
\Lambda = \begin{pmatrix}\lambda_1 & 0 \\ 0 &\lambda_2 \end{pmatrix}
\end{align}
with the ordering $\lambda_1 \geq \lambda_2$, and the rotation matrix
\begin{align}
R = \begin{pmatrix} \cos \theta & - \sin \theta \\ \sin \theta & \cos \theta \end{pmatrix} \in SO(2),
\end{align}
with the Euler angle $\theta$. The decomposition yields three identities for the components of the deformation tensor $T_{11},T_{22},T_{12}$ 
\begin{align}
T_{11} &= \lambda_1 \cos^2 \theta + \lambda_2 \sin^2 \theta\,,\\
T_{22} &= \lambda_1 \sin^2\theta + \lambda_2 \cos^2 \theta\,,\\
T_{12} &= (\lambda_2 - \lambda_1)\cos \theta \sin \theta\,.
\end{align}
Solving for the Euler angle $\theta$, we obtain a representation for the normalized eigenvector fields
\begin{align}
\bm{v}_1 &=\pm (\cos(\theta), -\sin(\theta))\,,\\
\bm{v}_2 &=\pm (\sin(\theta),\ \ \ \cos(\theta))\,,
\end{align}
where the sine and cosine are expressed in terms of the partial derivatives of the displacement potential
\begin{align}
\cos(\theta) &= \frac{\sqrt{T_{11} - T_{22} + \sqrt{4 T_{12}^2 +(T_{11}-T_{22})^2}}}{\sqrt{2} \sqrt[4]{4T_{12}^2 + (T_{11}-T_{22})^2}}\,,\\
\sin(\theta) &= \frac{\sqrt{2}T_{12}}{\sqrt[4]{4T_{12}^2+(T_{11}-T_{22})^2}\sqrt{T_{11}-T_{22} + \sqrt{4T_{12}^2 + (T_{11}-T_{22})^2}}}\,.
\end{align}
These relations enable us to evaluate several useful non-linear identities. We evaluate the derivatives of the eigenvalue field in a general frame and subsequently take the limit $T_{12} \to 0$ with the condition $T_{11}\geq T_{22}$. 

The first-order derivatives of the eigenvalue fields in the eigenframe satisfy the linear equations
\begin{align}
\nabla \lambda_1&= (T_{111}, T_{112})\,,\\
\nabla \lambda_2&= (T_{122}, T_{222})\,.
\end{align}
The second order derivatives receive a non-linear correction
\begin{align}
\frac{\partial^2}{\partial q_1^2} \lambda_1&= T_{1111} + \frac{2 T_{112}^2}{T_{11}-T_{22}}\,,\\
\frac{\partial^2}{\partial q_1 \partial q_2} \lambda_1&= T_{1112} + \frac{2 T_{112}T_{122}}{T_{11}-T_{22}}\,,\\
\frac{\partial^2}{\partial q_2^2} \lambda_1&= T_{1122} + \frac{2T_{122}^2}{T_{11}-T_{22}}\,,\\
\frac{\partial^2}{\partial q_1^2} \lambda_2&= T_{1122} - \frac{2 T_{112}^2}{T_{11}-T_{22}}\,,\\
\frac{\partial^2}{\partial q_1 \partial q_2} \lambda_2&= T_{1222} - \frac{2 T_{112}T_{122}}{T_{11}-T_{22}}\,,\\
\frac{\partial^2}{\partial q_2^2} \lambda_2&= T_{2222} - \frac{2T_{122}^2}{T_{11}-T_{22}}\,.
\end{align}

The first-order derivative in the direction of the eigenvalue field naturally coincides with the first-order derivatives described above
\begin{align}
\bm{v}_1 \cdot \nabla \lambda_1 &= \pm T_{111}\,,\quad
\bm{v}_2 \cdot \nabla \lambda_1 = \pm T_{112}\,,\\
\bm{v}_1 \cdot \nabla \lambda_2 &= \pm T_{122}\,,\quad
\bm{v}_2 \cdot \nabla \lambda_2 = \pm T_{222}\,.
\end{align}
Note that these combinations transform as scalars under rotations. The second-order derivatives receive an additional non-linear term due to the variation of the eigenvector fields. The second-order derivatives of the first eigenvalue field take the form
\begin{align}
\bm{v}_1 \cdot \nabla (\bm{v}_1 \cdot \nabla \lambda_1) &= T_{1111} + \frac{3T_{112}^2}{T_{11}-T_{22}}\,,\\
\bm{v}_2 \cdot \nabla (\bm{v}_1 \cdot \nabla \lambda_1) &= \pm \left[T_{1112} + \frac{3T_{112}T_{122}}{T_{11}-T_{22}}\right]\,,\\
\bm{v}_1 \cdot \nabla (\bm{v}_2 \cdot \nabla \lambda_1) &= \pm \left[T_{1112} - \frac{T_{112}(T_{111}-2T_{122})}{T_{11}-T_{22}}\right]\,,\\
\bm{v}_2 \cdot \nabla (\bm{v}_2 \cdot \nabla \lambda_1) &= T_{1122} - \frac{T_{122}(T_{111}-2T_{122})}{T_{11}-T_{22}}\,.
\end{align}
Note that the directional derivatives in the directions $\bm{v}_1$ and $\bm{v}_2$ do not commute. The second-order derivatives of the second eigenvalue field yield the analogous identities
\begin{align}
\bm{v}_1 \cdot \nabla (\bm{v}_1 \cdot \nabla \lambda_2) &= T_{1122} + \frac{T_{112}(T_{222}-2T_{112})}{T_{11}-T_{22}}\,,\\
\bm{v}_2 \cdot \nabla (\bm{v}_1 \cdot \nabla \lambda_2) &= \pm\left[T_{1222} + \frac{T_{122}(T_{222}-2T_{112})}{T_{11}-T_{22}}\right]\,,\\
\bm{v}_1 \cdot \nabla (\bm{v}_2 \cdot \nabla \lambda_2) &= \pm\left[T_{1222} - \frac{3T_{112}T_{122}}{T_{11}-T_{22}}\right]\,,\\
\bm{v}_2 \cdot \nabla (\bm{v}_2 \cdot \nabla \lambda_2) &= T_{2222} - \frac{3T_{122}^2}{T_{11}-T_{22}}\,.
\end{align}

%%%%%%%%%%%%%%%%%%%%%%%%%%%%%%%%%%%%%%%%%%%%%%%%%%%%%%%%%%%%%%%%
\section{Non-linear eigenvalue relations: an algebraic approach}\label{ap:eigenvalueRel_alt}
Unfortunately, the derivation of the non-linear eigenvalue relations presented in appendix \ref{ap:eigenvalueRel} does in practice not straightforwardly generalize to the three-dimensional case. It formally works but is not sufficiently efficient in practice. However, using an algebraic approach we can reproduce the results presented in section \ref{ap:eigenvalueRel} more efficiently. This method does generalize to the higher-dimensional setting. We present this derivation here for the two-dimensional case in anticipation of an upcoming paper on the three-dimensional caustic skeleton and constraint random fields \cite{Feldbrugge:2022}.

In two dimensions, the eigenvalue fields are related to the principle invariants of the deformation tensor
\begin{align}
\text{Tr}\ \bm{\psi} &= T_{11} + T_{22} = \lambda_1 + \lambda_2\,,\label{eq:principle_1}\\
\text{det}\ \bm{\psi} &=  T_{11}  T_{22} - T_{12}^2 = \lambda_1 \lambda_2\,.\label{eq:principle_2}
\end{align}
Assuming $T_{11} > T_{22}$ and the ordering $\lambda_1> \lambda_2$ these equations have a unique solution (see equations \eqref{eq:lambda_1} and \eqref{eq:lambda_2}). In the eigenframe, $T_{12}=0$, we will assume the orientations $\bm{v}_1=(1,0)$ and $\bm{v}_2=(0,1)$. 

Now, apply the differential operators $\partial_i$ and $\partial_{i}\partial_j$ for $i,j=1,2$ to equations \eqref{eq:principle_1} and \eqref{eq:principle_2} to obtain a set of five equations. Rotate to the eigenframe by substituting $\lambda_1 = T_{11}$, $\lambda_2=T_{22}$, and $T_{12}=0$ and solve for $\partial_i\lambda_j$ and $\partial_i \partial_j \lambda_k$. The equations have a unique solution that coincides with the partial derivatives of the eigenvalue fields presented in appendix \ref{ap:eigenvalueRel}.

To evaluate the directional derivatives of the eigenvalue fields in the direction of the eigenvectors, we use the chain rule 
\begin{align}
\bm{v}_i \cdot \nabla (\bm{v}_j \cdot \nabla \lambda_k)= (\bm{v}_i \cdot  \nabla )\bm{v}_j \cdot \nabla \lambda_k + \bm{v}_i (\mathcal{H} \lambda_k )\bm{v}_j \,,
\end{align}
with the Hessian $\mathcal{H}$. We already obtained identities for the derivatives of the eigenvalue fields in the eigenframe. What remains is the evaluation of the derivative of the eigenvector fields $(\bm{v}_i \cdot \nabla) \bm{v}_j$. First note that the normalization $\|\bm{v}_i\|=1$ implies the identity
\begin{align}
\bm{v}_i \cdot \partial_j \bm{v}_i=0\,.\label{eq:normal_v}
\end{align} 
The vector $\partial_i \bm{v}_j$ is normal to $\bm{v}_j$. In addition, application of the operator $\partial_j$ on the eigenequation $(\bm{\psi}-\lambda_i I)\bm{v}_i=\bm{0}$ yields the identity
\begin{align}
(\partial_j \bm{\psi} - \partial_j \lambda_i I)\bm{v}_i + (\bm{\psi}-\lambda_i I)\partial_j \bm{v}_i=\bm{0}\,. \label{eq:eigen_d}
\end{align}
Note that this equation does not yield a unique solution since the matrix $\bm{\psi}-\lambda_i I$ is singular. However, the combined system of equation \eqref{eq:normal_v} with \eqref{eq:eigen_d} is solvable. In the eigenframe, we obtain the gradient of the eigenvector fields
\begin{align}
\nabla \bm{v}_1 &= \begin{pmatrix}
0, & T_{112}/(T_{11}-T_{22})\\
0, & T_{122}/(T_{11}-T_{22})
\end{pmatrix},\\
\nabla \bm{v}_2 &= \begin{pmatrix}
-T_{112}/(T_{11}-T_{22}), & 0\\
-T_{122}/(T_{11}-T_{22}), & 0
\end{pmatrix}.
\end{align}
When combining the derivatives of the eigenvalue and eigenvector fields, we reproduce the remaining results presented in appendix \ref{ap:eigenvalueRel}.

%%%%%%%%%%%%%%%%%%%%%%%%%%%%%%%%%%%%%%%%%%%%%%%%%%%%%%%%%%%%%%%%
\section{Rotations of derivatives of the deformation tensor}\label{ap:rotations}
Under a rotation, the derivatives of the deformation tensor transform nontrivially. When applying the rotation matrix
\begin{align}
R_\theta = \begin{pmatrix} c  & s  \\- s  & c \end{pmatrix},
\end{align}
with $c=\cos\theta$ and $s=\sin \theta$, the first, second, third, and fourth-order derivatives transform into each other. The first-order derivatives transform as a vector
\begin{align}
\begin{pmatrix} T_1 \\ T_2 \end{pmatrix}
\mapsto
\begin{pmatrix}
c & -s \\
s &  c 
\end{pmatrix}
\begin{pmatrix} T_1 \\ T_2 \end{pmatrix}.
\end{align}
The second-order derivatives transform as
\begin{align}
\begin{pmatrix} T_{11} \\ T_{12} \\ T_{22} \end{pmatrix}
\mapsto
\begin{pmatrix}
c^2 & -2 c s    &  s^2 \\
c s &  c^2- s^2 & -c s \\
s^2 & 2c s      &  c^2 
\end{pmatrix}
\begin{pmatrix} T_{11} \\ T_{12} \\ T_{22} \end{pmatrix}.
\end{align}
The third-order derivatives transform as
\begin{align}
\begin{pmatrix} T_{111} \\ T_{112} \\ T_{122}\\ T_{222} \end{pmatrix}
\mapsto
\begin{pmatrix}
c^3   & -3 c^2 s        & 3 c s^2          & -s^3   \\
c^2 s &  c(c^2 - 2 s^2) & s( s^2- 2 c^2)   &  c s^2 \\
c s^2 &  s(2 c^2 - s^2) & c(c^2 -  2 s^2)  & -c^2 s \\
s^3   &  3 c s^2        & 3 c^2 s          & c^3 
\end{pmatrix}
\begin{pmatrix} T_{111} \\ T_{112} \\ T_{122}\\ T_{222} \end{pmatrix}.
\end{align}
The fourth-order derivatives transform as
\begin{align}
\begin{pmatrix} T_{1111} \\ T_{1112} \\ T_{1122}\\ T_{1222} \\ T_{2222} \end{pmatrix}
\mapsto
\begin{pmatrix}
c^4     & -4 c^3 s          & 6 c^2 s^2               &- 4 c s^3          &  s^4     \\
c^3 s   & c^2(c^2 - 3 s^2)  & 3cs(s^2 - c^2)     & s^2(3 c^2  - s^2) & -c s^3   \\
c^2 s^2 & 2cs( c^2 -  s^2)  & c^4  - 4 c^2 s^2  + s^4 & 2cs(  s^2 -  c^2) &  c^2 s^2 \\
c s^3   & s^2(3 c^2  - s^2) & 3cs( c^2 - s^2 )   & c^2(c^2  - 3 s^2) & -c^3 s   \\
s^4     & 4cs^3             & 6 c^2 s^2               & 4 c^3 s           &  c^4 
\end{pmatrix}
\begin{pmatrix} T_{1111} \\ T_{1112} \\ T_{1122}\\ T_{1222} \\ T_{2222} \end{pmatrix}.
\end{align}
\end{document}